\documentclass[aps,prd,nofootinbib,twocolumn,superscriptaddress,preprintnumbers,letterpaper,natbib,nofootinbib]{revtex4-2}
\pdfoutput = 1
\usepackage{amssymb}
\usepackage{amsmath}
\usepackage{epsfig}
\usepackage{hyperref}
\usepackage{color}
\usepackage{cleveref}
\usepackage{multirow}
\usepackage{capt-of}
\usepackage{siunitx}
\usepackage{graphicx}
\usepackage[caption=false]{subfig}
\usepackage{slashed}
\usepackage{tabularx}
\usepackage{enumitem}
\usepackage{multirow}
\usepackage{url}

\newcommand{\bea}{\begin{eqnarray}}
\newcommand{\eea}{\end{eqnarray}}
\newcommand{\al}[1]{\begin{align} #1 \end{align} }

\makeatletter
\def\simgt{\mathrel{\lower2.5pt\vbox{\lineskip=0pt\baselineskip=0pt
           \hbox{$>$}\hbox{$\sim$}}}}
\def\simlt{\mathrel{\lower2.5pt\vbox{\lineskip=0pt\baselineskip=0pt
           \hbox{$<$}\hbox{$\sim$}}}}
\makeatother

\newcommand{\B}{\mathcal{B}}
\newcommand{\Bp}{{\B^+}}
\newcommand{\Bpconj}{{\bar{\B}^-}}
\newcommand{\mes}{\mathcal{M}}
\newcommand{\ucbar}{\bar{u}^\text{c}}
\newcommand{\ccbar}{\bar{c}^\text{c}}
\newcommand{\ACP}{A_{\rm CP}}
\newcommand{\acpf}{a^f_{\rm CP}}
\newcommand{\ACPB}{{\tilde{A}_{\rm CP}}}
\newcommand{\acpBf}{\tilde{a}^f_{\rm CP}}
\newcommand{\Br}{\text{Br}}
\newcommand{\prn}[1]{ \left(  #1 \right) }

\newcommand\psiB{\psi_{\mathcal{B}}}
\newcommand\psiBbar{\bar{\psi}_{\mathcal{B}}}
\newcommand{\BctoB}{B_c^+ \to B^+ + f}
\newcommand{\BctoBconj}{B_c^- \to B^- + \bar{f}}
\newcommand{\BrBctoB}{\Br^f_{B_c^+}}
\newcommand{\Btopsi}{B^+ \to \psiBbar + \Bp}
\newcommand{\Btopsiconj}{B^- \to \psiB + \bar{\B}^-}
\newcommand{\BrBtopsi}{\Br_{B^+}^{\Bp}}
\newcommand{\Brmestoelld}{\Br_{\mes^+}^{\ell_d}}
\newcommand{\ztwo}{\mathbb{Z}_2}
\newcommand\phiB{\phi_{\mathcal{B}}}
\newcommand\phiBstar{\phi^*_{\mathcal{B}}}
\newcommand{\Neff}{N_{\rm eff}}
\newcommand{\ck}[1]{{\color{black}{#1}}}
\begin{document}

\title{
Charged \texorpdfstring{$B$}{B} Mesogenesis
}
\preprint{LCTP-21-24}
\preprint{MITP-21-041}
\author{Fatemeh Elahi}
\email{felahi@uni-mainz.de}
\affiliation{\footnotesize \sl PRISMA$^+$ Cluster of Excellence \& Mainz Institute for Theoretical Physics\\
Johannes Gutenberg University, 55099 Mainz, Germany\\}

\author{Gilly Elor}
\email{gelor@uni-mainz.de}
\affiliation{\footnotesize \sl PRISMA$^+$ Cluster of Excellence \& Mainz Institute for Theoretical Physics\\
Johannes Gutenberg University, 55099 Mainz, Germany\\}

\author{Robert McGehee}
\email{rmcgehee@umich.edu}
\affiliation{\footnotesize \sl Leinweber Center for Theoretical Physics, Department of Physics,\\ University of Michigan, Ann Arbor, MI 48109, USA}

\begin{abstract}
We leverage the CP violation in charged $B$ meson decays to generate the observed baryon asymmetry and dark matter at $\mathcal{O}(10 \text{ MeV})$ temperatures. We realize this in two scenarios: $B_c^+$ Mesogenesis and $B^+$ Mesogenesis. In the first, CP violating $B_c^\pm$ decays to $B^\pm$ mesons are followed by decays to dark and Standard Model baryons. In the second, CP violating $B^\pm$ decays to lighter charged mesons are accompanied by the latter's decays to dark and Standard Model leptons, which then scatter into the baryon asymmetry. $B_c^+$ Mesogenesis is actively being probed at Belle and LHCb, while $B^+$ Mesogenesis can be tested at colliders and sterile neutrino searches.  
\end{abstract}

\maketitle

\section{Introduction}
\label{sec:Intro}
How did we come to be here? Said quantitatively: what are the origins of the measured baryon asymmetry (BAU) and dark matter? The answer to this fundamental question still eludes us after decades of effort. 

Explanations of the BAU usually fall into one of two broad categories: \emph{electroweak baryogenesis}~\cite{Kuzmin:1985mm,Cohen:1990py,Cohen:1990it,Turok:1990in,Turok:1990zg,McLerran:1990zh,Dine:1990fj,Cohen:1991iu,Nelson:1991ab,Cohen:1992yh,Farrar:1993hn} or \emph{leptogenesis}~\cite{Fukugita:1986hr}. These ideas satisfy the three Sakharov conditions~\cite{Sakharov:1967dj} -- baryon number violation, C and CP Violation (CPV), and departure from thermal equilibrium -- in unique ways. Electroweak baryogenesis attempts to explain the BAU and satisfy these criteria using a strongly first order electroweak phase transition (EWPT), while leptogenesis uses out-of-equilibrium decays of heavy neutrinos already motivated by the seesaw mechanism~\cite{Yanagida:1980xy,Minkowski:1977sc,GellMann:1980vs}.

However, each of these answers suffers significant drawbacks. Many models of electroweak baryogenesis require fine tuning~\cite{Liebler:2015ddv}; construct extended Higgs sectors, but \emph{still can't} make the EWPT strongly first order~\cite{Kurup:2017dzf,Baum:2020vfl}; fail to actually produce the observed baryon asymmetry~\cite{Cline:2021dkf}; or are outright excluded by increasingly precise experimental results~\cite{Andreev:2018ayy}. Often, they simultaneously neglect dark matter.\footnote{Sometimes, dark matter may be explained ``after the fact'' (see \emph{e.g.}~\cite{Hall:2021zsk}).} Though the original formulations of electroweak baryogenesis were minimal, Nature increasingly seems to disfavor this now less-than-simple asymmetry generator, perhaps in favor of a similar mechanism in the dark sector~\cite{Shelton:2010ta,Servant:2013uwa,Cline:2017qpe,Carena:2018cjh,Hall:2019ank,Hall:2019rld}. Leptogenesis, in contrast, suffers not from experimental exclusion but from exclusion of experiment -- there is no hope of directly probing the heavy states in leptogenesis models~\cite{Buchmuller:2004nz}. It may be the true origin of the BAU, but humans may never know.

These substantial disadvantages of decades-old ideas should not be ignored. They portend eventual failure and sound a call to innovation. Answering this call, a new paradigm of low-scale baryogenesis has been proposed: \emph{Mesogenesis} \cite{Elor:2018twp,Elor:2020tkc}. In this framework, an out-of-equilibrium scalar decays to SM quarks which hadronize at low temperatures. The resulting SM mesons undergo known CP-violating processes and decay into dark sector particles carrying SM baryon or lepton number. The decays conserve baryon and lepton number and thus generate an equal and opposite baryon or lepton asymmetry between the dark and visible sectors. In the latter scenario, dark sector interactions then convert the lepton asymmetry into a baryon asymmetry. 

What is most compelling about Mesogenesis is that it revives an original, but long dead~\cite{Jarlskog:1985ht,Gavela:1993ts,Gavela:1994dt,Huet:1994jb}, hope of electroweak baryogenesis: that the requisite CP-violating processes already reside in the SM. In Mesogenesis, we look to the mesons. The BAU is directly proportional to CP-violating experimental observables, making Mesogenesis testable at current experiments. Furthermore, the dark sector typically contains a dark matter candidate whose abundance will be generated along with the baryon asymmetry. All of these mechanisms \emph{do not} violate $B$ or $L$, but rather, hide equal and opposite asymmetries in the dark sector. 

In neutral $B$ Mesogenesis \cite{Elor:2018twp}, the baryon asymmetry is generated by leveraging the CPV in $B^0_q$ particle/anti-particle oscillations, while in $D^+$ Mesogenesis \cite{Elor:2020tkc}, the CPV of $D^\pm$ meson decays is utilized. Both these flavors of Mesogenesis necessarily occur at low (5-20 MeV) scales, and are generically testable.\footnote{See \cite{Alonso-Alvarez:2021qfd} for a detailed study of experimental implications of decays into dark sector baryons and \cite{Hyperons} for additional indirect signals.} Indeed for the case of neutral $B$ Mesogenesis, some experimental searches and proposals are already in various stages at LHCb \cite{Borsato:2021aum,Rodriguez:2021urv} and Belle-I and -II \cite{Belle2}. 

But some of the simplest and most compelling Mesogenesis stories have, until now, been overlooked. Charged $B$ mesons contain a large amount of CPV in their decays to SM final states \cite{pdg}. In the present work, we introduce two scenarios of \emph{Charged $B$ Mesogenesis}, in which baryogenesis and dark matter production proceed by leveraging the CPV in charged $B_c^+$ or $B^+$ decays, which subsequently quickly decay into a dark sector state. What makes $B_c^+$ Mesogenesis so compelling is: 1) it is the simplest iteration of Mesogenesis to date and 2) it uses CPV in $B_c^+$ channels which are primed for exploration both experimentally and theoretically. On the other hand, the $B^+$ Mesogenesis scenario takes its inspiration from the aforementioned $D^+$ Mesogenesis. But out of these Mesogenesis possibilities, and indeed all other baryogenesis possibilities, $B^+$ Mesogenesis has the most hope of using the CPV \emph{in the SM alone} to generate the BAU. Both of these Charged $B$ Mesogenesis proposals motivate a host of new or improved experimental measurements of charged $B$ and lighter charged mesons. 

This paper is organized as follows. In Section \ref{sec:twoBs}, we summarize the common features of both charged $B$ Mesogenesis scenarios, as well as their differences. We move on to expound both $B_c^+$ and $B^+$ Mesogenesis\footnote{The ``$+$" in the naming convention is chosen for consistency with the definition of the charge asymmetry observable.} possibilities in Sections \ref{sec:Bcmeso} and \ref{sec:BMeso}, respectively. In Section \ref{sec:discovery}, we summarize the various experimental discovery prospects of Charged $B$ Mesogenesis.  We remark on future directions in Section \ref{sec:Outlook}. Appendices contain supplementary derivations and tables. 

\section{Two \texorpdfstring{$B$}{B}, or Not To Be}
\label{sec:twoBs}

All Mesogenesis constructions assume the existence of a $\sim 10-100$ GeV scalar field $\Phi$ which decays out of thermal equilibrium at late times into SM quarks $\Phi \to q_i \bar{q}_i$. The upper bound on the mass is to prevent a too small branching fraction of $\Phi$ to $b$ quarks (see \emph{e.g.} \cite{Djouadi:1997yw}). The lower bound is kinematic and permits $\Phi$'s decay to the pair of mesons under consideration in the particular Mesogenesis scenario. 

This decay occurs when the temperature of the Universe lies in the range
\bea
T_{\rm BBN} \,\, < \, T_R \, < \,\, T_{\rm QCD} \,.
\label{eq:TR}
\eea
The lower bound in Eq.~\eqref{eq:TR} ensures that the decay does not interfere with BBN, while the upper bound ensures that the produced quarks immediately hardronize into SM mesons. In charged $B$ Mesogenesis, $\Phi$ couples to $b$ quarks, resulting in the production of equal and opposite amounts of charged SM $B$ mesons.

\begin{figure*}[t]
\centering
\includegraphics[width=0.85 \textwidth]{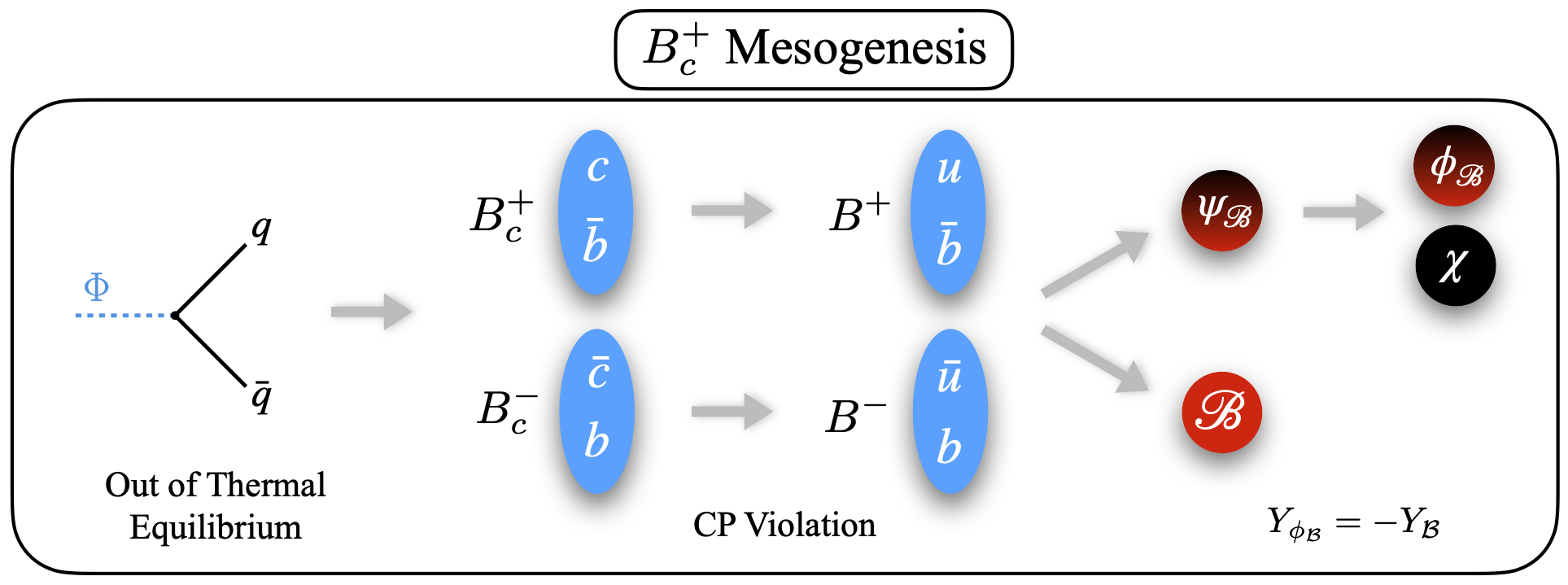}
\vspace{-0.3cm}
\caption{An illustration of the way in which $B_c^+$ Mesogenesis realizes the Sakharov conditions. Out-of-equilibrium $\Phi$ decays to $B_c^\pm$ mesons are followed by their CP-violating decays to $B^\pm$s. These in turn decay to both SM and dark baryons while preserving baryon number. The intermediate $\psiB$s quickly decay to $\ztwo$ odd $\chi$s and $\phiB$s which comprise up to $\sim 80 \%$ of dark matter. }
\label{fig:Bcillustr}
\end{figure*}

The evolution of $\Phi$ number density, as well as the radiation density, are governed by the interplay of the following Boltzmann equations and Hubble rate: 
\begin{align}
\label{eq:BERadPhiH}
 \frac{d n_\Phi}{d t} + 3 H n_\Phi &= -\, \Gamma_\Phi n_\Phi \,, \nonumber \\
\frac{d \rho_{\rm rad}}{dt} + 4 H \rho_{\rm rad} &= +\, \Gamma_\Phi m_\Phi n_\Phi \,,  \\
H^2  &= \frac{8 \pi}{3 M_{\rm Pl}^2}\prn{ \rho_{\rm rad} + m_\Phi n_\Phi } \,, \nonumber
\end{align}
where $\Gamma_\Phi$ is the decay width of $\Phi$. We assume that $\Phi$ was in equilibrium with the SM bath at some high temperature and has a full $T^3$ number density. The temperature at which $\Phi$ decays, i.e. the ``reheat temperature,'' can be defined in the usual way: $3 H(T_R) \equiv \Gamma_\Phi$. MeV scale reheat temperatures in Eq.~\eqref{eq:TR} then correspond to a decay width in the range $10^{-22} \, \text{GeV} \lesssim \Gamma_\Phi \lesssim 3 \times 10^{-21} \, \text{GeV}$. 

As $\Phi$ couples to quarks, one may wonder about any possible signature of $\Phi$ at the LHC. The production cross section of $\Phi$ at the LHC is roughly $\sigma_{qq \to \Phi} \sim g_{\Phi qq}^2/ m_\Phi^2$. Given the assumption above, that $\Gamma_\Phi \sim m_\Phi g_{\Phi qq}^2 \simeq 10^{-21}$ GeV, and that we expect $m_\Phi \sim O(10) $ GeV, we can roughly estimate the production cross section of $\Phi$ at the LHC: $\sigma_{qq \to \Phi} \lesssim 10^{-24} \, \text{GeV}^{-2}\sim 4 \times 10^{-10} \, \text{ab}$.
Therefore, even with $3 \, \text{ab}^{-1}$ of integrated luminosity, we do not expect $\Phi$ production at the LHC. Furthermore, even if some amount of $\Phi$s was produced, distinguishing it over background is inconceivable. The primary decay mode at the LHC of $\Phi$ is into two jets, making di-jet production with an invariant mass below $100$ GeV the dominant background. The cross section of this process at the LHC is enormous: $ 5.4 \times 10^{14} \text{ab}$~\cite{Alwall:2014,Alwall:2011}. Hence, $\Phi$ cannot be constrained at the LHC. 

At the low, MeV-scale reheat temperatures, the produced charged $B$ mesons will undergo SM decays into lighter charged and neutral mesons. Such decays contain CPV, parameterized by the experimentally observable charge asymmetry $\ACP$. Next, the charged meson daughter quickly decays into new dark sector states. We consider two possible scenarios: the second decay produces either 1) dark (and SM) baryons or 2) dark (and SM) leptons: 
\begin{itemize}
\item{\emph{Dark Baryons}: The daughter charged mesons undergo baryon-number-conserving decays into a dark anti-baryon and a SM baryon, directly generating equal and opposite dark and visible sector baryon asymmetries. This is kinematically possible if the first charged $B$ mesons are $B_c^+$, which then decay to $B^+$. We dub this scenario \emph{$B_c^+$ Mesogenesis}. }
\item{\emph{Dark Leptons}: Light daughter charged mesons can undergo lepton-number-conserving decays into a light dark anti-lepton and a SM lepton. The resulting generated lepton asymmetry is then transferred to the baryon asymmetry via dark sector scatterings. This scenario is interesting when the first charged $B$ mesons are $B^+$, which then decay to charged $D$ mesons, kaons, and pions. We dub this scenario \emph{$B^+$ Mesogenesis}. }
\end{itemize}

In both of these Charged $B$ Mesogenesis scenarios, there is a lingering dark sector baryon asymmetry equal and opposite to the BAU. Thanks to lower bounds on (dark) baryon masses, this dark baryon asymmetry is always guaranteed to comprise at least $\sim 20 \%$ of dark matter, perhaps even all of it, depending on the masses of the dark sector states. In what follows, we describe the mechanisms, parameter spaces, current constraints and signals of these two distinct Charged $B$ Mesogenesis frameworks.

\section{\texorpdfstring{$B^+_c$}{Bc} Mesogenesis}
\label{sec:Bcmeso}

In $B_c^+$ Mesogenesis, the BAU is generated from the decays: 
\begin{subequations}\label{eq:BcMech}
\begin{align}
B_c^+ \to \, & B^+ + f \,, \label{eq:BcMech1}\\
& B^+ \to \, \psiBbar + \mathcal{B}^+, \label{eq:BcMech2}
\end{align}
\end{subequations}
where $f$ is a neutral light meson, $\Bp$ is a charged SM baryon, and $\psiB$ is a dark sector Dirac fermion with baryon number $B = 1$. The CPV in the first decay satisfies one of Sakharov's conditions and could have both SM and new physics contributions. See \emph{e.g.} \cite{Choi:2009ym} for a list of the nine expected SM decays Eq.~\eqref{eq:BcMech1}. For a particular final state $f$, this CPV is parameterized by the charge asymmetry observable:
\begin{align}
\ACP^f = \frac{\Gamma(B^+_c \to f) - \Gamma(B^-_c \to \bar{f})}{\Gamma(B^+_c \to f) + \Gamma(B^-_c \to \bar{f})} \,.
\label{eq:ACPgen}
\end{align}

The produced $B^+$ quickly decays into a SM charged baryon $\Bp$ and dark sector anti-baryon $\psiBbar$. Note that \emph{this decay conserves baryon number}. The net result of both decays in Eq.~\eqref{eq:BcMech} is the generation of equal and opposite baryon asymmetries between the dark and SM sectors. In fact, the SM baryon yield, $Y_{\B}$, is proportional to experimental observables in $B_c^+$ and $B^+$ decays: 
\bea
\label{eq:YBexpobs}
Y_{\B} &\equiv& \frac{ n_{\B} - n_{\bar{\B}} }{s} \propto \sum_{f} \acpf \BrBctoB \times \sum_{\Bp} \BrBtopsi, \\
\acpf &\equiv& \ACP^f / \prn{1+\ACP^f}, \nonumber \\
\BrBctoB  &\equiv& \Br\prn{\BctoB}, \nonumber \\
\BrBtopsi &\equiv& \Br\prn{\Btopsi}. \nonumber
\eea
Above, $s$ is the entropy density in the SM bath.

To prevent proton decay, we require\footnote{Neutron stars may place a slightly tighter bound, but have inherent astrophysical and model uncertainties~\cite{McKeen:2018xwc}, so we ignore these for now. Another bound comes from the lack of the decay of $Be^9$ into dark baryons\cite{McKeen:2020oyr}, which is slightly stronger than the proton mass. }
\bea
m_{\psi_B} > m_p - m_e \simeq 937.8 \, \text{MeV} \,.
\label{eq:mpsikin}
\eea
This lower mass bound instead permits $\psiB$ to decay to a proton, electron, and neutrino.\footnote{Strictly speaking, there is a fine-tuned possibility that $\psiB$ satsfies Eq.~\eqref{eq:mpsikin}, but still cannot decay to a proton and electron. In this sliver of parameter space, $\psiB$ is stable, additional dark sector states are unnecessary, and $\psiB$ could cause neutron decays which may address its lifetime anomaly (see \emph{e.g.}~\cite{Cline:2018ami}). However, we do not consider this further.} This decay could washout the generated baryon asymmetry. To prevent this, we minimally expand the dark sector to allow $\psiB$ to rapidly decay into additional dark sector states (similar to the setup in \cite{Elor:2018twp}). We add two more dark sector particles: a Dirac fermion $\chi$ and a complex scalar $\phiB$ with $B=1$, to allow $\psiBbar$ to quickly decay:
\bea
\label{eq:DarkDecay}
\psiBbar \to \phiBstar + \bar{\chi} \,.
\eea

To stabilize $\phiB$, we introduce a $\ztwo$ symmetry under which $\chi$ and $\phiB$ are odd and $\psiB$ is even and require
\bea
\left|m_{\phiB} - m_\chi \right| \,<\, m_p + m_e\,.
\eea
The Lagrangian term
\bea
\mathcal{L}_d \, = \, y_d \, \psiBbar \, \phiB \, \chi \,,
\eea
is allowed by all the symmetries and mediates the decay Eq.~\eqref{eq:DarkDecay}.

Since the $\psiBbar$ decay occurs quickly, its dark anti-baryon asymmetry is simply transferred to $\phiBstar$. This fixed asymmetry in $\phiBstar$ (and $\bar{\chi}$) then comprises up to $\sim 80 \%$ of dark matter. The symmetric components of $\phiB$ and $\chi$ tend to be overproduced, but may be sufficiently depleted by dark sector annihilations. We assume this and don't comment further since it has no bearing on the Mesogenesis mechanism\footnote{For details on depleting the symmetric abundances, see \cite{Elor:2018twp}.}.

However, the asymmetries in $\phiBstar$ and $\bar{\chi}$ cannot account for the entirety of dark matter since $B^+$ doesn't have enough mass to decay to both $\sim 5 \text{ GeV}$ of asymmetric dark matter and a SM baryon simultaneously. Thus, between $\sim 20 - 80 \%$ of dark matter has to be outside of the asymmetric components of $\chi$ and $\phi_B$. The precise amount of other dark matter is solely a function of $m_{\phiB}$ and $m_{\chi}$, since their asymmetries are just opposite the BAU. There are two simple possibilities: 1) the rest of dark matter is from a \emph{symmetric} amount of $\chi$s and $\phi_B$s or 2) the rest of dark matter is just some other dark sector state(s), unrelated a priori to the $B_c^+$ Mesogenesis scenario. 

Since either of these dark matter choices is not essential to $B_c^+$ Mesogenesis, we relegate further discussion to App.~\ref{sec:restofDM}. Fig.~\ref{fig:Bcillustr} summarizes the mechanism. With this bird's eye view, we proceed to detail a simple UV model.

\subsection{UV Model}
\label{subsec:UVmodelBc}
The decay in Eq.~\eqref{eq:BcMech2} 
proceeds through a dimension six, four fermion operator. 
Following the UV model of \cite{Elor:2018twp}, we add a colored triplet scalar $\phi$ with electric charge  assignment $Q_{\rm EM} = -1/3$ and baryon number $B=-2/3$. The following Lagrangian is then allowed by all the symmetries: 
\bea
\hspace{-0.1in}
\mathcal{L}_\phi &=& 
 - \! \! \sum_{i,j} y_{ij} \phi^* \bar{u}_{iR} d_{jR}^c - \! \! \sum_k y_{\psiB k} \phi d_{kR}^c \psiB + \text{h.c.}, 
\label{eq:UVmodel}
\eea
where the flavor indices $i,j,k$ account for all flavorful  variations of this model, as there is no a priori reason to assume a specific flavor structure. Such a model has a simple Supersymmetric realization \cite{Alonso-Alvarez:2019fym} where the mediator $\phi$ can be identified with a right handed squark. As such, $\phi$ is constrained by collider searches for Supersymmetric particles and must be heavier than about 1 TeV (see \cite{Alonso-Alvarez:2021qfd} for detailed bounds from colliders and flavor observables). 

Integrating out the heavy $\phi$, we arrive at the following operator which mediates meson decays:
\bea
\mathcal{O} = \frac{y^2}{M_\phi^2} \psiBbar b \ucbar_i d_j + \text{h.c.}\,,
\label{eq:generalOp}
\eea
where $y^2 \equiv y_{ij} \, y_{\psiB 3}$. This particular flavor structure is all that is necessary for $B_c^+$ Mesogenesis, but could be part of a larger UV model with other non-zero Yukawas as in Eq.~\eqref{eq:UVmodel}. Note that this operator \emph{conserves baryon number}. It mediates the parton level decay $\bar{b}  \to \psiBbar u_i d_j$ within the meson decay Eq.~\eqref{eq:BcMech2}. There are four possible flavorful variations of Eq.~\eqref{eq:generalOp} leading to different final state SM baryons from the $B^+$ decay. Table \ref{tab:hadronmasses} summarizes these four possible decay modes. Eq.~\eqref{eq:generalOp} also gives rise to decays of neutral $B^0_{s,d}$ mesons and $b$-flavored baryons which can be used to indirectly probe the mechanism (see Table I of \cite{Alonso-Alvarez:2021qfd}).  

\begin{table}[t]
\renewcommand{\arraystretch}{1.4}
  \setlength{\arrayrulewidth}{.20mm}
\centering
\small
\setlength{\tabcolsep}{0.45 em}
\begin{tabular}{|c | c | c |}
    \hline
    $\,\,$ Interaction  $\,\,$ & $\,\,$Parton decay $\,\,$ & $B^+$ decay \\ \hline
    \hline
    $\psiBbar \, b\, \ucbar \, d$ & $\bar{b}\to \, \psiBbar \, u\, d$ &  $ B^+ \to \psiBbar + p^+ \,(uud) \,\, $ \\ \hline
    $\psiBbar \, b\, \ucbar \, s$ & $\bar{b}\to \psiBbar \, u\, s$ & $\, B^+ \to \, \psiBbar + \Sigma^+ \,(uus)\,\,$ \\ \hline
    $\psiBbar \, b\, \ccbar \, d$ & $\bar{b}\to \psiBbar \, c\, d$ & $ \, B^+ \to \, \psiBbar + \Lambda_c^+ \,(ucd)\,\,$ \\ \hline
    $\psiBbar \, b\, \ccbar \, s$ & $\bar{b}\to \psiBbar \, c\, s$ & $\, B^+ \to  \,  \psiBbar + \Xi^+_c \,(ucs)\,\,$ \\ \hline
\end{tabular}
\caption{Here we present the four different flavorful variations of the operator Eq.~\eqref{eq:generalOp}, and the corresponding parton-level decays and final state hadron decay products. Constraints on the branching fraction for each operator can be found in ~\cite{Alonso-Alvarez:2021qfd}.
}
\label{tab:hadronmasses}
\end{table}

\subsection{Results}
The Boltzmann equations for the BAU are greatly simplified since all the decays in Eq.~\eqref{eq:BcMech} occur very quickly at MeV temperatures. The evolution of the baryon asymmetry is then governed by
\bea
\label{eq:BEbaryonasym}
\frac{d}{dt}\prn{n_{\B} - n_{\bar{\B}}} &&+ 3 H \prn{n_{\B} - n_{\bar{\B}}} =  \\ \nonumber
&&- \, 2 \Gamma_{\Phi}^B n_\Phi \sum_{\Bp} \BrBtopsi \sum_{f} \acpf \BrBctoB \,,
\eea
where we have defined $\Gamma_\Phi^B \equiv \Gamma_\Phi \Br(\Phi \to q ) \Br(q \to B_c)$.
See App.~\ref{app:BoltzDerive} for details (as well as \cite{Elor:2020tkc}). 

We numerically integrate Eq.~\eqref{eq:BEbaryonasym} while tracking $\Phi$, Hubble (see Eq.~\eqref{eq:BERadPhiH}), and the particles in the decays of Eqs.~\eqref{eq:BcMech} and \eqref{eq:DarkDecay}. We allow the values of the experimental observables $\sum_{\Bp} \BrBtopsi$ and $\sum_f \acpf \BrBctoB$ to be free parameters and find:
\al{
\label{eq:YBscaling}
\frac{Y_{\mathcal{B}}}{Y_{\mathcal{B}}^{\rm obs}}   \simeq \frac{\sum_{\Bp} \BrBtopsi}{10^{-3}}   \frac{\sum_f \acpf  \BrBctoB}{ 6.45 \times 10^{-5}} \frac{T_R}{20 \, \text{ MeV}} \frac{2 m_{B_c^+}}{m_\Phi} \, ,
}
where $Y_{\mathcal{B}}^{\rm obs}=8.69 \times 10^{-11}$ is the observed baryon asymmetry today~\cite{Zyla:2020zbs}.

\begin{figure}[t!]
\centering
\includegraphics[width=\columnwidth]{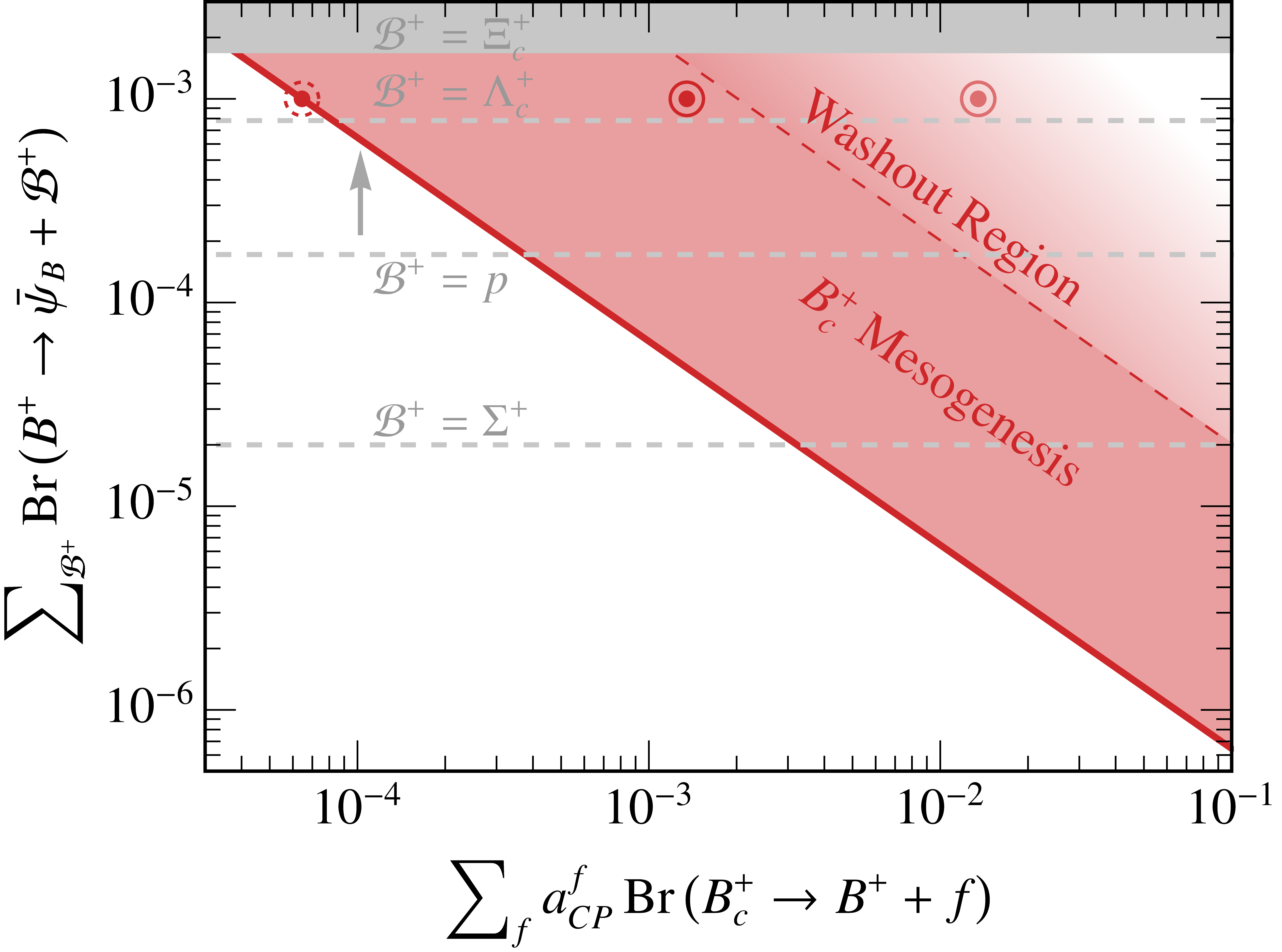}
\vspace{-0.4cm}
\caption{Viable parameter space for $B_c^+$ Mesogenesis in red. Current constraints for different final-state $\Bp$ are shown in gray. Three circled benchmark points are discussed more in the text and highlighted in Fig.~\ref{fig:Bcpbenchmark}.}
\label{fig:Bcparamsp}
\end{figure}

The viable parameter space where $B_c^+$ Mesogenesis successfully produces the observed BAU is shown in red in Fig.~\ref{fig:Bcparamsp} as a function of the experimental observables $\sum_{\Bp} \BrBtopsi$ and $\sum_f \acpf \BrBctoB$. The various dashed gray lines show the upper bounds on $\BrBtopsi$ for the different possible final-state SM baryons shown in Table~\ref{tab:hadronmasses}. The weakest bound corresponds to $\Bp = \Xi_c^+$ and is thus shaded gray. 

These same decays arise as a byproduct of the neutral $B$ Mesogenesis mechanism \cite{Elor:2018twp} and have been extensively studied in \cite{Alonso-Alvarez:2021qfd}. In particular, the most constraining limit on the branching fractions of charged $B$ mesons decaying into charged $p$, $\Lambda_c$, or $\Xi_c$ and missing energy, $\Br\prn{ B^+ \to \Bp + \text{MET} }$, was found in \cite{Alonso-Alvarez:2021qfd} by recasting an analysis of an old search by the ALEPH collaboration at LEP \cite{ALEPH:2000vvi}. The bound on decays to final state $\Sigma^+$ is the result of a recent study by the Belle collaboration \cite{Belle:2021gmc}. The maximal allowed branching fraction for each of the decay modes in Table~\ref{tab:hadronmasses} ranges from $10^{-4} - 10^{-2}$ depending on the dominating operator and $\psiB$ mass (see Fig.~5 of \cite{Alonso-Alvarez:2021qfd}). For concreteness, we have set $\psiB = 2 \text{ GeV}$, which only impacts the strength of the gray bounds in Fig.~\ref{fig:Bcparamsp}. 

There are currently no stringent constraints, nor robust SM or new physics predictions, for the observables in the $B_c^+$ decays: $\sum_f \acpf \, \BrBctoB$. We therefore emphasize that any measurement of these observables will be a critical step towards confirming $B_c^+$ Mesogenesis. We defer a detailed discussion of current status and future prospects of these observables to Sec.~\ref{subsec:BcSignals}.

We did not include any scattering or annihilation terms in Eq.~\eqref{eq:BEbaryonasym}. At high enough temperatures, both the $B_c^+$ and $B^+$ can annihilate which will washout some of the generated asymmetry. The lifetime of the $B_c^+$ meson is roughly $\tau_{B_c} = 7.9 \times 10^{8} \, \text{MeV}^{-1}$ while that of the $B^+$ meson is about $\tau_B = 2.4 \times 10^9 \, \text{MeV}^{-1}$. We thus find that meson decays dominate over annihilations as long as temperatures are $\lesssim 20 \, \text{MeV}$ \cite{Elor:2020tkc}. The viable parameter space in Fig.~\ref{fig:Bcparamsp} corresponds to a scan over $T_R$ with  $T_R^\text{max}=20 \text{ MeV}$ and $T_R^\text{min}=5 \text{ MeV}$. Likewise, we scan over the full range of possible $\Phi$ masses from $m_\Phi^\text{min} = 2 m_{B_c^+}$ to $m_\Phi^\text{max} = 100 \text{ GeV}$.

For reheat temperatures in the range $20 \, \text{MeV} \lesssim T_R \lesssim T_{\rm QCD}$, $B_c^+$ Mesogenesis can still explain the BAU. Indeed, the ``Washout Region'' in Fig.~\ref{fig:Bcparamsp} is viable parameter space in which the BAU is initially overproduced. This excess asymmetry can be depleted by washout effects simply by raising the reheat temperature. For $T_R \gtrsim 20 \, \text{MeV}$, $B^+$ mesons start scattering and annihilating significantly before they have the chance to decay to the dark sector, suppressing the initial asymmetry generation provided by the CP-violating $B_c^+$ decays. This causes the final dark sector baryon asymmetry, and consequently, the BAU, to be much smaller than approximated in Eq.~\eqref{eq:YBscaling}.

However, the validity of our simplified Boltzmann equations breaks down when $T_R \gtrsim 20 \, \text{MeV}$, since we've assumed such scatterings are negligible. A detailed numerical solution of the Boltzmann equations of Charged $B$ Mesogenesis in the presence of washout terms is beyond the scope of this work. We leave a quantitative  investigation of this part of parameter space to future work.

\begin{figure}[t!]
\centering
\includegraphics[width=\columnwidth]{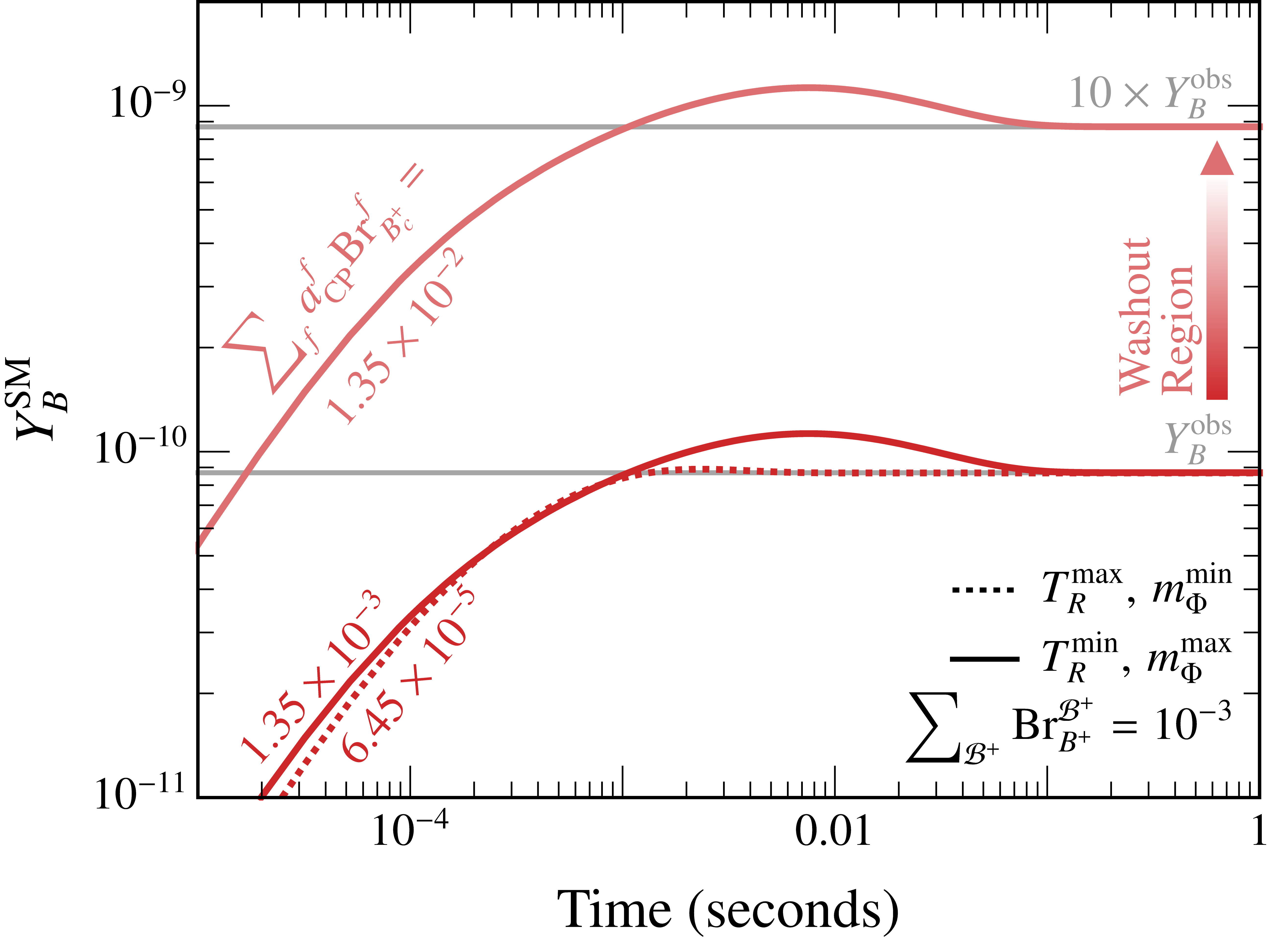}
\vspace{-0.3cm}
\caption{Benchmark points of $B_c^+$ Mesogenesis from the viable parameter space in Fig.~\ref{fig:Bcpbenchmark} which (over)generate the observed SM baryon asymmetry.}
\label{fig:Bcpbenchmark}
\end{figure}

For illustrative purposes, we also circle three representative benchmark points in Fig.~\ref{fig:Bcparamsp} and show the evolution of the BAU corresponding to each in Fig.~\ref{fig:Bcpbenchmark}. Two of these curves correspond to the extremal values of $(m_\Phi, T_R)$ with the free experimental observables set to achieve $Y_{\mathcal{B}}= Y_{\mathcal{B}}^{\rm obs}$. We also show a benchmark point with $\sum_f \acpf \BrBctoB$ larger by a factor of 10 which overproduces the BAU by a factor of 10, reinforcing the approximate scaling in Eq.~\eqref{eq:YBscaling}. This point is firmly in the ``Washout Region'' and demonstrates that an initial BAU in excess of the observed BAU is easily possible. To achieve the correct BAU for such a point, we need only increase $T_R$ enough for washout effects to sufficiently deplete the initially excessive BAU. The ``bump'' in all three yield curves result from the effects of $\Phi$ contributing sizeably to Hubble in Eq.~\eqref{eq:BERadPhiH}.  The larger $T_R$, the less pronounced the bump because $\Phi$ decays earlier in the evolution of the baryon asymmetry.

\subsection{Signals, Searches, and Prospects}
\label{subsec:BcSignals}
The experimental observables of $B_c^+$ Mesogenesis which parametrize the $B_c^+$ and $B^+$ decays and control the BAU generation are defined in Eq.~\eqref{eq:YBexpobs}. We discuss the current prospects for observing them, as well as theoretical progress computing their values in the SM and beyond. 

\vspace{0.1in}
{\bf $\mathbf{B^+_c}$ Decays:} The relevant observables in the $B^+_c$ decays of Eq.~\eqref{eq:BcMech1} are: the CPV, $A^f_{\rm CP}$, and branching ratio for different decay modes, $\BrBctoB$.  

The LHC experiments ATLAS, CMS, and LHCb have all measured the $B_c^+$ mass and lifetime \cite{Anderlini:2014dha}. LHCb, in particular, is well suited for conducting searches of $B_c^+$ decays \cite{Yuan:2014mya,Tuning:2013nio}. There have been numerous measurements at LHCb of the branching fraction of fully hadronic $B_c^+$ decays which involve weak transitions of a $b$ to a $c$-quark \emph{e.g.} \cite{LHCb:2012ihf,LHCb:2012ag,LHCb:2013kwl,LHCb:2017lpu}. The decays relevant for $B_c^+$ Mesogenesis involve a $B^+$ meson in the final state i.e. the $b$-quark acts as a spectator. LHCb has made such a measurement, of $B_c^+ \to B^0_s \, \pi^+$ \cite{LHCb:2013xlg}. 

But to date, no such searches exist for similar modes more directly relevant for $B_c^+$ Mesogenesis, and both $\acpf$ and $\BrBctoB$ have not yet been measured. Hopefully, these will become possible with increased luminosity at LHCb \cite{Gouz:2002kk}. Additionally, an electron Future Circular Collider would be well equipped to conduct such searches as part of their broader $B_c^+$ physics program \cite{FCC:2018byv}. $B_c^+$ Mesogenesis directly links the generation of the BAU to these observables and thus strongly motivates looking for these yet undiscovered CP-violating $B_c^+$ decays. Regarding $\ACP^f$, most studies of the CPV within SM decays consider decays to $D^+$ and $K^+$, which are expected to have sizable CPV \cite{PhysRevD.62.057503}, rather than $B^+$. There are currently no measurements of $\ACP^f$ for the decay modes relevant for $B_c^+$ Mesogenesis. For this reason, our parameter space in Fig.~\ref{fig:Bcparamsp} was fully unconstrained in the $\acpf \BrBctoB$ direction.

On the theory side, there are SM predictions for the branching fraction of $B_c^+ \to B^+ + f$; see \cite{Choi:2009ym} and references therein. These predictions can be sizable (as large as  $3\%$), but they also vary greatly in the literature. While the theoretical uncertainty warrants caution, for illustrative purposes, note that fixing the $\BrBctoB$ to be $3\%$ predicts $\ACP \gtrsim 10^{-4}-10^{-2}$ for successful $B_c^+$ Mesogenesis. While there are no SM predictions for $\ACP^f$ in the decay modes of interest, such a value for $\ACP$ is not unreasonable to expect. For instance, in the decay $B_c^+ \to B^+ + \bar{K}^0$, the $b$ quark acts as a spectator. So, the CPV is expected to be on the order of other processes in the charm sector which can be sizeable. Furthermore, there could be contributions from new physics to such decays. A calculation of the SM CPV and branching fractions, as well as possible contributions from new physics, is interesting and highly motivated by $B_c^+$ Mesogenesis. We leave this to future work. 

\begin{figure*}[t]
\centering
\includegraphics[width=0.9 \textwidth]{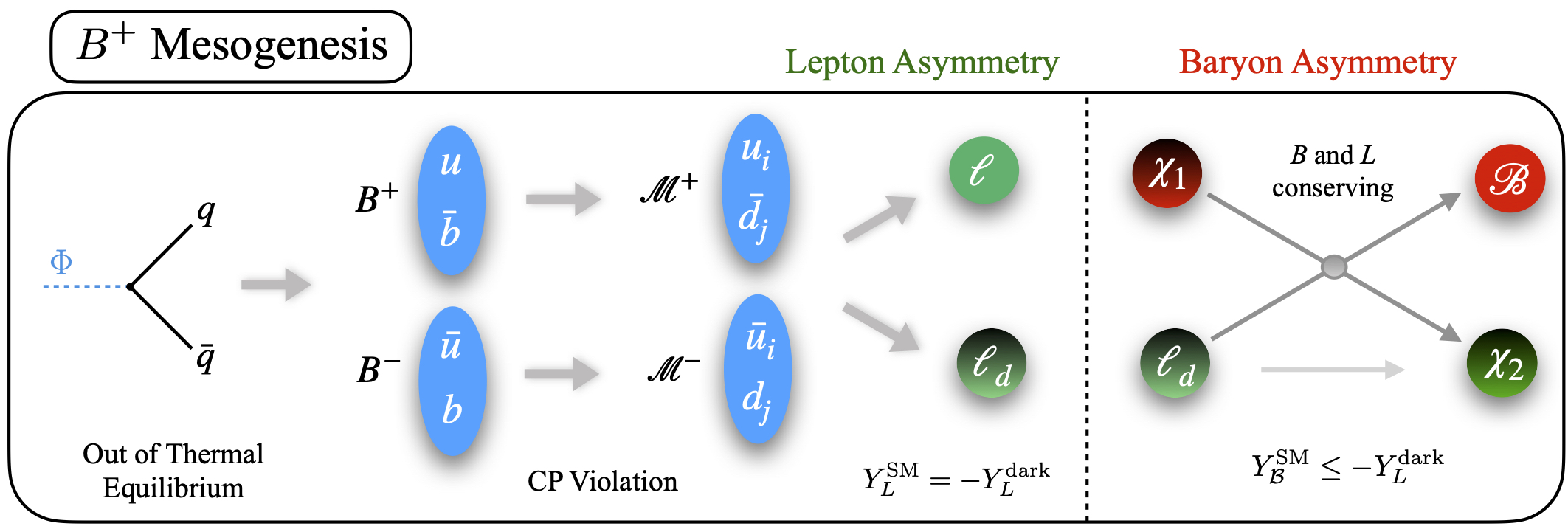}
\vspace{-0.3cm}
\caption{An illustration of the way in which $B^+$ Mesogenesis realizes the Sakharov conditions. At MeV scales, $B^\pm$ mesons are produced and undergo CP violating SM decays to charged mesons $\mes^\pm = \left\{\pi^\pm, K^\pm, D^\pm, D^\pm_s, K^{\ast +} \right\}$. The charged mesons subsequently decay into a dark lepton generating an equal and opposite dark and visible lepton asymmetry. Dark sector scatterings involving dark states carrying lepton and baryon number then transfer the lepton asymmetry into an equal and opposite dark and SM baryon asymmetry. 
}
\label{fig:Billustr}
\end{figure*}

\vspace{0.1in}
{\bf $\mathbf{B^+}$ Decays:}
The decay of the $B^+$ meson into a SM and dark baryon pair arises through the operator in Eq.~\eqref{eq:generalOp}. Four different final-state SM baryons, corresponding to different flavorful variations of Eq.~\eqref{eq:generalOp}, are summarized in Table \ref{tab:hadronmasses}. These are exactly the same decays that arise as a byproduct of the neutral $B$ Mesogenesis mechanism \cite{Elor:2018twp}, for which signals have been extensively studied in \cite{Alonso-Alvarez:2021qfd}. As such, we conclude that measuring the branching fraction $\Br_{B^+}$ of the decays in Table~\ref{tab:hadronmasses} is currently within reach of $B$ factories and hadron colliders. In fact, some of the operators are already being probed by Belle-II and may soon also be tested at LHCb. In particular, the Belle-II collaboration has developed a search \cite{Belle2} for the process $B_s^0 \to \psiBbar + \Lambda^0$ which arises from the $bud$ operator. This represents an indirect probe of $B^+_c$ Mesogenesis, as the same operator gives rise to $B^+ \to \, \psiBbar \, p^+ + \text{MET}$ which directly controls the BAU in our current setup.\footnote{See Table I of \cite{Alonso-Alvarez:2021qfd} for all possible direct and indirect decays arising from Eq.~\eqref{eq:generalOp}.} 

The searches for (apparent) baryon-number-violating decays of neutral $B^0$ are currently underway. However, the analogous searches for $B^+$ decays to directly probe $B$-Mesogenesis are not only within reach of current experiments, but are likely easier to trigger on due to the charged track, making the $B^+$ decays easier to reconstruct than their neutral counterparts. Other indirect probes of both $B_c^+$ and $B^0$ Mesogenesis are the decays of $b$-flavored hadrons to light mesons and missing energy (see Table I of \cite{Alonso-Alvarez:2021qfd}). A search for such decay modes has been developed by LHCb \cite{Rodriguez:2021urv}, and will hopefully be implemented soon. 

Additionally, the UV model in Eq.~\eqref{eq:UVmodel} can give rise to fully invisible decays of $b$-flavored baryons or their decays to pions and missing energy or photons and missing energy. Such decays have been explored in \cite{Hyperons} and can probe a complementary region of parameter space for both $B^0$ and $B_c^+$ Mesogenesis. Similarly  new hyperon decay modes can serve as indirect probes \cite{Alonso-Alvarez:2021oaj}.

\vspace{0.1in}
{\bf Summary:} There is not a plethora of current constraints on $\BctoB$ decays, nor robust SM or beyond-SM predictions for the CP asymmetry and branching fraction of these processes--- which directly control the BAU in $B^+_c$ Mesogenesis. We also have no reason to expect these observables to be too small to generate the observed BAU and the required value of the product of $\ACP \times \Br$ in Fig.~\ref{fig:Bcparamsp} seems reasonable. As such, $B^+_c$ Mesogenesis highly motivates both the search for and theoretical computation of these decays. Regarding the branching fractions of $B^+$ decays, such measurements are within reach of current hadron colliders and $B$ factories. In particular, the same UV model that gives rise to neutral $B$ Mesogenesis also gives rise to $B_c^+$ Mesogenesis. As such, it is noteworthy that these ongoing searches are currently exploring this new mechanism, at no added charge!

\section{\texorpdfstring{$B^+$}{B} Mesogenesis}
\label{sec:BMeso}

In $B^+$ Mesogenesis, a lepton asymmetry is first generated from the decays: 
\begin{subequations}\label{eq:BMech}
\begin{align}
 B^+ \, \to \, &\mes^+ \,+\,  \mes \,, \label{eq:BMech1}\\
&\mes^+  \, \to \, \ell_d \,+\, \ell^+ \,, \label{eq:BMech2} 
\end{align}
\end{subequations}
where $\mes^+$ is a charged SM meson: $\pi^+$, $K^+$, $D^+$, $D_s^+$ or a resonant meson $K^{\ast +}$, $D^{\ast +}$; $\ell_d$ is a dark lepton with SM lepton number $L = 1$ and mass $m_{\ell_d} < m_{\mes^+} - m_\ell$; and the SM charged lepton $\ell^+$ can be a positron, antimuon, or antitau (in the case of $D^+$ and $D_s^+$ decays). This is an analogous setup to the $D^+$ Mesogenesis mechanism of \cite{Elor:2020tkc}. 

The initial SM decay of the $B^+$ meson in Eq.~\eqref{eq:BMech1} contains CPV, measured by the charge asymmetry observable: 
\bea
\ACPB^f = \frac{\Gamma \prn{B^+ \to f } - \Gamma \prn{ B^- \to f }}{\Gamma \prn{B^+ \to f } + \Gamma \prn{ B^- \to f }} \,.
\eea
This is analogous to the CP-violating observable from $B_c^+$ Mesogenesis, defined in Eq.~\eqref{eq:ACPgen}. To help distinguish the two, we refer to the CPV relevant for $B^+$ Mesogenesis with a ``\, $\tilde{\null}$ \,". We define $\Br_{B^+}^f \equiv \Br \prn{ B^+ \to f }$. The relevant decay modes are summarized in the tables of App.~\ref{app:BdecayModes} which also include the current limits on $\ACPB^f$ and $\Br_{B^+}^f$.

Given a sizable $\ACPB$ in Eq.~\eqref{eq:BMech1}, the subsequent decay of $\mes^+$ into a dark lepton $\ell_d$ in Eq.~\eqref{eq:BMech2} results in the  generation of a dark lepton asymmetry $Y_{\ell_d} \equiv \prn{n_{\ell_d} - n_{\bar{\ell_d}} } / s$ that is equal and opposite to a SM lepton asymmetry $Y_L^{\rm SM} = - Y_{\ell_d}$. Note that this process does not violate lepton number. The generated lepton asymmetry is then related to experimental observables as follows 
\bea
\label{eq:BplusYell}
Y_{\ell_d} \,\, \propto \,\, \sum_{\mes^+} \Brmestoelld \, \sum_f \ACPB^f \, \Br_{B^+}^f \,,
\eea
where $\Brmestoelld \equiv \Br \prn{ \mes^+ \to \ell_d + \ell^+ }$. 

The generated lepton asymmetry may then be transferred to a baryon asymmetry via dark sector scatterings off two additional states in the dark sector, $\chi_1$ and $\chi_2$, 
\al{
\label{eq:DarkScat}
\ell_d + \chi_1 \, \rightarrow \, \chi_2 + \mathcal{B}.
}
$\chi_1$ and $\chi_2$ are appropriately charged under baryon and lepton number so that this scatter conserves both. We assume an initial $\chi_1$ number density is produced from $\Phi$ decays. We additionally require that the scattering rate $\langle \sigma v \rangle$ for this process is sufficiently large to efficiently transfer the lepton asymmetry at $T_R$. The possible charge assignments and models giving rise to Eq.~\eqref{eq:DarkScat} were studied in \cite{Elor:2020tkc}. The same assignments and models work equally well for $B^+$ Mesogenesis so we do not comment on them further. We simply require 
\bea
\label{eq:LtoBob}
Y_L / Y_B^{\rm obs} \geq 1\,.
\eea
$B^+$ Mesogenesis is summarized in Fig.~\ref{fig:Billustr}. 

\begin{figure}[t]
\centering
\includegraphics[width=\columnwidth]{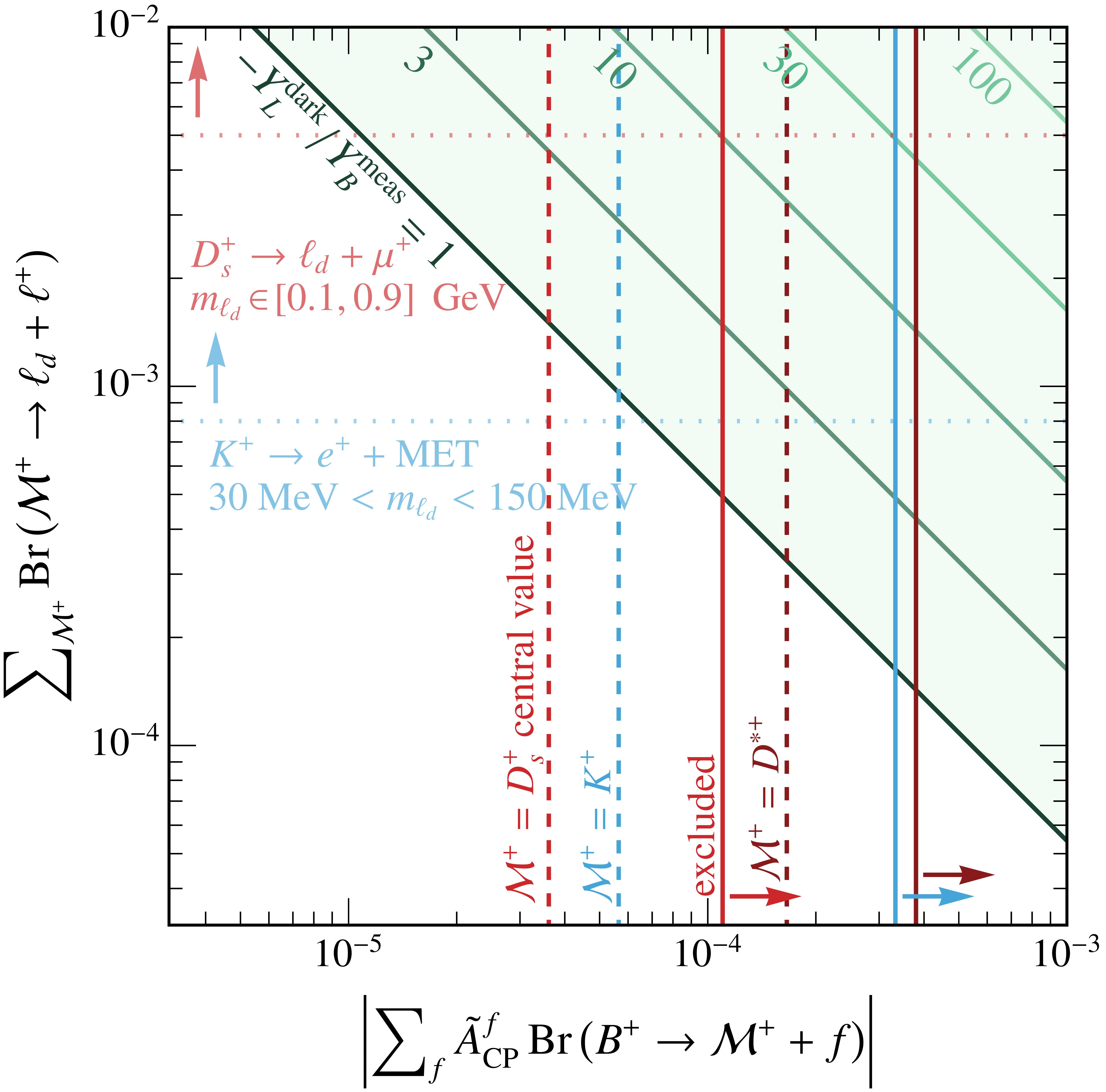}
\vspace{-0.4cm}
\caption{Same as Fig.~\ref{fig:B+ParamSpace1} for $\mes^+ = K^+$, $D_s^+$, as well as the resonance $D^{*+}$. Channels shown here have a net negative central value for the summed CPV.}
\label{fig:B+ParamSpace2}
\end{figure}

Just as in the $B_c^+$ scenario above, a baryon asymmetry equal and opposite to the BAU will remain in the dark sector in whichever of $\chi_1$ or $\chi_2$ has baryon number. It is guaranteed to be at least $\sim 20 \%$ of dark matter due to lower bounds on the mass of baryons. Unlike the $B_c^+$ scenario above though, these dark baryons are being sourced directly from $\Phi$s and may have masses as large as $5 \text{ GeV}$, and can therefore comprise all of dark matter. Given a point in parameter space that achieves the correct baryon asymmetry, the correct dark matter abundance can be produced by adjusting $(m_{\chi_1}, m_{\chi_2}, \langle \sigma v \rangle)$. The discussion of the dark sector parallels that of \cite{Elor:2020tkc} and we refer the reader to that work for more details. 

The decay of $\mes^\pm$ proceeds through an effective operator of the form 
\bea
\mathcal{O} = \frac{c_{ij}}{\Lambda^2} \Bigl[ \bar{d}_i \Gamma^\mu u_j\Bigr] \Bigl[ \bar{\ell}_d \Gamma_\mu \ell^- \Bigr] + \text{h.c.}\,,
\eea
where $\Gamma^\mu$ represents all possible Lorentz contractions. The UV model from which this operator arises was discussed extensively in \cite{Elor:2020tkc,Dror:2019onn,Dror:2019dib}. For a given UV model, the scale of the operator $\Lambda$ is experimentally constrained. Typically, $\Lambda$ must be larger than a few hundreds of GeV to a few TeV. As discussed in \cite{Elor:2020tkc}, $\Lambda$ within this range is still small enough to achieve a large enough $\Brmestoelld$ to generate the lepton asymmetry. We leave a more detailed study of the UV models, and in particular their flavorful variations, to future work. 

\begin{figure}[t]
\centering
\includegraphics[width=\columnwidth]{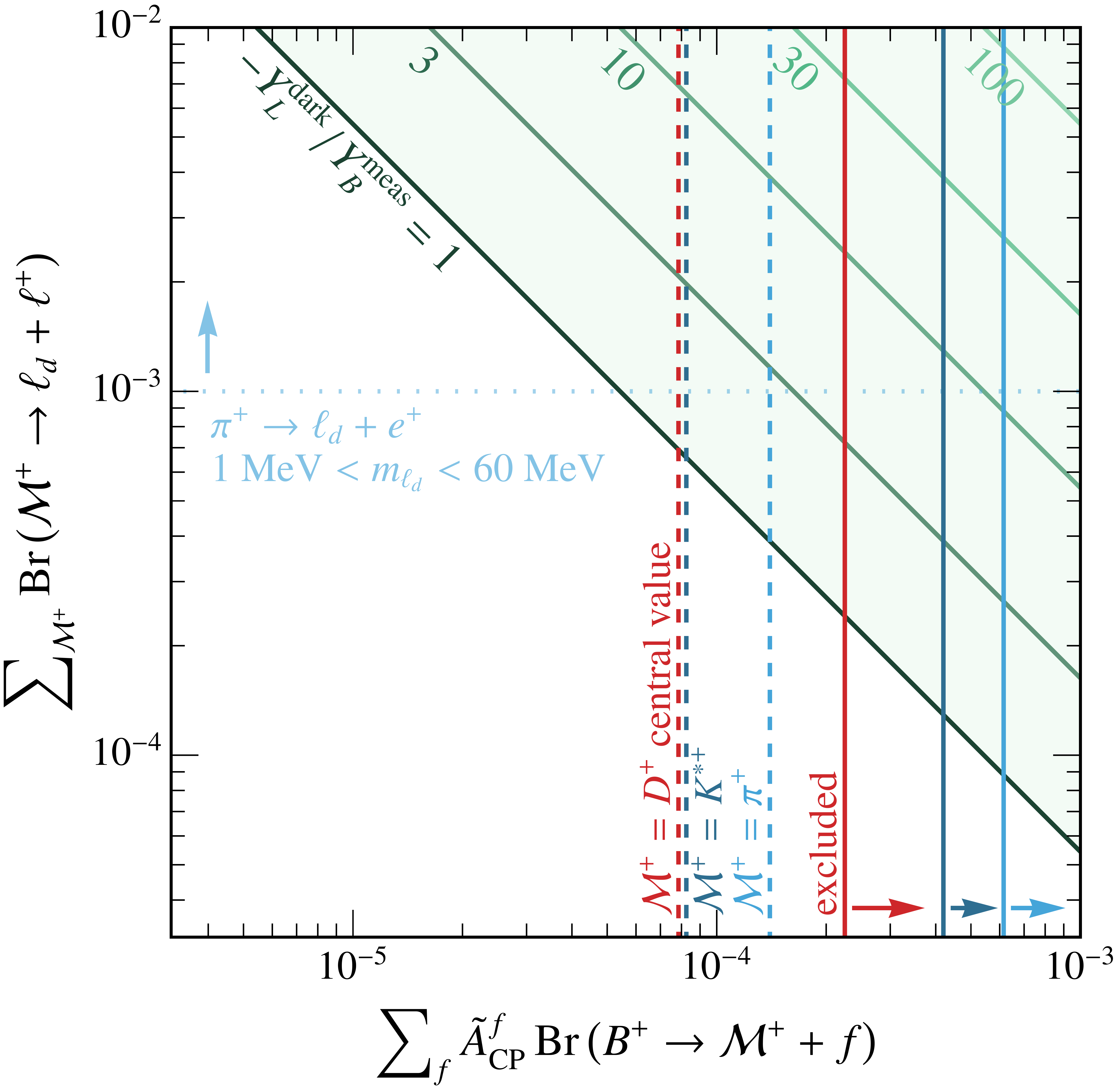}
\vspace{-0.4cm}
\caption{Parameter space plot for $\mes^+ = \pi^+$, and $D^+$, as well as the resonance $K^{*+}$. The region of parameter space where Eq.\eqref{eq:LtoBob} is satisfied is shown in green, along with central values and bounds on $\sum_f \ACPB^f \Br_{B^+}^f$ and a bound on $\Br \prn{\pi^+ \to \ell_d + e^+}$. Channels shown here have a net positive central value for the summed CPV.}
\label{fig:B+ParamSpace1}
\end{figure}

\subsection{Results}
\label{subsec:BResults}

The evolution of the generated lepton asymmetry is
\al{
\label{eq:BmesoLasym}
&\frac{d}{dt} \prn{ n_{\ell_d} - n_{\bar{\ell}_d} } + 3 H \prn{ n_{\ell_d} - n_{\bar{\ell}_d} } = - \langle \sigma v \rangle n_{\chi_1} \prn{ n_{\ell_d} - n_{\bar{\ell}_d}} \nonumber \\
&+2 \Gamma_\Phi^B n_\Phi \sum_{\mes^+} \Brmestoelld \sum_f N_{\mes^+}^f \acpBf \Br_{B^+}^f ,
}
where $N_{\mes^+}^f$ counts the multiplicity of $\mes^+$ in the final state and must be odd. Here $\Gamma_\Phi^B \equiv \Gamma_\Phi \Br (\Phi \to b) \Br (b \to B)$, and $\acpBf \equiv \ACPB^f / (1+\ACPB^f) \approx \ACPB^f$ (see \cite{Elor:2020tkc} for details). For simplicity of notation, we approximate $\acpBf \approx \ACPB^f$, which is true in our parameter space of interest. Once again, we assume throughout that $T_R \lesssim 20$ MeV, such that all processes occur fast enough relative to Hubble so that we can ignore SM scatterings that would washout the asymmetry. 

We also must track the abundance of $\chi_1$, which we assume is simply produced from $\Phi$ decays, 
\bea
\frac{d n_{\chi_1}}{dt} + 3 H n_{\chi_1} && = \Gamma_\Phi n_\Phi \Br \prn{ \Phi \to \chi_1 \bar{\chi}_1 } \\ \nonumber
&& \qquad - \langle \sigma v \rangle n_{\bar{\ell}_d} n_{\chi_1} \,.
\eea
The scattering term, corresponding to the process in Eq.~\eqref{eq:DarkScat}, transfers the lepton to the baryon asymmetry such that the BAU of the Universe is found by solving the following evolution equation 
\bea
\frac{d}{dt} \prn{ n_{\mathcal{B}} - n_{\bar{\mathcal{B}}} } && + 3 H\prn{ n_{\mathcal{B}} - n_{\bar{\mathcal{B}}} } = \\ \nonumber
&& \qquad - \langle \sigma v \rangle n_{\chi_1} \prn{ n_{\ell_d} - n_{\bar{\ell}_d} } \,.
\eea
The produced lepton asymmetry efficiently transfers when the scattering rate dominates Hubble, 
\bea
\frac{n_{\chi_1} \langle \sigma v \rangle}{H(T)} \Big|_{T_R} \geq \frac{Y_B^{\rm obs}}{Y_L^{\rm dark}}\,.
\eea
This requires a sizable, though feasible \cite{Elor:2020tkc}, dark sector $\langle \sigma v \rangle$ --- we leave a detailed study of the dark sector dynamics and its UV completion to future work.

We numerically solve the Boltzmann equations for the lepton asymmetry in Eq.~\eqref{eq:BmesoLasym}, neglecting the scattering term, to verify that a large enough lepton asymmetry can be generated consistently within current bounds of the experimental observables. Integrating Eq.~\eqref{eq:BmesoLasym}, we find that the generated lepton asymmetry is approximately
\bea
\hspace{-0.2in} \frac{Y_L^{\rm dark}}{Y_B^{\rm obs}} \simeq \frac{\sum_{\mes^+} \Brmestoelld}{10^{-3}} \frac{\sum_f \ACPB^f \Br_{B^+}^f}{5.4 \times 10^{-5}} \frac{T_R}{20 \, \text{MeV}} \frac{2 m_{B^+}}{m_\Phi}.
\eea

The viable parameter space which satisfies Eq.~\eqref{eq:LtoBob} is green in  Figs.~\ref{fig:B+ParamSpace1} and Fig.~\ref{fig:B+ParamSpace2}. The former is relevant for $\mes^+ = \pi^+$ and $D^+$ along with the resonance  $K^{\ast +}$, which correspond to a net positive central value of $\sum_f \ACPB^f \Br_{B^+}^f$; the latter, for $\mes^+ = K^+$, $D_s^+$ and the resonance $D^{\ast +}$, which correspond to a net negative central value of $\sum_f \ACPB^f \Br_{B^+}^f$. In both figures, the dashed vertical lines correspond to the measured central values of $\ACPB^f \, \times \Br_{B^+}^f$ for each $\mes^+$, while the solid lines correspond to their maximum values consistent with current measurements (see Sec.~\ref{subsec:BSignals} for a discussion of these measurements). The horizontal dotted lines correspond to limits on the branching fraction of $\mes^+$ to SM charged leptons and anomalous missing energy. As indicated, these bounds depend on $m_{\ell_d}$ and may therefore be easily avoided. For all flavors of $\mes^+$ considered, there is viable parameter space in which the generated lepton asymmetries are over ten times larger than the observed baryon asymmetry. In particular, for some cases, the lepton asymmetry can be as great as one hundred times larger.

\subsection{Current Constraints, Signals, and Prospects}
\label{subsec:BSignals}
The experimental observables of $B^+$ Mesogenesis defined in Eq.~\eqref{eq:BplusYell} parametrize the $B^+$ and $\mes^+$ decays and directly control the generated BAU. Below, we enumerate their current measurements, commenting on their theoretical predictions within the SM, and discuss prospects for improving or developing searches to measure these observables.

\vspace{0.1in}
{\bf $\mathbf{B^+}$ Decays:}
We first consider the decays $B^+ \to \mes^+ + f$, where $\mes^+ = \{ \pi^+ \,, K^+ \,, D^+ \,, D_s^+\}$ represents an odd number of final-state $\mes^+$ particles. The current experimentally measured values for the branching fractions and $\ACPB^f$ \cite{pdg} for these decays are shown in Tables \ref{table:Btopimodes}, \ref{table:BtoKimodes}, and \ref{table:BtoDmodes}. Summing the central values and uncertainties for the decays of $B^+$ to each $\mes^+$ we consider, we find
\begin{subequations}
\begin{align}
\sum_{\mes^+ = \, \pi^+} \! \! \! \! \ACPB^f  \times \Br_{B^+}^f &= \prn{1.4 \times 10^{-4}}^{+ 4.8 \times 10^{-4}}_{-6.1 \times 10^{-4}}, \label{eq:Bpi}\\
\sum_{\mes^+ = \, K^+} \! \! \! \! \ACPB^f  \times \Br_{B^+}^f &= \prn{-5.6 \times 10^{-5}}^{+2.25 \times 10^{-4} }_{-2.77 \times 10^{-4}}, \label{eq:BK} \\
\sum_{\mes^+ = \, D^+} \! \! \! \! \ACPB^f  \times \Br_{B^+}^f &= \prn{7.85 \times 10^{-5}}^{+ 1.5 \times 10^{-4}}_{-1.3 \times 10^{-4}}, \label{eq:BD} \\
\sum_{\mes^+ = \, D_s^+} \! \! \! \! \ACPB^f  \times \Br_{B^+}^f &= \prn{-3.61 \times 10^{-5}}^{+6.6 \times 10^{-5} }_{- 7.4 \times 10^{-5}} . \label{eq:BDs}
\end{align}
\end{subequations}
The upper and lower uncertainties for each $\mes^+$ are computed by making each decay channel of $B^+ \to \mes^+ + f$ as positive and negative as possible, respectively, while allowing one standard deviation from the central values of both $\ACPB^f$ and $\Br_{B^+}^f$. These central values are the aforementioned dashed vertical lines in Figs.~\ref{fig:B+ParamSpace1} and \ref{fig:B+ParamSpace2}, while their deviations correspond to the solid vertical lines.

An intriguing possibility is to achieve $B^+$ Mesogenesis using the predicted CPV and branching fractions in the SM alone. To this end, one would need calculations of $\ACPB^f$ and $\Br_{B^+}^f$ of the exclusive decay modes listed in Tables~\ref{table:Btopimodes},~\ref{table:BtoKimodes}, and ~\ref{table:BtoDmodes}. Comprehensive calculations for these within the SM are challenging and do not currently exist. However, there are hints that suggest it may be feasible to achieve $B^+$ Mesogenesis using only SM CPV. In particular, some of the SM computations that do exist in the literature of the branching fractions of $B^+$ decays are found to be of the same order of magnitude as their experimental limits e.g. \cite{Beneke:2005vv}. It would be very interesting to additionally pursue calculations of $\ACPB^f$ within the SM. If they were sufficiently large, one of the original motivations of electroweak baryogenesis would then hold for $B^+$ Mesogenesis: the requisite CPV to explain the BAU could reside within the SM alone.

\vspace{0.1in}
{\bf $\mathbf{\mes^+}$ Decays:}
Current limits on the branching fraction for the decay of the charged meson into a SM charged lepton and anomalous missing energy, $\mes^+ \to \ell_d + \ell^+$, can be recast from peak searches for sterile neutrinos. Such bounds are entirely kinematic and do not make assumptions about the sterile neutrino model. In particular, given a bound on the lepton mixing $|U_{N \ell}|^2$, one can extract a bound on the branching fraction \cite{PhysRevD.24.1232} for the decay for a given dark lepton mass. 
Current limits on sterile neutrino mixing can be extracted from \cite{Bryman:2019bjg} (and references therein). Note that given the kinematic nature of peak searches, such bounds are strongly dependent on $m_{\ell_d}$. Furthermore, none of the bounds quoted below hold for $m_{\ell_d} < 1 \, \text{MeV}$, such that this region of parameter space is totally unconstrained.

For charged pion decays into electrons, recasting current limits from peak searches 
~\cite{Aguilar-Arevalo:2017vlf,Aguilar-Arevalo:2019owf}, one arrives at the constraint:
\begin{align}
\label{eq:pionlimitsE}
\nonumber
\Br(\pi^\pm \to e^\pm + \ell_d)  \,&< \, 10^{-4} - 10^{-3}  \,,\\ \nonumber
 \quad \text{for} \,\, 1 \, \text{MeV} \, &< \,\, m_{\ell_d}< 60 \, \text{MeV} \,, \\ \nonumber
   \Br(\pi^\pm \to e^\pm + \ell_d)  \,&< \, 10^{-4} - 10^{-5}  \,,\\ 
 \quad \text{for} \,\, 60 \, \text{MeV} \, &< \,\, m_{\ell_d} < 130 \, \text{MeV} \,,
\end{align}
where the range in Eq.~\eqref{eq:pionlimitsE} reflects the possible variation of $m_{\ell_d}$ (see Fig 5 of \cite{Aguilar-Arevalo:2017vlf}). For final-state muons, the bound is \cite{Aguilar-Arevalo:2019owf}
\begin{align}
\label{eq:pionlimitsM}
\nonumber
\Br(\pi^\pm \to \mu^\pm + \ell_d)  \,&<\, 3 \times 10^{-5} \,, \\ 
\quad \text{for} \,\, \, 15.7\, \text{MeV}\, &< m_{\ell_d}\, < 33.8 \,\text{MeV} \, .
\end{align}
Next-generation experiments which
would improve the limit on the branching fraction are being proposed \cite{Doug}.

For Kaons decaying into electrons, current limits are recast from an NA62 search \cite{NA62:2017qcd}. 
\begin{align}
\nonumber
\Br(K^\pm \to e^\pm + \text{MET})  \,&<\, 1.1 \times 10^{-7} \,, \\ 
\quad \text{for} \,\, \, 50 \, \text{MeV}\, &< m_{\ell_d}\, < 350 \,\text{MeV} \, .
\end{align}
While for Kaons decaying into muons, current limits are recast from KEK \cite{Hayano:1982wu}, BNL \cite{E949:2014gsn} and NA62 \cite{NA62:2021bji}
\begin{align}
\nonumber
\Br(K^\pm \to \mu^\pm + \text{MET})  \,&<\,  8 \times 10^{-8} -  8 \times 10^{-4} \,, \\ 
\quad \text{for} \,\, \, 30 \, \text{MeV}\, &< m_{\ell_d}\, < 350 \,\text{MeV} \, .
\end{align}
Note that given the mass sensitivity, it is possible to evade any one of them by judicious choice of $m_{\ell_d}$. These bounds are shown as dotted horizontal lines in Figs.~\ref{fig:B+ParamSpace1} and \ref{fig:B+ParamSpace2}.

Regarding the decay $D^+ \to \ell_d + \ell^+$, where $\ell = e\,, \mu$, while it is possible to place correlated limits on $|U_{\ell 4}|^2$ as a function of $m_{\nu_4}$ \cite{Bryman:2019bjg}, such limits are also model dependent. Thus, we do not consider them robust in constraining $B^+$ Mesogenesis. However, Belle has performed a peak search for similar decays of charged $B$s \cite{Belle:2016nvh} and one may therefore conclude that similar studies for $D$ mesons should be possible and would probe the parameter space relevant here. 

For $D_s$ decays to electrons and anomalous missing energy, while a limit has been set on the branching fraction for the decay $D_s^+ \to e^+ \nu_e$ \cite{CLEO:2009lvj,BaBar:2010ixw,Belle:2013isi}, no relevant limit on the branching fraction of $D_s$ to electrons and anomalous missing energy has yet been produced \cite{Bryman:2019bjg}. The work of \cite{Bryman:2019bjg} did set a lepton mixing matrix coefficient involving muons. The recast limit is: 
\begin{align}
\nonumber
\Br(D_s^+ \to \mu^+ + \ell_d)  \,&<\,  1 \times 10^{-4} - 5 \times 10^{-3}\,, \\ 
\quad \text{for} \,\, \, 100 \, \text{MeV}\, &< m_{\ell_d}\, < 900 \,\text{MeV} \, .
\end{align}
Both $D$ and $D_s$ are heavy enough to decay into $\tau$s and missing energy. While limits do exist on $|U_{\tau 4}|^2$ for a given sterile neutrino model, to date, no peak search has been performed. 

\newpage
\vspace{0.1in}
{\bf Resonant $\mathcal{M}^{\ast +}$:} 
The possibility that $\mes^{\ast +}$ is a resonant state is particularly interesting from an experimental perspective. Searches at LHCb would be able to trigger on the decay vertex of $\mes^{\ast +}$, thereby gaining a handle on reconstruction\footnote{We thanks Xabier  Cid  Vidal for discussions regarding this possibility.}. Decays of $B^+$ mesons to $K^{\ast +}$ or $D^{\ast +}$ are summarized in Tables.~\ref{table:BtoDmodes} and \ref{table:BtoOtherimodes}, and the central values are given by;
\begin{subequations}
\begin{align}
\hspace{-0.08in} \sum_{\mes^{\ast +} =\, K^{*+}} \! \! \! \! \ACPB^f  \times \Br_{B^+}^f &= \prn{8.3 \times 10^{-5}}^{+3.4 \times 10^{-4} }_{-4.0 \times 10^{-4}}\,, \label{eq:BKs}\\
\hspace{-0.08in}  \sum_{\mes^{\ast +} =\, D_s^{*+}} \! \! \! \! \ACPB^f  \times \Br_{B^+}^f &= \prn{-1.7 \times 10^{-4}}^{+1.3 \times 10^{-4} }_{- 2.2 \times 10^{-4}} \,. \label{eq:BDss}
\end{align}
\end{subequations}
These values are shown in Figs. \ref{fig:B+ParamSpace1} and \ref{fig:B+ParamSpace2}, and permit a sizable parameter space for generating the lepton asymmetry. 

Resonant states $\mes^{*+}$, typically quickly decay into their respective $\mes^+$, but with some branching fraction to other final states. 
We therefore do not include the $\mes^{\ast +}$ channels when summing over the $\mes^+$ modes, since it is possible that $\mes^{\ast +}$ does not quickly decay to a $\mes^+$ (which could then in turn decay quickly to a dark lepton). Indeed, it is possible for $\mes^{\ast +}$ itself to decay to a dark and SM lepton pair. For these reasons, we have considered $\mes^{\ast +}$ separately from $\mes^+$. However, note that for a given UV model with a chosen flavor structure, the operator that would allow $K^+$ or $D^+$ to quickly decay to a dark lepton would also permit $K^{\ast +} \to K^+$ or $D^{\ast +} \to D^+$ to do the same. Further measurements of the decay modes of $\mes^{\ast +}$ will be helpful in distinguishing between these two possibilities. 

\vspace{0.1in}
{\bf Summary:} Experimental limits on the branching fractions and CPV in $B^+$ decays currently leave a large swath of viable parameter space for $B^+$ Mesogenesis as shown in Figs.~\ref{fig:B+ParamSpace1} and \ref{fig:B+ParamSpace2}. Improved measurements of these observables at $B$ factories and hadron colliders will thus further probe this scenario. The scenario in which $B^+$ first decays into a resonance meson is particularly appealing for reconstructability at LHCb. Meanwhile, experiments searching for sterile neutrinos have and will continue to set constraints on charged mesons decaying into SM leptons and anomalous missing energy. Peak searches are particularly well suited to place model-independent bounds on such branching fractions and are further motivated by this scenario.

\newpage
\section{Discovery Prospects}
\label{sec:discovery}
We have presented two new mechanisms for generating the BAU using known SM processes; namely the decays of SM charged $B$ mesons. These mechanisms are testable and motivate a variety of experimental and theoretical studies.  We now highlight the action items required to make progress towards discovering Charged $B$ Mesogenesis.

\vspace{0.2in}
{\bf $B_c^+$ Mesogenesis:}
\begin{itemize}
    \item Experimental measurements of the branching fraction and CPV by LHCb or at a FCC of $B_c$ decays to final states involving $B$ mesons 
    \item Continued theoretical work towards a SM prediction for the branching fraction and CPV for the relevant decay modes; should the SM CPV prove insufficient to generate the BAU, a study of new physics contributions would be of interest.
    \item Continued searches for the decays of (charged and neutral) $B$ mesons and $b$-flavored baryons into SM states and missing energy; several such studies are already underway in an effort to observe neutral $B$ Mesogenesis.
\end{itemize}

\vspace{0.2in}
{\bf $B^+$ Mesogenesis:}
\begin{itemize}
    \item Searches at hadron colliders and $B$ factories to improve the measurements of the branching fraction and CPV in charged $B$ decays to final state pions, kaons, and $D$ mesons
    \item Continued theoretical work to calculate $\ACPB$ of these decay modes in the SM since it is likely the SM has sufficient CPV to explain the BAU; if so, $B^+$ Mesogenesis would be the only known mechanism of Mesogenesis (or baryogenesis in general) \emph{whose sole source of CPV resides in the SM}
    \item The scenario in which $B^+$ first decays into a charged resonance is well suited for searches at LHCb
    \item Improved searches for charged mesons decaying into leptons and anomalous missing energy; a peak search for charged $D$ and $D_s$ decaying into $\tau$s would be particularly useful; Belle-II would be an ideal experiment to conduct such a search; possibilities exist at BESIII as well.
\end{itemize}

By making the above measurements, experiments will be able to hone in on the parameter space for charged $B$ Mesogenesis and either discover or constrain it. Improved calculations of the CPV and branching fractions of charged $B$ mesons, while highly non-trivial, could demonstrate that Charged $B$ Mesogenesis is the only viable way to explain the BAU using CPV within the SM alone. 

\section{Outlook}
\label{sec:Outlook}
It is possible that our very existence arises from the decay of charged $B$ mesons at temperatures slightly above 20 MeV. To gain a quantitative understanding of how to make the Universe above 20 MeV requires a detailed numerical solution of the Boltzmann equations in the washout regime --- a subject of ongoing and future work.   

It may also be that our explorations of the different ``flavors'' of Mesogenesis are far from complete. In particular, a large amount of CPV is allowed in the top sector which leads one to ponder if a mechanism could be discovered which leverages it to generate the BAU. Of course, in order for a parton level process to directly feed into the BAU requires it to be active above the QCD scale. While scattering and washout would play an important role here, one could still consider the scenario of ``$t$-genesis". At lower, MeV scales the formation of Top Mesons are possible. These states are theoretically possible within the SM, but have thus far evaded detection as they are expected to be very short lived. However, generating the BAU from such states decaying into dark baryons is an intriguing possibility that we leave for future work.

Additional  variations of Charged $B$ Mesogenesis may exist. One may pose the challenge of constructing a mechanism in which $B^\pm$ mesons decay directly into dark baryons. One may attempt to consider decays through higher dimension operators and CKM insertions. For instance, $B^+$ could decay into $D_s^+$ and one may consider the kinematically allowed decays $D_s^+ \to \psiBbar \, n \, e^+ \nu_e$ and $D_s^+ \to \psiBbar \, p$. However, the CPV and branching fraction in the $B^+$ decays (the product of which is $~10^{-4}$) would require the branching fraction to be greater than $10^{-4}$. But the expected branching fraction -- given the CKM insertions, phase space suppression, and collider constraints on $y^2/M_\phi^2$ -- is order of magnitudes smaller than what is required to achieve Mesogenesis. Thus,  more exotic constructions may be required. One could consider a UV model involving fractionally charged dark baryons or a setup in which $B^+$ undergo scatterings off dark baryons (rather than decays into them). We leave such exercises to future work. 

This work, as well as the existing Mesogenesis literature, has remained agnostic about the nature of the $\Phi$ field and of a detailed model of the dark sector. 
Exploring the possible models and associated signals on both these fronts is an interesting future direction that will play a pivotal role in nailing down the details of Mesogenesis once the action items discussed in the present work have been addressed.

We return to our original question: how did we come to be? Was it two $B$s, or not two $B$s? This paper has strongly motivated the former by introducing two simple Mesogenesis scenarios using charged $B$s. Whether 'tis $B_c^+$ or $B^+$, only time and nobler experiment will tell.

\begin{acknowledgments}
We thank Olcyr Sumensari for early collaboration, useful discussions, and comments on the draft.
We thank Doug Bryman, Xabier Cid Vidal and Robert Shrock for useful discussions and comments on the draft. 
We thank Liupan An and Javier Fuentes Martin for useful discussions.
The research of F.E. and G.E. is supported by the Cluster of Excellence {\em Precision Physics, Fundamental Interactions and Structure of Matter\/} (PRISMA${}^+$ -- EXC~2118/1) within the German Excellence Strategy (project ID 39083149). F.E. is also supported by by grant 05H18UMCA1 of the German Federal Ministry for Education and Research (BMBF)
R.M. is supported in part by the DoE grant DE-SC0007859.

\end{acknowledgments}

\onecolumngrid
\appendix

\section*{Appendices}

\section{Derivation of Boltzmann Equations}
\label{app:BoltzDerive}

Here we present a detailed derivation of the Boltzmann equations for $B_c^+$ Mesogenesis. The Boltzmann equations for $B^+$ Mesogenesis are identical to those of $D^+$ Mesogenesis (see the derivation in the appendix of \cite{Elor:2020tkc} for details).

\subsection{Evolution of the Baryon Asymmetry}
\label{sec:Basymm}

The number density Boltzmann equations for $B_c^\pm$ mesons are
\begin{align}
\frac{dn_{B_c^+}}{dt} + 3 H n_{B_c^+} &= \Gamma^{B_c}_\Phi n_\Phi - \Gamma_{B_c} n_{B_c^+} \,, \nonumber\\
\frac{dn_{B_c^-}}{dt} + 3 H n_{B_c^-} &= \Gamma^{B_c}_\Phi n_\Phi - \Gamma_{B_c} n_{B_c^-}\,,
\label{eq:Bc}
\end{align}
where $\Gamma^{B_c}_\Phi$ is the rate of decay of $\Phi$ to a pair of $B_c^\pm$ mesons and $\Gamma_{B_c}$ is the total $B_c^\pm$ decay rate. We further assume that there is no CPV in $\Phi$ decays.

The number density Boltzmann equations for the charged B mesons are:
\begin{align}
\frac{dn_{B^+}}{dt} + 3 H n_{B^+} &= n_{B_c^+}  \Gamma_{B_c}  \sum_f \Br\prn{\BctoB} - \Gamma_{B} n_{B^+} \,,\nonumber\\
\frac{dn_{B^-}}{dt} + 3 H n_{B^-} &= n_{B_c^-} \Gamma_{B_c}  \sum_f \Br\prn{\BctoBconj} - \Gamma_{B} n_{B^-}\,,
\label{eq:Bpm}
\end{align}
where $\Gamma_{B}$ is the total $B^\pm$ decay rate. All SM annihilation terms that would appear are negligible relative to the decay terms thanks to the low reheating temperature. 

For intuition and simplicity, we first solve for the baryon asymmetry generated in the SM sector:
\begin{align}
\label{Bassym}
\frac{dn_\Bp}{dt} + 3 H n_\Bp &= \Gamma_B \Br\prn{\Btopsi} n_{B^+}\,, \nonumber\\
\frac{dn_\Bpconj}{dt} + 3 H n_\Bpconj &= \Gamma_B \Br\prn{\Btopsiconj} n_{B^-}\,.
\end{align}
Taking $\Br\prn{\Btopsi} = \Br\prn{\Btopsiconj}$ and assuming no CPV in this interaction, we get
\begin{equation}
\frac{d}{dt} \prn{n_\Bp  - n_\Bpconj} + 3 H \prn{n_\Bp  - n_\Bpconj} = \Gamma_B \Br\prn{\Btopsi} \prn{n_{B^+} - n_{B^-}}\,.
\label{eq:baras}
\end{equation}
Let us assume that the decays happen almost instantaneously: $H \ll \Gamma_\Phi, \Gamma_{B_c}, \Gamma_B$. Thus, we can use the following approximations:
\begin{align*}
 \Gamma_{B_c} n_{B_c^+} =  \Gamma_{B_c} n_{B_c^-} \simeq  \Gamma_\Phi^{B_c} n_\Phi \hspace{1.2 in} &\text{From Eq.~} \ref{eq:Bc}\\
 \Gamma_B n_{B^+} \simeq \Gamma_{B_c} n_{B_c^+} \Br\prn{\BctoB} \hspace{0.5 in} &\text{From Eq.~} \ref{eq:Bpm}\\
 \Gamma_B n_{B^-} \simeq \Gamma_{B_c} n_{B_c^-} \Br\prn{\BctoBconj} \hspace{0.5 in} &\text{From Eq.~} \ref{eq:Bpm}.
\end{align*}
Hence, writing the sum over possible $\B^+$ explicitly, the right hand side of Eq.~\ref{eq:baras} becomes
\begin{align}
\Gamma_B \Br\prn{\Btopsiconj} \prn{n_{B^+} - n_{B^-}}  =  &\Gamma_\Phi^{B_c} n_\Phi \sum_{\B^+} \Br\prn{\Btopsi}\nonumber\\
&\times \sum_f \left[  \Br\prn{\BctoB} - \Br\prn{\BctoBconj}\right] \,.
\label{eq:Basym}
\end{align}
Using the definition of $\ACP^f$ from Eq.~\ref{eq:ACPgen}, we find
\begin{equation}
\Br\prn{\BctoB} - \Br\prn{\BctoBconj} = \Br\prn{\BctoB} \frac{ 2 \ACP^f}{1+ \ACP^f} \equiv 2 \BrBctoB \acpf \,.
\end{equation}
Finally, Eq.~\ref{eq:baras} becomes 
\begin{equation}
\frac{d}{dt} \prn{n_{\Bp}  - n_{\Bpconj}} + 3 H \prn{n_{\Bp}  - n_{\Bpconj}} = 2 \Gamma_\Phi^{B_c} n_\Phi \sum_{\B^+} \BrBtopsi \sum_f  \BrBctoB \acpf \,.
\end{equation}

\subsection{Evolution of Dark Matter}
 Now let us extend the dark sector allowing $\psiB \to \chi \phiB$ with $100\%$ branching ratio. The Boltzmann equation for $\psiB$, is 
 \begin{align}
\label{phiBE}
\frac{dn_{\psiB}}{dt} + 3 H n_{\psiB} &= \Gamma_{B} \Br\prn{\Btopsi} n_{B^+} - \Gamma_{\psiB} n_{\psiB}\,, \nonumber\\
\frac{dn_{\psiBbar}}{dt} + 3 H n_{\psiBbar} &= \Gamma_{B} \Br\prn{\Btopsiconj} n_{B^-} - \Gamma_{\psiBbar} n_{\psiBbar} \,. 
\end{align}
Again, since the decays are faster than Hubble, we can approximately say  $\Gamma_{\psiB} n_{\psiB} \simeq \Gamma_{B} \Br\prn{\Btopsi} n_{B^+}$. Similarly, $\Gamma_{\psiBbar} n_{\psiBbar} \simeq  \Gamma_{B} \Br\prn{\Btopsiconj} n_{B^-}$. 

The Boltzmann Equation for $\phiB + \phiBstar $ is 
\begin{equation}
\frac{d}{dt} \prn{n_{\phiB } + n_{ \phiBstar} } + 3 H \prn{n_{\phiB } + n_{ \phiBstar} }  = \Gamma_{\psiB} \Br (\psiB \to \chi \phiB) ( n_{\psiB} + n_{\psiBbar}) - 2 \langle \sigma v\rangle_d   \prn{ n^2_{\phiB + \phiBstar} - n^2_{\text{eq}, \phiB + \phiBstar} }\,, 
\end{equation}
 where the first term on the right hand side is the production of $\phiB +\phiBstar$ from $\Phi$ decay, and the second term describes the annihilation of $\phiB \phiBstar $ to SM particles. We can follow the same procedure as in Sec.~\ref{sec:Basymm} and simplify $ \Gamma_{\psiB} ( n_{\psiB} + n_{\psiBbar})$:
\begin{align*}
 \Gamma_{\psiB} ( n_{\psiB} + n_{\psiBbar}) & = \Gamma_{B} \Br\prn{\Btopsi} ( n_{B^+}+ n_{B^-}) \\
& = \Gamma_\Phi^{B_c} n_\Phi \sum_{\B^+} \BrBtopsi  \sum_f \left[  \Br\prn{\BctoB} + \Br\prn{\BctoBconj}\right]\\
& = 2 \Gamma_\Phi^{B_c} n_\Phi \sum_{\B^+} \BrBtopsi\sum_f  \frac{\BrBctoB}{1+ \ACP^f}\,.
\end{align*}
Therefore, the evolution of $\phiB + \phiBstar  $ is governed by the following equation: 
\begin{equation}
\frac{d}{dt} \prn{n_{\phiB } + n_{ \phiBstar} } + 3 H \prn{n_{\phiB } + n_{ \phiBstar} } = 2 \Gamma_\Phi^{B_c} n_\Phi \sum_{\B^+} \BrBtopsi\sum_f  \frac{\BrBctoB}{1+ \ACP^f}- 2 \langle \sigma v\rangle_d   \prn{ n^2_{\phiB + \phiBstar} - n^2_{\text{eq}, \phiB + \phiBstar} }\,. 
\end{equation}
The Boltzmann equation for $\chi$ is essentially identical.

\subsection{The ``Rest'' of Dark Matter}
\label{sec:restofDM}
As mentioned in Sec.~\ref{sec:Bcmeso}, there are two simple possibilities for the rest of dark matter not accounted for by the requisite asymmetry in $\phiBstar$s and $\bar{\chi}s$. The first is simply that the rest of dark matter is comprised of a symmetric abundance of $\chi$s and $\phi_B$s and their conjugates. This option is appealing since the rest of dark matter is made out of stuff that's already in the model: just more $\chi$s and $\phi_B$s (and their conjugates). However, there are a few challenges to this possibility. First, depleting the symmetric abundances requires WIMP-scale (or larger) annihilation cross sections. Since dark matter is GeV-scale, this would immediately be ruled out by indirect detection unless the annihilations are solely to SM neutrinos, or into more exotic fractionally charged dark baryons \cite{Elor:2018twp}. Second, these annihilations have to fully stop prior to SM neutrino decoupling in order to prevent reheating them after decoupling, which would increase $\Neff$. This is a non-trivial constraint since the freezeout time of $\sim \text{GeV}$ dark matter is near the $\sim 3 \text{ MeV}$ neutrino decoupling temperature. An additional aesthetic deficit of this possibility is that the annihilation cross sections, lighter dark sector states into which $\chi$s and $\phi_B$s annihilate, and those lighter states' subsequent (and quick) annihilations to SM neutrinos would all be additional input parameters into the $B_c^+$ Mesogenesis scenario, totally decoupled from the fundamentals of the scenario itself. However, the above constraints are not insurmountable and it is possible to realize situations in which all of dark matter is made of both a symmetric and asymmetric abundance of $\chi$s and $\phi_B$s.

The second possibility is that the remaining $\mathcal{O}(1/2)$ of dark matter is made up of new dark sector states. This seems quite simpler and requires fewer input parameters. The rest of dark matter could be a single additional dark sector state with a single interaction with the SM to produce its requisite relic abundance. It could be a frozen-in GeV scale dark matter evading (in)direct detection bounds, as well as $\Neff$ bounds, or it could be a TeV scale classic WIMP. Any usual dark matter scenario could be ``pasted'' onto the rest of our model to account for the final $\sim 20 \%$ of dark matter that's not in the asymmetries of $\chi$s and $\phi_B$s.

\section{Decay modes for \texorpdfstring{$B^+$}{B+} Mesogenesis}
\label{app:BdecayModes}
Here we present detailed tables containing the relevant decay modes for $B^+$ Mesogenesis.
\vspace{-1in}
\begin{table}[t]
\renewcommand{\arraystretch}{1.0}
\setlength{\arrayrulewidth}{.1mm}
\centering
\small
\setlength{\tabcolsep}{0.26 em}
\setlength{\arrayrulewidth}{.25mm}
\begin{tabular}{ |c | c | c |  c | c| }
    \hline
    $B^{+}$ Decay Mode to $\pi^+$ &     $\ACP$     & Branching Fraction  $\Gamma_i / \Gamma$ & Ref  \\ \hline \hline
     \ck{$\bar D^0 \, \pi^+$} &   $0.100 \pm 0.032$  &    $( 4.68 \pm 0.13 ) \times 10^{-3}$     & \cite{Belle:2017psv,BaBar:2006rof} \\ \hline
     \ck{$\bar D^{* \, 0}_{\rm CP(+1)} \, \pi^+$} &  $0.016 \pm 0.010$   &    $(2.7 \pm0.6 ) \times 10^{-3}$     &  \cite{Belle:2006cuz} \\ \hline 
      \ck{$D_{\rm CP(-1)} \, \pi^+$} & $0.017 \pm 0.026$ &   $(2.0 \pm0.4 ) \times 10^{-3}$   &  \cite{Belle:2006cuz} \\ \hline  
  \ck{$K^0 \pi^0 \pi^+$} &  $0.07\pm 0.06$  &    $< 6.6  \times 10^{-5}$     & \cite{BaBar:2015pwa,CLEO:2002jwu}   \\ \hline    
   \ck{  $f_0(1370) \pi^+ $ } &     $0.72 \pm 0.22$  &    $ <4.0  \times 10^{-6}$     & \cite{BaBar:2009vfr}  \\ \hline 
    \ck{$K_0^* (1430)^0  \pi^+$} &  $0.061\pm 0.032$    &    $(3.9^{+0.6}_{-0.5}) \times 10^{-5}$     &\cite{BaBar:2015pwa,BaBar:2008lpx,Belle:2005rpz}   \\ \hline    
 \ck{  $ \chi_{c1}(1P)\, \pi^+$} &  $0.07 \pm 0.18$   &   $(2.2 \pm0.5 ) \times 10^{-5}$    & \cite{Belle:2006eqz} \\ \hline    
      \ck{  $f_2 (1270) \pi^+ $ } &  $0.40 \pm 0.06$     &    $( 2.2^{+0.7}_{-0.4} ) \times 10^{-6}$     &\cite{LHCb:2019xmb,BaBar:2009vfr,BaBar:2005jqu}   \\ \hline      
    \ck{$\psi (2S) \, \pi^+$} &   $0.03 \pm 0.06$    &   $(2.44 \pm 0.30 ) \times 10^{-5}$     &\cite{Belle:2008ztp}\\ \hline  
 \ck{  $\pi^+ \pi^- \pi^+ $ } &   $-0.14^{+0.23}_{-0.16}$   &    $( 1.52\pm0.14 ) \times 10^{-5}$     &   \cite{LHCb:2020xcz} \\ \hline  
 \ck{$K_2^* (1430)^0  \pi^+$} &   $-0.6\pm 0.07$             &    $( 5.6^{+2.2}_{-1.5} ) \times 10^{-6}$     & \cite{BaBar:2008lpx}   \\ \hline
 \ck{$\bar{D}^{* \, 0} \, \pi^+$} &                                               $-0.0010 \pm 0.0028$             &    $(5.7 \pm 1.2 ) \times 10^{-3}$     & \cite{BaBar:2006zod}   \\ \hline
  \ck{$D_{\rm CP(+1)} \, \pi^+$}  & $-0.0080 \pm 0.0026$     &   $(1.38 \pm  0.13) \times 10^{-3}$   &  \cite{LHCb:2017wbt,LHCb:2016gpc}\\ \hline 
  \ck{$\bar{D}^0 \, \pi^+$} &  $-0.007 \pm 0.007$   &   $(4.68\pm 0.13) \times 10^{-3}$   &  \cite{Belle:2017psv,BaBar:2006rof}  \\ \hline
    \ck{$D^{* \, 0}_{\rm CP(-1)} \, \pi^+$} &   $-0.09 \pm 0.05$    &    $(2.4 \pm 0.9 ) \times 10^{-3}$     & \cite{Belle:2006cuz}   \\ \hline 
      \end{tabular}
\vspace{5mm}
\caption{The decay modes of $B^+$ mesons into final states involving an odd number $\pi^+$ mesons, along with the $A_{\rm CP}$ and branching fraction for each.}
\label{table:Btopimodes}
\end{table}

\begin{table}[t]
\renewcommand{\arraystretch}{1.0}
\setlength{\arrayrulewidth}{.1mm}
\centering
\small
\setlength{\tabcolsep}{0.26 em}
\setlength{\arrayrulewidth}{.25mm}
\begin{tabular}{ |c | c | c |  c | c | }
    \hline
    $B^{+}$ Decay Mode to $K^{+}$ &  $\ACP$  & Branching Fraction  $\Gamma_i / \Gamma$ & Ref  \\ \hline \hline
 \ck{$K^0_s \, K^+$} &  $-0.21\pm 0.14$    &    $( 1.51^{+0.15}_{-0.13} ) \times 10^{-4}$     &  \cite{1308.1277}   \\ \hline 
\ck{$\bar D^{* \, 0}_{\rm CP(+1)} \, K^+$} &   $-0.11 \pm 0.08$   &    $( 2.60\pm0.33 ) \times 10^{-4}$     & \cite{0807.2408}  \\ \hline  
   \ck{$D_{\rm CP(-1)} \, K^+$} &   $-0.10 \pm 0.07$   &    $( 1.96\pm0.18 ) \times 10^{-4}$     &\cite{1007.0504,Belle:2006cuz}   \\ \hline  
\ck{$\bar{D}^0 \, K^+$} &    $-0.017 \pm 0.005$         &   $( 3.63 \pm 0.12 ) \times 10^{-4}$    & \cite{1007.0504,1708.06370,1603.08993} \\ \hline   
  \ck{$\chi_{c1} \, K^+$} &   $-0.009 \pm 0.033$         &   $(4.74 \pm0.22 ) \times 10^{-4}$    & \cite{1911.11740,1105.0177} \\ \hline  
  \ck{$b_1^0 \, K^+$ } &   $-0.46\pm 0.20$             &    $( 9.1\pm 2.0) \times 10^{-6}$     &\cite{0707.4561}   \\ \hline  
   \ck{$D^0 \, K^+$} &         $-0.40 \pm 0.06$             &    $( 3.57\pm0.35 ) \times 10^{-6}$     & \cite{LHCb:2016gpc,LHCb:2012xkq}   \\ \hline 
  \ck{$f(980)^0 \, K^+$} &   $-0.08 \pm 0.09$             &    $(2.03 \pm 0.14^{+0.15}_{-0.002}  ) \times 10^{-5}$     &  \cite{BaBar:2008lpx}    \\ \hline    
 \ck{  $\eta \gamma K^+ $ } &   $-0.12\pm 0.07$             &    $( 7.9 \pm 0.9  ) \times 10^{-6}$     & \cite{BaBar:2008cqt,Belle:2004oww}  \\ \hline
         \ck{$f_2(1270)\, K^+$} &  $-0.68^{+0.19}_{-0.17}$             &    $(1.07 \pm 0.27 ) \times 10^{-6}$     &  \cite{BaBar:2008lpx}  \\ \hline 
  \ck{$\pi^0 \, K^+$} &    $0.037 \pm 0.021$             &    $(1.29 \pm 0.05) \times 10^{-5}$     & \cite{1210.1348}   \\ \hline 
 \ck{$f_0(1500)\, K^+$} &   $0.28 \pm 0.30$             &    $(3.7 \pm2.2 ) \times 10^{-6}$     & \cite{BaBar:2012iuj}   \\ \hline 
 \ck{$\rho^0\, K^+$} &   $0.37\pm 0.10$             &    $( 3.7\pm 0.5) \times 10^{-6}$     &  \cite{BaBar:2008lpx}   \\ \hline
    \ck{$K^+ \pi^+ \pi^-$(non-resonant)} &   $0.027  \pm 0.008$             &    $( 1.63^{+0.21}_{-0.15}) \times 10^{-5}$     &  \cite{0803.4451,Belle:2005rpz} \\ \hline 
  \ck{$K^+ K^- K^+$ (non-resonant)} &    $0.06 \pm 0.05$             &    $( 2.38^{+0.28}_{-0.5} ) \times 10^{-5}$     & \cite{1201.5897,Belle:2004drb}  \\ \hline
\ck{$J/\psi (1S) \, K^+$} &   $0.0018 \pm 0.0030$      &   $(1.020 \pm0.019 ) \times 10^{-3}$    & \cite{CLEO:1997ilq,BaBar:2005sdl,1908.01848,1903.06414,1709.06108,BaBar:2005pcw,BaBar:2004htr} \\ \hline 
\ck{$K^{*+} K^- K^+$} &  $0.11\pm 0.09$             &    $( 3.62\pm0.49 ) \times 10^{-5}$     & \cite{BaBar:2006qhm}  \\ \hline 
 \ck{$\psi (2S) \,  K^+$} &   $0.012 \pm 0.020$         &   $( 6.24\pm0.20 ) \times 10^{-4}$    &\cite{1709.06108} \\ \hline 
 \ck{$\eta_c \, K^+$} &    $0.01 \pm 0.07$    &   $( 1.09\pm0.08 ) \times 10^{-3}$    &\cite{1911.11740,1709.06108}  \\ \hline
\ck{$D^{* \, 0}_{\rm CP(-1)} \, K^+$} &   $0.07 \pm 0.10$             &    $( 2.19\pm0.3 ) \times 10^{-4}$     & \cite{BaBar:2008qcq}  \\ \hline
  \ck{$D_{\rm CP(+1)} \, K^+$} & $0.120 \pm 0.014$   &    $( 1.80\pm 0.07 ) \times 10^{-4}$     & \cite{LHCb:2017wbt,Belle:2006cuz}  \\ \hline
       
\end{tabular}
\vspace{5mm}
\caption{The dominant decay modes of $B^+$ mesons into final states involving an odd number $K^+$ mesons, along with the $A_{\rm CP}$ and branching fraction for each.}
\label{table:BtoKimodes}
\end{table}

\begin{table}[t]
\renewcommand{\arraystretch}{1.0}
\setlength{\arrayrulewidth}{.1mm}
\centering
\small
\setlength{\tabcolsep}{0.26 em}
\setlength{\arrayrulewidth}{.25mm}
\begin{tabular}{ |c | c | c |  c | c| }
    \hline
    $B^{+}$ Decay Mode to $D^+$/$D_s^+$&  $\ACP$    & Branching Fraction  $\Gamma_i / \Gamma$ & Ref  \\ \hline \hline
   \ck{$D_s^+ \, \bar{D}^0$} &  $-0.004 \pm 0.007$             &    $(9.0 \pm 0.9) \times 10^{-3}$     & \cite{LHCb:2018uli}   \\ \hline 
 \ck{$D_s^+ \, \phi$} &   $0.0 \pm 0.4$                  &  $< 4.2 \times 10^{-7}$     & \cite{LHCb:2017vtv}   \\ \hline    
 \ck{$D^{+} \, \bar{D}^{*0}$} &                         $0.13 \pm 0.18$             &    $(6.3 \pm 2.4) \times 10^{-4}$     &  \cite{BaBar:2006uih} \\ \hline    
 \ck{$D^{+} \, \bar{D}^{0}$} &                         $0.016 \pm 0.025$             &    $(3.8 \pm1.1 ) \times 10^{-4}$     & \cite{LHCb:2018uli,BaBar:2006uih,Belle:2008doh}    \\ \hline    
 $D^{*+} \, \bar{D}^{*0}$ &        $-0.15 \pm 0.11$             &    $(8.1 \pm 1.2 ) \times 10^{-4}$     &   \cite{BaBar:2006uih} \\ \hline    
 $D^{*+} \, \bar{D}^{0}$ &     $-0.06 \pm 0.13$             &    $( 3.6\pm 0.8 ) \times 10^{-4}$     & \cite{BaBar:2006uih}  \\ \hline    

\end{tabular}
\vspace{5mm}
\caption{The dominant decay modes of $B^+$ mesons into final states involving $D^+$ and $D_s^+$ mesons, along with the $A_{\rm CP}$ and branching fraction for each.}
\label{table:BtoDmodes}
\end{table}

\begin{table}[t]
\renewcommand{\arraystretch}{1.0}
\setlength{\arrayrulewidth}{.1mm}
\centering
\setlength{\tabcolsep}{0.26 em}
\setlength{\arrayrulewidth}{.25mm}
\begin{tabular}{ |c || c | c |  c | c| }
    \hline
    $B^{+}$ Decay Mode &                                                            $A_{CP}$                         & Branching Fraction  $\Gamma_i / \Gamma$ & Ref  \\ \hline \hline

     $\chi_{c1} \, K^*(892)^+$ &                                                      $0.5 \pm 0.5$                  &   $(3.0 \pm 0.6 ) \times 10^{-4}$   & \cite{BaBar:2004htr,BaBar:2008flx,Belle:2005eoz}\\ \hline  
      $\psi (2S) \, K^*(892)^+$ &                                                      $0.08 \pm 0.21$              &   $(6.7 \pm 1.4) \times 10^{-4}$     & \cite{BaBar:2004htr,CLEO:2000ped} \\ \hline       
     $D_{\rm CP(+1)} \, K^*(892)^+$ &                                          $0.08 \pm 0.06$             &    $(6.2 \pm 0.7) \times 10^{-4}$     & \cite{LHCb:2017egy,BaBar:2009dzx}  \\ \hline   
     $\eta^{'} \, K^*_2 (1430)^+$ &                                                 $0.15  \pm 0.13$             &    $( 2.8\pm0.5 ) \times 10^{-5}$     &  \cite{BaBar:2010cpi} \\ \hline    
      $K_0^* (1430)^+  \pi^0$ &                                                  $0.26^{+0.18}_{-0.14}$             &    $( 1.19^{+ 0.20}_{-0.23} ) \times 10^{-5}$     & \cite{BaBar:2015pwa} \\ \hline    
       $\omega \, K_2^* (1430)^+$ &                                             $0.14  \pm 0.15$             &    $( 2.1\pm 0.4 ) \times 10^{-5}$     & \cite{BaBar:2009mcf}  \\ \hline    
   $\eta^{'} \, K^*(892)^+$ &                                                        $-0.26  \pm 0.27$             &    $(4.8^{+1.8}_{-1.6} ) \times 10^{-6}$     & \cite{BaBar:2010cpi} \\ \hline     
    $K^{*}(892)^+ \, \pi^0$ &                                                      $-0.39  \pm 0.21$             &    $(6.8 \pm 0.9) \times 10^{-6}$     &  \cite{BaBar:2015pwa,BaBar:2011cmh} \\ \hline   
      $ \bar{D}^0 K^* (892)^+$ &                                                      $-0.007 \pm 0.0019$         &   $(5.3 \pm 0.4) \times 10^{-4}$   &  \cite{BaBar:2006kjz,CLEO:2001wbc} \\ \hline
      $\eta \, K^*_2 (1430)^+$ &                                                     $-0.45  \pm 0.30$             &    $(9.1 \pm 3.0 ) \times 10^{-6}$     & \cite{BaBar:2006ltn}  \\ \hline   
  $D_{\rm CP(-1)} \, K^*(892)^+$ &                                          $-0.23 \pm 0.22$             &    $(2.7 \pm 0.8 ) \times 10^{-4}$     &  \cite{BaBar:2009dzx} \\ \hline              
       $J/\psi  \, K^{*}(892)^+$ &                                                          $-0.048 \pm 0.033$        &   $(1.43 \pm0.08 ) \times 10^{-3}$  &  \cite{BaBar:2004htr,CLEO:1997ilq,Belle:2002otd,BaBar:2007esv}  \\ \hline   
\end{tabular}
\vspace{5mm}
\caption{Summary of dominant $B^{\pm}$ decay modes to resonance states.}
\label{table:BtoOtherimodes}
\end{table}

\clearpage

\twocolumngrid

\bibliography{Refs}

%merlin.mbs apsrev4-1.bst 2010-07-25 4.21a (PWD, AO, DPC) hacked
%Control: key (0)
%Control: author (72) initials jnrlst
%Control: editor formatted (1) identically to author
%Control: production of article title (-1) disabled
%Control: page (0) single
%Control: year (1) truncated
%Control: production of eprint (0) enabled
\begin{thebibliography}{138}%
\makeatletter
\providecommand \@ifxundefined [1]{%
 \@ifx{#1\undefined}
}%
\providecommand \@ifnum [1]{%
 \ifnum #1\expandafter \@firstoftwo
 \else \expandafter \@secondoftwo
 \fi
}%
\providecommand \@ifx [1]{%
 \ifx #1\expandafter \@firstoftwo
 \else \expandafter \@secondoftwo
 \fi
}%
\providecommand \natexlab [1]{#1}%
\providecommand \enquote  [1]{``#1''}%
\providecommand \bibnamefont  [1]{#1}%
\providecommand \bibfnamefont [1]{#1}%
\providecommand \citenamefont [1]{#1}%
\providecommand \href@noop [0]{\@secondoftwo}%
\providecommand \href [0]{\begingroup \@sanitize@url \@href}%
\providecommand \@href[1]{\@@startlink{#1}\@@href}%
\providecommand \@@href[1]{\endgroup#1\@@endlink}%
\providecommand \@sanitize@url [0]{\catcode `\\12\catcode `\$12\catcode
  `\&12\catcode `\#12\catcode `\^12\catcode `\_12\catcode `\%12\relax}%
\providecommand \@@startlink[1]{}%
\providecommand \@@endlink[0]{}%
\providecommand \url  [0]{\begingroup\@sanitize@url \@url }%
\providecommand \@url [1]{\endgroup\@href {#1}{\urlprefix }}%
\providecommand \urlprefix  [0]{URL }%
\providecommand \Eprint [0]{\href }%
\providecommand \doibase [0]{http://dx.doi.org/}%
\providecommand \selectlanguage [0]{\@gobble}%
\providecommand \bibinfo  [0]{\@secondoftwo}%
\providecommand \bibfield  [0]{\@secondoftwo}%
\providecommand \translation [1]{[#1]}%
\providecommand \BibitemOpen [0]{}%
\providecommand \bibitemStop [0]{}%
\providecommand \bibitemNoStop [0]{.\EOS\space}%
\providecommand \EOS [0]{\spacefactor3000\relax}%
\providecommand \BibitemShut  [1]{\csname bibitem#1\endcsname}%
\let\auto@bib@innerbib\@empty
%</preamble>
\bibitem [{\citenamefont {Kuzmin}\ \emph {et~al.}(1985)\citenamefont {Kuzmin},
  \citenamefont {Rubakov},\ and\ \citenamefont {Shaposhnikov}}]{Kuzmin:1985mm}%
  \BibitemOpen
  \bibfield  {author} {\bibinfo {author} {\bibfnamefont {V.~A.}\ \bibnamefont
  {Kuzmin}}, \bibinfo {author} {\bibfnamefont {V.~A.}\ \bibnamefont {Rubakov}},
  \ and\ \bibinfo {author} {\bibfnamefont {M.~E.}\ \bibnamefont
  {Shaposhnikov}},\ }\href {\doibase 10.1016/0370-2693(85)91028-7} {\bibfield
  {journal} {\bibinfo  {journal} {Phys. Lett.}\ }\textbf {\bibinfo {volume}
  {155B}},\ \bibinfo {pages} {36} (\bibinfo {year} {1985})}\BibitemShut
  {NoStop}%
%%CITATION = PHLTA,155B,36;%%
\bibitem [{\citenamefont {Cohen}\ \emph {et~al.}(1990)\citenamefont {Cohen},
  \citenamefont {Kaplan},\ and\ \citenamefont {Nelson}}]{Cohen:1990py}%
  \BibitemOpen
  \bibfield  {author} {\bibinfo {author} {\bibfnamefont {A.~G.}\ \bibnamefont
  {Cohen}}, \bibinfo {author} {\bibfnamefont {D.~B.}\ \bibnamefont {Kaplan}}, \
  and\ \bibinfo {author} {\bibfnamefont {A.~E.}\ \bibnamefont {Nelson}},\
  }\href {\doibase 10.1016/0370-2693(90)90690-8} {\bibfield  {journal}
  {\bibinfo  {journal} {Phys. Lett.}\ }\textbf {\bibinfo {volume} {B245}},\
  \bibinfo {pages} {561} (\bibinfo {year} {1990})}\BibitemShut {NoStop}%
%%CITATION = PHLTA,B245,561;%%
\bibitem [{\citenamefont {Cohen}\ \emph
  {et~al.}(1991{\natexlab{a}})\citenamefont {Cohen}, \citenamefont {Kaplan},\
  and\ \citenamefont {Nelson}}]{Cohen:1990it}%
  \BibitemOpen
  \bibfield  {author} {\bibinfo {author} {\bibfnamefont {A.~G.}\ \bibnamefont
  {Cohen}}, \bibinfo {author} {\bibfnamefont {D.~B.}\ \bibnamefont {Kaplan}}, \
  and\ \bibinfo {author} {\bibfnamefont {A.~E.}\ \bibnamefont {Nelson}},\
  }\href {\doibase 10.1016/0550-3213(91)90395-E} {\bibfield  {journal}
  {\bibinfo  {journal} {Nucl. Phys.}\ }\textbf {\bibinfo {volume} {B349}},\
  \bibinfo {pages} {727} (\bibinfo {year} {1991}{\natexlab{a}})}\BibitemShut
  {NoStop}%
%%CITATION = NUPHA,B349,727;%%
\bibitem [{\citenamefont {Turok}\ and\ \citenamefont
  {Zadrozny}(1990)}]{Turok:1990in}%
  \BibitemOpen
  \bibfield  {author} {\bibinfo {author} {\bibfnamefont {N.}~\bibnamefont
  {Turok}}\ and\ \bibinfo {author} {\bibfnamefont {J.}~\bibnamefont
  {Zadrozny}},\ }\href {\doibase 10.1103/PhysRevLett.65.2331} {\bibfield
  {journal} {\bibinfo  {journal} {Phys. Rev. Lett.}\ }\textbf {\bibinfo
  {volume} {65}},\ \bibinfo {pages} {2331} (\bibinfo {year}
  {1990})}\BibitemShut {NoStop}%
%%CITATION = PRLTA,65,2331;%%
\bibitem [{\citenamefont {Turok}\ and\ \citenamefont
  {Zadrozny}(1991)}]{Turok:1990zg}%
  \BibitemOpen
  \bibfield  {author} {\bibinfo {author} {\bibfnamefont {N.}~\bibnamefont
  {Turok}}\ and\ \bibinfo {author} {\bibfnamefont {J.}~\bibnamefont
  {Zadrozny}},\ }\href {\doibase 10.1016/0550-3213(91)90356-3} {\bibfield
  {journal} {\bibinfo  {journal} {Nucl. Phys.}\ }\textbf {\bibinfo {volume}
  {B358}},\ \bibinfo {pages} {471} (\bibinfo {year} {1991})}\BibitemShut
  {NoStop}%
%%CITATION = NUPHA,B358,471;%%
\bibitem [{\citenamefont {McLerran}\ \emph {et~al.}(1991)\citenamefont
  {McLerran}, \citenamefont {Shaposhnikov}, \citenamefont {Turok},\ and\
  \citenamefont {Voloshin}}]{McLerran:1990zh}%
  \BibitemOpen
  \bibfield  {author} {\bibinfo {author} {\bibfnamefont {L.~D.}\ \bibnamefont
  {McLerran}}, \bibinfo {author} {\bibfnamefont {M.~E.}\ \bibnamefont
  {Shaposhnikov}}, \bibinfo {author} {\bibfnamefont {N.}~\bibnamefont {Turok}},
  \ and\ \bibinfo {author} {\bibfnamefont {M.~B.}\ \bibnamefont {Voloshin}},\
  }\href {\doibase 10.1016/0370-2693(91)91794-V} {\bibfield  {journal}
  {\bibinfo  {journal} {Phys. Lett.}\ }\textbf {\bibinfo {volume} {B256}},\
  \bibinfo {pages} {451} (\bibinfo {year} {1991})}\BibitemShut {NoStop}%
%%CITATION = PHLTA,B256,451;%%
\bibitem [{\citenamefont {Dine}\ \emph {et~al.}(1991)\citenamefont {Dine},
  \citenamefont {Huet}, \citenamefont {Singleton},\ and\ \citenamefont
  {Susskind}}]{Dine:1990fj}%
  \BibitemOpen
  \bibfield  {author} {\bibinfo {author} {\bibfnamefont {M.}~\bibnamefont
  {Dine}}, \bibinfo {author} {\bibfnamefont {P.}~\bibnamefont {Huet}}, \bibinfo
  {author} {\bibfnamefont {R.~L.}\ \bibnamefont {Singleton}, \bibfnamefont
  {Jr}}, \ and\ \bibinfo {author} {\bibfnamefont {L.}~\bibnamefont
  {Susskind}},\ }\href {\doibase 10.1016/0370-2693(91)91905-B} {\bibfield
  {journal} {\bibinfo  {journal} {Phys. Lett.}\ }\textbf {\bibinfo {volume}
  {B257}},\ \bibinfo {pages} {351} (\bibinfo {year} {1991})}\BibitemShut
  {NoStop}%
%%CITATION = PHLTA,B257,351;%%
\bibitem [{\citenamefont {Cohen}\ \emph
  {et~al.}(1991{\natexlab{b}})\citenamefont {Cohen}, \citenamefont {Kaplan},\
  and\ \citenamefont {Nelson}}]{Cohen:1991iu}%
  \BibitemOpen
  \bibfield  {author} {\bibinfo {author} {\bibfnamefont {A.~G.}\ \bibnamefont
  {Cohen}}, \bibinfo {author} {\bibfnamefont {D.~B.}\ \bibnamefont {Kaplan}}, \
  and\ \bibinfo {author} {\bibfnamefont {A.~E.}\ \bibnamefont {Nelson}},\
  }\href {\doibase 10.1016/0370-2693(91)91711-4} {\bibfield  {journal}
  {\bibinfo  {journal} {Phys. Lett.}\ }\textbf {\bibinfo {volume} {B263}},\
  \bibinfo {pages} {86} (\bibinfo {year} {1991}{\natexlab{b}})}\BibitemShut
  {NoStop}%
%%CITATION = PHLTA,B263,86;%%
\bibitem [{\citenamefont {Nelson}\ \emph {et~al.}(1992)\citenamefont {Nelson},
  \citenamefont {Kaplan},\ and\ \citenamefont {Cohen}}]{Nelson:1991ab}%
  \BibitemOpen
  \bibfield  {author} {\bibinfo {author} {\bibfnamefont {A.~E.}\ \bibnamefont
  {Nelson}}, \bibinfo {author} {\bibfnamefont {D.~B.}\ \bibnamefont {Kaplan}},
  \ and\ \bibinfo {author} {\bibfnamefont {A.~G.}\ \bibnamefont {Cohen}},\
  }\href {\doibase 10.1016/0550-3213(92)90440-M} {\bibfield  {journal}
  {\bibinfo  {journal} {Nucl. Phys.}\ }\textbf {\bibinfo {volume} {B373}},\
  \bibinfo {pages} {453} (\bibinfo {year} {1992})}\BibitemShut {NoStop}%
%%CITATION = NUPHA,B373,453;%%
\bibitem [{\citenamefont {Cohen}\ \emph {et~al.}(1992)\citenamefont {Cohen},
  \citenamefont {Kaplan},\ and\ \citenamefont {Nelson}}]{Cohen:1992yh}%
  \BibitemOpen
  \bibfield  {author} {\bibinfo {author} {\bibfnamefont {A.~G.}\ \bibnamefont
  {Cohen}}, \bibinfo {author} {\bibfnamefont {D.~B.}\ \bibnamefont {Kaplan}}, \
  and\ \bibinfo {author} {\bibfnamefont {A.~E.}\ \bibnamefont {Nelson}},\
  }\href {\doibase 10.1016/0370-2693(92)91640-U} {\bibfield  {journal}
  {\bibinfo  {journal} {Phys. Lett.}\ }\textbf {\bibinfo {volume} {B294}},\
  \bibinfo {pages} {57} (\bibinfo {year} {1992})},\ \Eprint
  {http://arxiv.org/abs/hep-ph/9206214} {arXiv:hep-ph/9206214 [hep-ph]}
  \BibitemShut {NoStop}%
%%CITATION = HEP-PH/9206214;%%
\bibitem [{\citenamefont {Farrar}\ and\ \citenamefont
  {Shaposhnikov}(1994)}]{Farrar:1993hn}%
  \BibitemOpen
  \bibfield  {author} {\bibinfo {author} {\bibfnamefont {G.~R.}\ \bibnamefont
  {Farrar}}\ and\ \bibinfo {author} {\bibfnamefont {M.~E.}\ \bibnamefont
  {Shaposhnikov}},\ }\href {\doibase 10.1103/PhysRevD.50.774} {\bibfield
  {journal} {\bibinfo  {journal} {Phys. Rev.}\ }\textbf {\bibinfo {volume}
  {D50}},\ \bibinfo {pages} {774} (\bibinfo {year} {1994})},\ \Eprint
  {http://arxiv.org/abs/hep-ph/9305275} {arXiv:hep-ph/9305275 [hep-ph]}
  \BibitemShut {NoStop}%
%%CITATION = HEP-PH/9305275;%%
\bibitem [{\citenamefont {Fukugita}\ and\ \citenamefont
  {Yanagida}(1986)}]{Fukugita:1986hr}%
  \BibitemOpen
  \bibfield  {author} {\bibinfo {author} {\bibfnamefont {M.}~\bibnamefont
  {Fukugita}}\ and\ \bibinfo {author} {\bibfnamefont {T.}~\bibnamefont
  {Yanagida}},\ }\href {\doibase 10.1016/0370-2693(86)91126-3} {\bibfield
  {journal} {\bibinfo  {journal} {Phys. Lett.}\ }\textbf {\bibinfo {volume}
  {B174}},\ \bibinfo {pages} {45} (\bibinfo {year} {1986})}\BibitemShut
  {NoStop}%
%%CITATION = PHLTA,B174,45;%%
\bibitem [{\citenamefont {Sakharov}(1967)}]{Sakharov:1967dj}%
  \BibitemOpen
  \bibfield  {author} {\bibinfo {author} {\bibfnamefont {A.~D.}\ \bibnamefont
  {Sakharov}},\ }\href {\doibase 10.1070/PU1991v034n05ABEH002497} {\bibfield
  {journal} {\bibinfo  {journal} {Pisma Zh. Eksp. Teor. Fiz.}\ }\textbf
  {\bibinfo {volume} {5}},\ \bibinfo {pages} {32} (\bibinfo {year}
  {1967})}\BibitemShut {NoStop}%
\bibitem [{\citenamefont {Yanagida}(1980)}]{Yanagida:1980xy}%
  \BibitemOpen
  \bibfield  {author} {\bibinfo {author} {\bibfnamefont {T.}~\bibnamefont
  {Yanagida}},\ }\href {\doibase 10.1143/PTP.64.1103} {\bibfield  {journal}
  {\bibinfo  {journal} {Prog. Theor. Phys.}\ }\textbf {\bibinfo {volume}
  {64}},\ \bibinfo {pages} {1103} (\bibinfo {year} {1980})}\BibitemShut
  {NoStop}%
\bibitem [{\citenamefont {Minkowski}(1977)}]{Minkowski:1977sc}%
  \BibitemOpen
  \bibfield  {author} {\bibinfo {author} {\bibfnamefont {P.}~\bibnamefont
  {Minkowski}},\ }\href {\doibase 10.1016/0370-2693(77)90435-X} {\bibfield
  {journal} {\bibinfo  {journal} {Phys. Lett.}\ }\textbf {\bibinfo {volume}
  {67B}},\ \bibinfo {pages} {421} (\bibinfo {year} {1977})}\BibitemShut
  {NoStop}%
%%CITATION = PHLTA,67B,421;%%
\bibitem [{\citenamefont {Gell-Mann}\ \emph {et~al.}(1979)\citenamefont
  {Gell-Mann}, \citenamefont {Ramond},\ and\ \citenamefont
  {Slansky}}]{GellMann:1980vs}%
  \BibitemOpen
  \bibfield  {author} {\bibinfo {author} {\bibfnamefont {M.}~\bibnamefont
  {Gell-Mann}}, \bibinfo {author} {\bibfnamefont {P.}~\bibnamefont {Ramond}}, \
  and\ \bibinfo {author} {\bibfnamefont {R.}~\bibnamefont {Slansky}},\
  }\bibfield  {booktitle} {\emph {\bibinfo {booktitle} {{Supergravity Workshop
  Stony Brook, New York, September 27-28, 1979}}},\ }\href@noop {} {\bibfield
  {journal} {\bibinfo  {journal} {Conf. Proc.}\ }\textbf {\bibinfo {volume}
  {C790927}},\ \bibinfo {pages} {315} (\bibinfo {year} {1979})},\ \Eprint
  {http://arxiv.org/abs/1306.4669} {arXiv:1306.4669 [hep-th]} \BibitemShut
  {NoStop}%
%%CITATION = ARXIV:1306.4669;%%
\bibitem [{\citenamefont {Liebler}\ \emph {et~al.}(2016)\citenamefont
  {Liebler}, \citenamefont {Profumo},\ and\ \citenamefont
  {Stefaniak}}]{Liebler:2015ddv}%
  \BibitemOpen
  \bibfield  {author} {\bibinfo {author} {\bibfnamefont {S.}~\bibnamefont
  {Liebler}}, \bibinfo {author} {\bibfnamefont {S.}~\bibnamefont {Profumo}}, \
  and\ \bibinfo {author} {\bibfnamefont {T.}~\bibnamefont {Stefaniak}},\ }\href
  {\doibase 10.1007/JHEP04(2016)143} {\bibfield  {journal} {\bibinfo  {journal}
  {JHEP}\ }\textbf {\bibinfo {volume} {04}},\ \bibinfo {pages} {143} (\bibinfo
  {year} {2016})},\ \Eprint {http://arxiv.org/abs/1512.09172} {arXiv:1512.09172
  [hep-ph]} \BibitemShut {NoStop}%
\bibitem [{\citenamefont {Kurup}\ and\ \citenamefont
  {Perelstein}(2017)}]{Kurup:2017dzf}%
  \BibitemOpen
  \bibfield  {author} {\bibinfo {author} {\bibfnamefont {G.}~\bibnamefont
  {Kurup}}\ and\ \bibinfo {author} {\bibfnamefont {M.}~\bibnamefont
  {Perelstein}},\ }\href {\doibase 10.1103/PhysRevD.96.015036} {\bibfield
  {journal} {\bibinfo  {journal} {Phys. Rev. D}\ }\textbf {\bibinfo {volume}
  {96}},\ \bibinfo {pages} {015036} (\bibinfo {year} {2017})},\ \Eprint
  {http://arxiv.org/abs/1704.03381} {arXiv:1704.03381 [hep-ph]} \BibitemShut
  {NoStop}%
\bibitem [{\citenamefont {Baum}\ \emph {et~al.}(2021)\citenamefont {Baum},
  \citenamefont {Carena}, \citenamefont {Shah}, \citenamefont {Wagner},\ and\
  \citenamefont {Wang}}]{Baum:2020vfl}%
  \BibitemOpen
  \bibfield  {author} {\bibinfo {author} {\bibfnamefont {S.}~\bibnamefont
  {Baum}}, \bibinfo {author} {\bibfnamefont {M.}~\bibnamefont {Carena}},
  \bibinfo {author} {\bibfnamefont {N.~R.}\ \bibnamefont {Shah}}, \bibinfo
  {author} {\bibfnamefont {C.~E.~M.}\ \bibnamefont {Wagner}}, \ and\ \bibinfo
  {author} {\bibfnamefont {Y.}~\bibnamefont {Wang}},\ }\href {\doibase
  10.1007/JHEP03(2021)055} {\bibfield  {journal} {\bibinfo  {journal} {JHEP}\
  }\textbf {\bibinfo {volume} {03}},\ \bibinfo {pages} {055} (\bibinfo {year}
  {2021})},\ \Eprint {http://arxiv.org/abs/2009.10743} {arXiv:2009.10743
  [hep-ph]} \BibitemShut {NoStop}%
\bibitem [{\citenamefont {Cline}\ and\ \citenamefont
  {Laurent}(2021)}]{Cline:2021dkf}%
  \BibitemOpen
  \bibfield  {author} {\bibinfo {author} {\bibfnamefont {J.~M.}\ \bibnamefont
  {Cline}}\ and\ \bibinfo {author} {\bibfnamefont {B.}~\bibnamefont
  {Laurent}},\ }\href@noop {} {\  (\bibinfo {year} {2021})},\ \Eprint
  {http://arxiv.org/abs/2108.04249} {arXiv:2108.04249 [hep-ph]} \BibitemShut
  {NoStop}%
\bibitem [{\citenamefont {Andreev}\ \emph {et~al.}(2018)\citenamefont {Andreev}
  \emph {et~al.}}]{Andreev:2018ayy}%
  \BibitemOpen
  \bibfield  {author} {\bibinfo {author} {\bibfnamefont {V.}~\bibnamefont
  {Andreev}} \emph {et~al.} (\bibinfo {collaboration} {ACME}),\ }\href
  {\doibase 10.1038/s41586-018-0599-8} {\bibfield  {journal} {\bibinfo
  {journal} {Nature}\ }\textbf {\bibinfo {volume} {562}},\ \bibinfo {pages}
  {355} (\bibinfo {year} {2018})}\BibitemShut {NoStop}%
%%CITATION = NATUA,562,355;%%
\bibitem [{\citenamefont {Hall}\ \emph {et~al.}(2021)\citenamefont {Hall},
  \citenamefont {McGehee}, \citenamefont {Murayama},\ and\ \citenamefont
  {Suter}}]{Hall:2021zsk}%
  \BibitemOpen
  \bibfield  {author} {\bibinfo {author} {\bibfnamefont {E.}~\bibnamefont
  {Hall}}, \bibinfo {author} {\bibfnamefont {R.}~\bibnamefont {McGehee}},
  \bibinfo {author} {\bibfnamefont {H.}~\bibnamefont {Murayama}}, \ and\
  \bibinfo {author} {\bibfnamefont {B.}~\bibnamefont {Suter}},\ }\href@noop {}
  {\  (\bibinfo {year} {2021})},\ \Eprint {http://arxiv.org/abs/2107.03398}
  {arXiv:2107.03398 [hep-ph]} \BibitemShut {NoStop}%
\bibitem [{\citenamefont {Shelton}\ and\ \citenamefont
  {Zurek}(2010)}]{Shelton:2010ta}%
  \BibitemOpen
  \bibfield  {author} {\bibinfo {author} {\bibfnamefont {J.}~\bibnamefont
  {Shelton}}\ and\ \bibinfo {author} {\bibfnamefont {K.~M.}\ \bibnamefont
  {Zurek}},\ }\href {\doibase 10.1103/PhysRevD.82.123512} {\bibfield  {journal}
  {\bibinfo  {journal} {Phys. Rev.}\ }\textbf {\bibinfo {volume} {D82}},\
  \bibinfo {pages} {123512} (\bibinfo {year} {2010})},\ \Eprint
  {http://arxiv.org/abs/1008.1997} {arXiv:1008.1997 [hep-ph]} \BibitemShut
  {NoStop}%
%%CITATION = ARXIV:1008.1997;%%
\bibitem [{\citenamefont {Servant}\ and\ \citenamefont
  {Tulin}(2013)}]{Servant:2013uwa}%
  \BibitemOpen
  \bibfield  {author} {\bibinfo {author} {\bibfnamefont {G.}~\bibnamefont
  {Servant}}\ and\ \bibinfo {author} {\bibfnamefont {S.}~\bibnamefont
  {Tulin}},\ }\href {\doibase 10.1103/PhysRevLett.111.151601} {\bibfield
  {journal} {\bibinfo  {journal} {Phys. Rev. Lett.}\ }\textbf {\bibinfo
  {volume} {111}},\ \bibinfo {pages} {151601} (\bibinfo {year} {2013})},\
  \Eprint {http://arxiv.org/abs/1304.3464} {arXiv:1304.3464 [hep-ph]}
  \BibitemShut {NoStop}%
%%CITATION = ARXIV:1304.3464;%%
\bibitem [{\citenamefont {Cline}\ \emph {et~al.}(2017)\citenamefont {Cline},
  \citenamefont {Kainulainen},\ and\ \citenamefont
  {Tucker-Smith}}]{Cline:2017qpe}%
  \BibitemOpen
  \bibfield  {author} {\bibinfo {author} {\bibfnamefont {J.~M.}\ \bibnamefont
  {Cline}}, \bibinfo {author} {\bibfnamefont {K.}~\bibnamefont {Kainulainen}},
  \ and\ \bibinfo {author} {\bibfnamefont {D.}~\bibnamefont {Tucker-Smith}},\
  }\href {\doibase 10.1103/PhysRevD.95.115006} {\bibfield  {journal} {\bibinfo
  {journal} {Phys. Rev. D}\ }\textbf {\bibinfo {volume} {95}},\ \bibinfo
  {pages} {115006} (\bibinfo {year} {2017})},\ \Eprint
  {http://arxiv.org/abs/1702.08909} {arXiv:1702.08909 [hep-ph]} \BibitemShut
  {NoStop}%
\bibitem [{\citenamefont {Carena}\ \emph {et~al.}(2019)\citenamefont {Carena},
  \citenamefont {Quir\'os},\ and\ \citenamefont {Zhang}}]{Carena:2018cjh}%
  \BibitemOpen
  \bibfield  {author} {\bibinfo {author} {\bibfnamefont {M.}~\bibnamefont
  {Carena}}, \bibinfo {author} {\bibfnamefont {M.}~\bibnamefont {Quir\'os}}, \
  and\ \bibinfo {author} {\bibfnamefont {Y.}~\bibnamefont {Zhang}},\ }\href
  {\doibase 10.1103/PhysRevLett.122.201802} {\bibfield  {journal} {\bibinfo
  {journal} {Phys. Rev. Lett.}\ }\textbf {\bibinfo {volume} {122}},\ \bibinfo
  {pages} {201802} (\bibinfo {year} {2019})},\ \Eprint
  {http://arxiv.org/abs/1811.09719} {arXiv:1811.09719 [hep-ph]} \BibitemShut
  {NoStop}%
\bibitem [{\citenamefont {Hall}\ \emph {et~al.}(2020)\citenamefont {Hall},
  \citenamefont {Konstandin}, \citenamefont {McGehee}, \citenamefont
  {Murayama},\ and\ \citenamefont {Servant}}]{Hall:2019ank}%
  \BibitemOpen
  \bibfield  {author} {\bibinfo {author} {\bibfnamefont {E.}~\bibnamefont
  {Hall}}, \bibinfo {author} {\bibfnamefont {T.}~\bibnamefont {Konstandin}},
  \bibinfo {author} {\bibfnamefont {R.}~\bibnamefont {McGehee}}, \bibinfo
  {author} {\bibfnamefont {H.}~\bibnamefont {Murayama}}, \ and\ \bibinfo
  {author} {\bibfnamefont {G.}~\bibnamefont {Servant}},\ }\href {\doibase
  10.1007/JHEP04(2020)042} {\bibfield  {journal} {\bibinfo  {journal} {JHEP}\
  }\textbf {\bibinfo {volume} {04}},\ \bibinfo {pages} {042} (\bibinfo {year}
  {2020})},\ \Eprint {http://arxiv.org/abs/1910.08068} {arXiv:1910.08068
  [hep-ph]} \BibitemShut {NoStop}%
\bibitem [{\citenamefont {Hall}\ \emph {et~al.}(2019)\citenamefont {Hall},
  \citenamefont {Konstandin}, \citenamefont {McGehee},\ and\ \citenamefont
  {Murayama}}]{Hall:2019rld}%
  \BibitemOpen
  \bibfield  {author} {\bibinfo {author} {\bibfnamefont {E.}~\bibnamefont
  {Hall}}, \bibinfo {author} {\bibfnamefont {T.}~\bibnamefont {Konstandin}},
  \bibinfo {author} {\bibfnamefont {R.}~\bibnamefont {McGehee}}, \ and\
  \bibinfo {author} {\bibfnamefont {H.}~\bibnamefont {Murayama}},\ }\href@noop
  {} {\  (\bibinfo {year} {2019})},\ \Eprint {http://arxiv.org/abs/1911.12342}
  {arXiv:1911.12342 [hep-ph]} \BibitemShut {NoStop}%
\bibitem [{\citenamefont {Buchmuller}\ \emph {et~al.}(2005)\citenamefont
  {Buchmuller}, \citenamefont {Di~Bari},\ and\ \citenamefont
  {Plumacher}}]{Buchmuller:2004nz}%
  \BibitemOpen
  \bibfield  {author} {\bibinfo {author} {\bibfnamefont {W.}~\bibnamefont
  {Buchmuller}}, \bibinfo {author} {\bibfnamefont {P.}~\bibnamefont {Di~Bari}},
  \ and\ \bibinfo {author} {\bibfnamefont {M.}~\bibnamefont {Plumacher}},\
  }\href {\doibase 10.1016/j.aop.2004.02.003} {\bibfield  {journal} {\bibinfo
  {journal} {Annals Phys.}\ }\textbf {\bibinfo {volume} {315}},\ \bibinfo
  {pages} {305} (\bibinfo {year} {2005})},\ \Eprint
  {http://arxiv.org/abs/hep-ph/0401240} {arXiv:hep-ph/0401240} \BibitemShut
  {NoStop}%
\bibitem [{\citenamefont {Elor}\ \emph {et~al.}(2019)\citenamefont {Elor},
  \citenamefont {Escudero},\ and\ \citenamefont {Nelson}}]{Elor:2018twp}%
  \BibitemOpen
  \bibfield  {author} {\bibinfo {author} {\bibfnamefont {G.}~\bibnamefont
  {Elor}}, \bibinfo {author} {\bibfnamefont {M.}~\bibnamefont {Escudero}}, \
  and\ \bibinfo {author} {\bibfnamefont {A.}~\bibnamefont {Nelson}},\ }\href
  {\doibase 10.1103/PhysRevD.99.035031} {\bibfield  {journal} {\bibinfo
  {journal} {Phys. Rev.}\ }\textbf {\bibinfo {volume} {D99}},\ \bibinfo {pages}
  {035031} (\bibinfo {year} {2019})},\ \Eprint
  {http://arxiv.org/abs/1810.00880} {arXiv:1810.00880 [hep-ph]} \BibitemShut
  {NoStop}%
%%CITATION = ARXIV:1810.00880;%%
\bibitem [{\citenamefont {Elor}\ and\ \citenamefont
  {McGehee}(2021)}]{Elor:2020tkc}%
  \BibitemOpen
  \bibfield  {author} {\bibinfo {author} {\bibfnamefont {G.}~\bibnamefont
  {Elor}}\ and\ \bibinfo {author} {\bibfnamefont {R.}~\bibnamefont {McGehee}},\
  }\href {\doibase 10.1103/PhysRevD.103.035005} {\bibfield  {journal} {\bibinfo
   {journal} {Phys. Rev. D}\ }\textbf {\bibinfo {volume} {103}},\ \bibinfo
  {pages} {035005} (\bibinfo {year} {2021})},\ \Eprint
  {http://arxiv.org/abs/2011.06115} {arXiv:2011.06115 [hep-ph]} \BibitemShut
  {NoStop}%
\bibitem [{\citenamefont {Jarlskog}(1985)}]{Jarlskog:1985ht}%
  \BibitemOpen
  \bibfield  {author} {\bibinfo {author} {\bibfnamefont {C.}~\bibnamefont
  {Jarlskog}},\ }\href {\doibase 10.1103/PhysRevLett.55.1039} {\bibfield
  {journal} {\bibinfo  {journal} {Phys. Rev. Lett.}\ }\textbf {\bibinfo
  {volume} {55}},\ \bibinfo {pages} {1039} (\bibinfo {year}
  {1985})}\BibitemShut {NoStop}%
\bibitem [{\citenamefont {Gavela}\ \emph
  {et~al.}(1994{\natexlab{a}})\citenamefont {Gavela}, \citenamefont
  {Hernandez}, \citenamefont {Orloff},\ and\ \citenamefont
  {Pene}}]{Gavela:1993ts}%
  \BibitemOpen
  \bibfield  {author} {\bibinfo {author} {\bibfnamefont {M.~B.}\ \bibnamefont
  {Gavela}}, \bibinfo {author} {\bibfnamefont {P.}~\bibnamefont {Hernandez}},
  \bibinfo {author} {\bibfnamefont {J.}~\bibnamefont {Orloff}}, \ and\ \bibinfo
  {author} {\bibfnamefont {O.}~\bibnamefont {Pene}},\ }\href {\doibase
  10.1142/S0217732394000629} {\bibfield  {journal} {\bibinfo  {journal} {Mod.
  Phys. Lett. A}\ }\textbf {\bibinfo {volume} {9}},\ \bibinfo {pages} {795}
  (\bibinfo {year} {1994}{\natexlab{a}})},\ \Eprint
  {http://arxiv.org/abs/hep-ph/9312215} {arXiv:hep-ph/9312215} \BibitemShut
  {NoStop}%
\bibitem [{\citenamefont {Gavela}\ \emph
  {et~al.}(1994{\natexlab{b}})\citenamefont {Gavela}, \citenamefont
  {Hernandez}, \citenamefont {Orloff}, \citenamefont {Pene},\ and\
  \citenamefont {Quimbay}}]{Gavela:1994dt}%
  \BibitemOpen
  \bibfield  {author} {\bibinfo {author} {\bibfnamefont {M.~B.}\ \bibnamefont
  {Gavela}}, \bibinfo {author} {\bibfnamefont {P.}~\bibnamefont {Hernandez}},
  \bibinfo {author} {\bibfnamefont {J.}~\bibnamefont {Orloff}}, \bibinfo
  {author} {\bibfnamefont {O.}~\bibnamefont {Pene}}, \ and\ \bibinfo {author}
  {\bibfnamefont {C.}~\bibnamefont {Quimbay}},\ }\href {\doibase
  10.1016/0550-3213(94)00410-2} {\bibfield  {journal} {\bibinfo  {journal}
  {Nucl. Phys. B}\ }\textbf {\bibinfo {volume} {430}},\ \bibinfo {pages} {382}
  (\bibinfo {year} {1994}{\natexlab{b}})},\ \Eprint
  {http://arxiv.org/abs/hep-ph/9406289} {arXiv:hep-ph/9406289} \BibitemShut
  {NoStop}%
\bibitem [{\citenamefont {Huet}\ and\ \citenamefont
  {Sather}(1995)}]{Huet:1994jb}%
  \BibitemOpen
  \bibfield  {author} {\bibinfo {author} {\bibfnamefont {P.}~\bibnamefont
  {Huet}}\ and\ \bibinfo {author} {\bibfnamefont {E.}~\bibnamefont {Sather}},\
  }\href {\doibase 10.1103/PhysRevD.51.379} {\bibfield  {journal} {\bibinfo
  {journal} {Phys. Rev. D}\ }\textbf {\bibinfo {volume} {51}},\ \bibinfo
  {pages} {379} (\bibinfo {year} {1995})},\ \Eprint
  {http://arxiv.org/abs/hep-ph/9404302} {arXiv:hep-ph/9404302} \BibitemShut
  {NoStop}%
\bibitem [{\citenamefont {Alonso-\'Alvarez}\ \emph
  {et~al.}(2021{\natexlab{a}})\citenamefont {Alonso-\'Alvarez}, \citenamefont
  {Elor},\ and\ \citenamefont {Escudero}}]{Alonso-Alvarez:2021qfd}%
  \BibitemOpen
  \bibfield  {author} {\bibinfo {author} {\bibfnamefont {G.}~\bibnamefont
  {Alonso-\'Alvarez}}, \bibinfo {author} {\bibfnamefont {G.}~\bibnamefont
  {Elor}}, \ and\ \bibinfo {author} {\bibfnamefont {M.}~\bibnamefont
  {Escudero}},\ }\href {\doibase 10.1103/PhysRevD.104.035028} {\bibfield
  {journal} {\bibinfo  {journal} {Phys. Rev. D}\ }\textbf {\bibinfo {volume}
  {104}},\ \bibinfo {pages} {035028} (\bibinfo {year} {2021}{\natexlab{a}})},\
  \Eprint {http://arxiv.org/abs/2101.02706} {arXiv:2101.02706 [hep-ph]}
  \BibitemShut {NoStop}%
\bibitem [{\citenamefont {Alonso-Alvarex}\ \emph {et~al.}(2021)\citenamefont
  {Alonso-Alvarex}, \citenamefont {Escudero}, \citenamefont {Elor},
  \citenamefont {Fornal}, \citenamefont {Grinstein},\ and\ \citenamefont
  {Camalich}}]{Hyperons}%
  \BibitemOpen
  \bibfield  {author} {\bibinfo {author} {\bibfnamefont {G.}~\bibnamefont
  {Alonso-Alvarex}}, \bibinfo {author} {\bibfnamefont {M.}~\bibnamefont
  {Escudero}}, \bibinfo {author} {\bibfnamefont {G.}~\bibnamefont {Elor}},
  \bibinfo {author} {\bibfnamefont {B.}~\bibnamefont {Fornal}}, \bibinfo
  {author} {\bibfnamefont {B.}~\bibnamefont {Grinstein}}, \ and\ \bibinfo
  {author} {\bibfnamefont {J.~M.}\ \bibnamefont {Camalich}},\ }\href@noop {} {\
   (\bibinfo {year} {2021})},\ \Eprint {http://arxiv.org/abs/2110.XXXXX}
  {arXiv:2110.XXXXX [hep-ph]} \BibitemShut {NoStop}%
\bibitem [{\citenamefont {Borsato}\ \emph {et~al.}(2021)\citenamefont {Borsato}
  \emph {et~al.}}]{Borsato:2021aum}%
  \BibitemOpen
  \bibfield  {author} {\bibinfo {author} {\bibfnamefont {M.}~\bibnamefont
  {Borsato}} \emph {et~al.},\ }\href@noop {} {\  (\bibinfo {year} {2021})},\
  \Eprint {http://arxiv.org/abs/2105.12668} {arXiv:2105.12668 [hep-ph]}
  \BibitemShut {NoStop}%
\bibitem [{\citenamefont {Rodr\'\i{}guez}\ \emph {et~al.}(2021)\citenamefont
  {Rodr\'\i{}guez}, \citenamefont {Chobanova}, \citenamefont {Cid~Vidal},
  \citenamefont {Soli\~no}, \citenamefont {Santos}, \citenamefont
  {Momb\"acher}, \citenamefont {Prouv\'e}, \citenamefont {Fern\'andez},\ and\
  \citenamefont {V\'azquez~Sierra}}]{Rodriguez:2021urv}%
  \BibitemOpen
  \bibfield  {author} {\bibinfo {author} {\bibfnamefont {A.~B.}\ \bibnamefont
  {Rodr\'\i{}guez}}, \bibinfo {author} {\bibfnamefont {V.}~\bibnamefont
  {Chobanova}}, \bibinfo {author} {\bibfnamefont {X.}~\bibnamefont
  {Cid~Vidal}}, \bibinfo {author} {\bibfnamefont {S.~L.}\ \bibnamefont
  {Soli\~no}}, \bibinfo {author} {\bibfnamefont {D.~M.}\ \bibnamefont
  {Santos}}, \bibinfo {author} {\bibfnamefont {T.}~\bibnamefont {Momb\"acher}},
  \bibinfo {author} {\bibfnamefont {C.}~\bibnamefont {Prouv\'e}}, \bibinfo
  {author} {\bibfnamefont {E.~X.~R.}\ \bibnamefont {Fern\'andez}}, \ and\
  \bibinfo {author} {\bibfnamefont {C.}~\bibnamefont {V\'azquez~Sierra}},\
  }\href@noop {} {\  (\bibinfo {year} {2021})},\ \Eprint
  {http://arxiv.org/abs/2106.12870} {arXiv:2106.12870 [hep-ph]} \BibitemShut
  {NoStop}%
\bibitem [{\citenamefont {with Belle-II}(2021)}]{Belle2}%
  \BibitemOpen
  \bibfield  {author} {\bibinfo {author} {\bibfnamefont {C.}~\bibnamefont {with
  Belle-II}},\ }\href@noop {} {\  (\bibinfo {year} {2021})},\ \Eprint
  {http://arxiv.org/abs/2110.XXXXX} {arXiv:2110.XXXXX [hep-ph]} \BibitemShut
  {NoStop}%
\bibitem [{\citenamefont {Tanabashi}\ \emph {et~al.}(2018)\citenamefont
  {Tanabashi} \emph {et~al.}}]{pdg}%
  \BibitemOpen
  \bibfield  {author} {\bibinfo {author} {\bibfnamefont {M.}~\bibnamefont
  {Tanabashi}} \emph {et~al.} (\bibinfo {collaboration} {ParticleDataGroup}),\
  }\href {\doibase 10.1103/PhysRevD.98.030001} {\bibfield  {journal} {\bibinfo
  {journal} {Phys. Rev.}\ }\textbf {\bibinfo {volume} {D98}},\ \bibinfo {pages}
  {030001} (\bibinfo {year} {2018})}\BibitemShut {NoStop}%
%%CITATION = PHRVA,D98,030001;%%
\bibitem [{\citenamefont {Djouadi}\ \emph {et~al.}(1998)\citenamefont
  {Djouadi}, \citenamefont {Kalinowski},\ and\ \citenamefont
  {Spira}}]{Djouadi:1997yw}%
  \BibitemOpen
  \bibfield  {author} {\bibinfo {author} {\bibfnamefont {A.}~\bibnamefont
  {Djouadi}}, \bibinfo {author} {\bibfnamefont {J.}~\bibnamefont {Kalinowski}},
  \ and\ \bibinfo {author} {\bibfnamefont {M.}~\bibnamefont {Spira}},\ }\href
  {\doibase 10.1016/S0010-4655(97)00123-9} {\bibfield  {journal} {\bibinfo
  {journal} {Comput. Phys. Commun.}\ }\textbf {\bibinfo {volume} {108}},\
  \bibinfo {pages} {56} (\bibinfo {year} {1998})},\ \Eprint
  {http://arxiv.org/abs/hep-ph/9704448} {arXiv:hep-ph/9704448 [hep-ph]}
  \BibitemShut {NoStop}%
%%CITATION = HEP-PH/9704448;%%
\bibitem [{\citenamefont {Alwall}\ \emph {et~al.}(2014)\citenamefont {Alwall},
  \citenamefont {Frederix}, \citenamefont {Frixione}, \citenamefont {Hirschi},
  \citenamefont {Maltoni}, \citenamefont {Mattelaer}, \citenamefont {Shao},
  \citenamefont {Stelzer}, \citenamefont {Torrielli},\ and\ \citenamefont
  {Zaro}}]{Alwall:2014}%
  \BibitemOpen
  \bibfield  {author} {\bibinfo {author} {\bibfnamefont {J.}~\bibnamefont
  {Alwall}}, \bibinfo {author} {\bibfnamefont {R.}~\bibnamefont {Frederix}},
  \bibinfo {author} {\bibfnamefont {S.}~\bibnamefont {Frixione}}, \bibinfo
  {author} {\bibfnamefont {V.}~\bibnamefont {Hirschi}}, \bibinfo {author}
  {\bibfnamefont {F.}~\bibnamefont {Maltoni}}, \bibinfo {author} {\bibfnamefont
  {O.}~\bibnamefont {Mattelaer}}, \bibinfo {author} {\bibfnamefont {H.-S.}\
  \bibnamefont {Shao}}, \bibinfo {author} {\bibfnamefont {T.}~\bibnamefont
  {Stelzer}}, \bibinfo {author} {\bibfnamefont {P.}~\bibnamefont {Torrielli}},
  \ and\ \bibinfo {author} {\bibfnamefont {M.}~\bibnamefont {Zaro}},\ }\href
  {\doibase 10.1007/jhep07(2014)079} {\bibfield  {journal} {\bibinfo  {journal}
  {Journal of High Energy Physics}\ }\textbf {\bibinfo {volume} {2014}}
  (\bibinfo {year} {2014}),\ 10.1007/jhep07(2014)079}\BibitemShut {NoStop}%
\bibitem [{\citenamefont {Alwall}\ \emph {et~al.}(2011)\citenamefont {Alwall},
  \citenamefont {Herquet}, \citenamefont {Maltoni}, \citenamefont {Mattelaer},\
  and\ \citenamefont {Stelzer}}]{Alwall:2011}%
  \BibitemOpen
  \bibfield  {author} {\bibinfo {author} {\bibfnamefont {J.}~\bibnamefont
  {Alwall}}, \bibinfo {author} {\bibfnamefont {M.}~\bibnamefont {Herquet}},
  \bibinfo {author} {\bibfnamefont {F.}~\bibnamefont {Maltoni}}, \bibinfo
  {author} {\bibfnamefont {O.}~\bibnamefont {Mattelaer}}, \ and\ \bibinfo
  {author} {\bibfnamefont {T.}~\bibnamefont {Stelzer}},\ }\href {\doibase
  10.1007/jhep06(2011)128} {\bibfield  {journal} {\bibinfo  {journal} {Journal
  of High Energy Physics}\ }\textbf {\bibinfo {volume} {2011}} (\bibinfo {year}
  {2011}),\ 10.1007/jhep06(2011)128}\BibitemShut {NoStop}%
\bibitem [{\citenamefont {Choi}\ and\ \citenamefont {Ji}(2009)}]{Choi:2009ym}%
  \BibitemOpen
  \bibfield  {author} {\bibinfo {author} {\bibfnamefont {H.-M.}\ \bibnamefont
  {Choi}}\ and\ \bibinfo {author} {\bibfnamefont {C.-R.}\ \bibnamefont {Ji}},\
  }\href {\doibase 10.1103/PhysRevD.80.114003} {\bibfield  {journal} {\bibinfo
  {journal} {Phys. Rev. D}\ }\textbf {\bibinfo {volume} {80}},\ \bibinfo
  {pages} {114003} (\bibinfo {year} {2009})},\ \Eprint
  {http://arxiv.org/abs/0909.5028} {arXiv:0909.5028 [hep-ph]} \BibitemShut
  {NoStop}%
\bibitem [{\citenamefont {McKeen}\ \emph {et~al.}(2018)\citenamefont {McKeen},
  \citenamefont {Nelson}, \citenamefont {Reddy},\ and\ \citenamefont
  {Zhou}}]{McKeen:2018xwc}%
  \BibitemOpen
  \bibfield  {author} {\bibinfo {author} {\bibfnamefont {D.}~\bibnamefont
  {McKeen}}, \bibinfo {author} {\bibfnamefont {A.~E.}\ \bibnamefont {Nelson}},
  \bibinfo {author} {\bibfnamefont {S.}~\bibnamefont {Reddy}}, \ and\ \bibinfo
  {author} {\bibfnamefont {D.}~\bibnamefont {Zhou}},\ }\href {\doibase
  10.1103/PhysRevLett.121.061802} {\bibfield  {journal} {\bibinfo  {journal}
  {Phys. Rev. Lett.}\ }\textbf {\bibinfo {volume} {121}},\ \bibinfo {pages}
  {061802} (\bibinfo {year} {2018})},\ \Eprint
  {http://arxiv.org/abs/1802.08244} {arXiv:1802.08244 [hep-ph]} \BibitemShut
  {NoStop}%
\bibitem [{\citenamefont {McKeen}\ \emph {et~al.}(2021)\citenamefont {McKeen},
  \citenamefont {Pospelov},\ and\ \citenamefont {Raj}}]{McKeen:2020oyr}%
  \BibitemOpen
  \bibfield  {author} {\bibinfo {author} {\bibfnamefont {D.}~\bibnamefont
  {McKeen}}, \bibinfo {author} {\bibfnamefont {M.}~\bibnamefont {Pospelov}}, \
  and\ \bibinfo {author} {\bibfnamefont {N.}~\bibnamefont {Raj}},\ }\href
  {\doibase 10.1103/PhysRevD.103.115002} {\bibfield  {journal} {\bibinfo
  {journal} {Phys. Rev. D}\ }\textbf {\bibinfo {volume} {103}},\ \bibinfo
  {pages} {115002} (\bibinfo {year} {2021})},\ \Eprint
  {http://arxiv.org/abs/2012.09865} {arXiv:2012.09865 [hep-ph]} \BibitemShut
  {NoStop}%
\bibitem [{\citenamefont {Cline}\ and\ \citenamefont
  {Cornell}(2018)}]{Cline:2018ami}%
  \BibitemOpen
  \bibfield  {author} {\bibinfo {author} {\bibfnamefont {J.~M.}\ \bibnamefont
  {Cline}}\ and\ \bibinfo {author} {\bibfnamefont {J.~M.}\ \bibnamefont
  {Cornell}},\ }\href {\doibase 10.1007/JHEP07(2018)081} {\bibfield  {journal}
  {\bibinfo  {journal} {JHEP}\ }\textbf {\bibinfo {volume} {07}},\ \bibinfo
  {pages} {081} (\bibinfo {year} {2018})},\ \Eprint
  {http://arxiv.org/abs/1803.04961} {arXiv:1803.04961 [hep-ph]} \BibitemShut
  {NoStop}%
\bibitem [{\citenamefont {Alonso-\'{A}lvarez}\ \emph
  {et~al.}(2020)\citenamefont {Alonso-\'{A}lvarez}, \citenamefont {Elor},
  \citenamefont {Nelson},\ and\ \citenamefont {Xiao}}]{Alonso-Alvarez:2019fym}%
  \BibitemOpen
  \bibfield  {author} {\bibinfo {author} {\bibfnamefont {G.}~\bibnamefont
  {Alonso-\'{A}lvarez}}, \bibinfo {author} {\bibfnamefont {G.}~\bibnamefont
  {Elor}}, \bibinfo {author} {\bibfnamefont {A.~E.}\ \bibnamefont {Nelson}}, \
  and\ \bibinfo {author} {\bibfnamefont {H.}~\bibnamefont {Xiao}},\ }\href
  {\doibase 10.1007/JHEP03(2020)046} {\bibfield  {journal} {\bibinfo  {journal}
  {JHEP}\ }\textbf {\bibinfo {volume} {03}},\ \bibinfo {pages} {046} (\bibinfo
  {year} {2020})},\ \Eprint {http://arxiv.org/abs/1907.10612} {arXiv:1907.10612
  [hep-ph]} \BibitemShut {NoStop}%
%%CITATION = ARXIV:1907.10612;%%
\bibitem [{\citenamefont {Zyla}\ \emph {et~al.}(2020)\citenamefont {Zyla} \emph
  {et~al.}}]{Zyla:2020zbs}%
  \BibitemOpen
  \bibfield  {author} {\bibinfo {author} {\bibfnamefont {P.}~\bibnamefont
  {Zyla}} \emph {et~al.} (\bibinfo {collaboration} {Particle Data Group}),\
  }\href {\doibase 10.1093/ptep/ptaa104} {\bibfield  {journal} {\bibinfo
  {journal} {PTEP}\ }\textbf {\bibinfo {volume} {2020}},\ \bibinfo {pages}
  {083C01} (\bibinfo {year} {2020})}\BibitemShut {NoStop}%
\bibitem [{\citenamefont {Barate}\ \emph {et~al.}(2001)\citenamefont {Barate}
  \emph {et~al.}}]{ALEPH:2000vvi}%
  \BibitemOpen
  \bibfield  {author} {\bibinfo {author} {\bibfnamefont {R.}~\bibnamefont
  {Barate}} \emph {et~al.} (\bibinfo {collaboration} {ALEPH}),\ }\href
  {\doibase 10.1007/s100520100612} {\bibfield  {journal} {\bibinfo  {journal}
  {Eur. Phys. J. C}\ }\textbf {\bibinfo {volume} {19}},\ \bibinfo {pages} {213}
  (\bibinfo {year} {2001})},\ \Eprint {http://arxiv.org/abs/hep-ex/0010022}
  {arXiv:hep-ex/0010022} \BibitemShut {NoStop}%
\bibitem [{\citenamefont {Hadjivasiliou}\ \emph {et~al.}(2021)\citenamefont
  {Hadjivasiliou} \emph {et~al.}}]{Belle:2021gmc}%
  \BibitemOpen
  \bibfield  {author} {\bibinfo {author} {\bibfnamefont {C.}~\bibnamefont
  {Hadjivasiliou}} \emph {et~al.} (\bibinfo {collaboration} {Belle}),\
  }\href@noop {} {\  (\bibinfo {year} {2021})},\ \Eprint
  {http://arxiv.org/abs/2110.14086} {arXiv:2110.14086 [hep-ex]} \BibitemShut
  {NoStop}%
\bibitem [{\citenamefont {Anderlini}(2014)}]{Anderlini:2014dha}%
  \BibitemOpen
  \bibfield  {author} {\bibinfo {author} {\bibfnamefont {L.}~\bibnamefont
  {Anderlini}} (\bibinfo {collaboration} {LHCb, CMS, ATLAS}),\ }in\ \href@noop
  {} {\emph {\bibinfo {booktitle} {{12th Conference on Flavor Physics and CP
  Violation}}}}\ (\bibinfo {year} {2014})\ \Eprint
  {http://arxiv.org/abs/1407.8066} {arXiv:1407.8066 [hep-ex]} \BibitemShut
  {NoStop}%
\bibitem [{\citenamefont {Yuan}(2014)}]{Yuan:2014mya}%
  \BibitemOpen
  \bibfield  {author} {\bibinfo {author} {\bibfnamefont {X.}~\bibnamefont
  {Yuan}} (\bibinfo {collaboration} {LHCb}),\ }\href {\doibase
  10.1088/1742-6596/556/1/012053} {\bibfield  {journal} {\bibinfo  {journal}
  {J. Phys. Conf. Ser.}\ }\textbf {\bibinfo {volume} {556}},\ \bibinfo {pages}
  {012053} (\bibinfo {year} {2014})}\BibitemShut {NoStop}%
\bibitem [{\citenamefont {Tuning}(2013)}]{Tuning:2013nio}%
  \BibitemOpen
  \bibfield  {author} {\bibinfo {author} {\bibfnamefont {N.}~\bibnamefont
  {Tuning}} (\bibinfo {collaboration} {LHCb}),\ }\href {\doibase
  10.22323/1.180.0379} {\bibfield  {journal} {\bibinfo  {journal} {PoS}\
  }\textbf {\bibinfo {volume} {EPS-HEP2013}},\ \bibinfo {pages} {379} (\bibinfo
  {year} {2013})}\BibitemShut {NoStop}%
\bibitem [{\citenamefont {Aaij}\ \emph
  {et~al.}(2012{\natexlab{a}})\citenamefont {Aaij} \emph
  {et~al.}}]{LHCb:2012ihf}%
  \BibitemOpen
  \bibfield  {author} {\bibinfo {author} {\bibfnamefont {R.}~\bibnamefont
  {Aaij}} \emph {et~al.} (\bibinfo {collaboration} {LHCb}),\ }\href {\doibase
  10.1103/PhysRevLett.109.232001} {\bibfield  {journal} {\bibinfo  {journal}
  {Phys. Rev. Lett.}\ }\textbf {\bibinfo {volume} {109}},\ \bibinfo {pages}
  {232001} (\bibinfo {year} {2012}{\natexlab{a}})},\ \Eprint
  {http://arxiv.org/abs/1209.5634} {arXiv:1209.5634 [hep-ex]} \BibitemShut
  {NoStop}%
\bibitem [{\citenamefont {Aaij}\ \emph
  {et~al.}(2012{\natexlab{b}})\citenamefont {Aaij} \emph
  {et~al.}}]{LHCb:2012ag}%
  \BibitemOpen
  \bibfield  {author} {\bibinfo {author} {\bibfnamefont {R.}~\bibnamefont
  {Aaij}} \emph {et~al.} (\bibinfo {collaboration} {LHCb}),\ }\href {\doibase
  10.1103/PhysRevLett.108.251802} {\bibfield  {journal} {\bibinfo  {journal}
  {Phys. Rev. Lett.}\ }\textbf {\bibinfo {volume} {108}},\ \bibinfo {pages}
  {251802} (\bibinfo {year} {2012}{\natexlab{b}})},\ \Eprint
  {http://arxiv.org/abs/1204.0079} {arXiv:1204.0079 [hep-ex]} \BibitemShut
  {NoStop}%
\bibitem [{\citenamefont {Aaij}\ \emph
  {et~al.}(2013{\natexlab{a}})\citenamefont {Aaij} \emph
  {et~al.}}]{LHCb:2013kwl}%
  \BibitemOpen
  \bibfield  {author} {\bibinfo {author} {\bibfnamefont {R.}~\bibnamefont
  {Aaij}} \emph {et~al.} (\bibinfo {collaboration} {LHCb}),\ }\href {\doibase
  10.1103/PhysRevD.87.112012} {\bibfield  {journal} {\bibinfo  {journal} {Phys.
  Rev. D}\ }\textbf {\bibinfo {volume} {87}},\ \bibinfo {pages} {112012}
  (\bibinfo {year} {2013}{\natexlab{a}})},\ \bibinfo {note} {[Addendum:
  Phys.Rev.D 89, 019901 (2014)]},\ \Eprint {http://arxiv.org/abs/1304.4530}
  {arXiv:1304.4530 [hep-ex]} \BibitemShut {NoStop}%
\bibitem [{\citenamefont {Aaij}\ \emph
  {et~al.}(2017{\natexlab{a}})\citenamefont {Aaij} \emph
  {et~al.}}]{LHCb:2017lpu}%
  \BibitemOpen
  \bibfield  {author} {\bibinfo {author} {\bibfnamefont {R.}~\bibnamefont
  {Aaij}} \emph {et~al.} (\bibinfo {collaboration} {LHCb}),\ }\href {\doibase
  10.1103/PhysRevLett.118.111803} {\bibfield  {journal} {\bibinfo  {journal}
  {Phys. Rev. Lett.}\ }\textbf {\bibinfo {volume} {118}},\ \bibinfo {pages}
  {111803} (\bibinfo {year} {2017}{\natexlab{a}})},\ \Eprint
  {http://arxiv.org/abs/1701.01856} {arXiv:1701.01856 [hep-ex]} \BibitemShut
  {NoStop}%
\bibitem [{\citenamefont {Aaij}\ \emph
  {et~al.}(2013{\natexlab{b}})\citenamefont {Aaij} \emph
  {et~al.}}]{LHCb:2013xlg}%
  \BibitemOpen
  \bibfield  {author} {\bibinfo {author} {\bibfnamefont {R.}~\bibnamefont
  {Aaij}} \emph {et~al.} (\bibinfo {collaboration} {LHCb}),\ }\href {\doibase
  10.1103/PhysRevLett.111.181801} {\bibfield  {journal} {\bibinfo  {journal}
  {Phys. Rev. Lett.}\ }\textbf {\bibinfo {volume} {111}},\ \bibinfo {pages}
  {181801} (\bibinfo {year} {2013}{\natexlab{b}})},\ \Eprint
  {http://arxiv.org/abs/1308.4544} {arXiv:1308.4544 [hep-ex]} \BibitemShut
  {NoStop}%
\bibitem [{\citenamefont {Gouz}\ \emph {et~al.}(2004)\citenamefont {Gouz},
  \citenamefont {Kiselev}, \citenamefont {Likhoded}, \citenamefont
  {Romanovsky},\ and\ \citenamefont {Yushchenko}}]{Gouz:2002kk}%
  \BibitemOpen
  \bibfield  {author} {\bibinfo {author} {\bibfnamefont {I.~P.}\ \bibnamefont
  {Gouz}}, \bibinfo {author} {\bibfnamefont {V.~V.}\ \bibnamefont {Kiselev}},
  \bibinfo {author} {\bibfnamefont {A.~K.}\ \bibnamefont {Likhoded}}, \bibinfo
  {author} {\bibfnamefont {V.~I.}\ \bibnamefont {Romanovsky}}, \ and\ \bibinfo
  {author} {\bibfnamefont {O.~P.}\ \bibnamefont {Yushchenko}},\ }\href
  {\doibase 10.1134/1.1788046} {\bibfield  {journal} {\bibinfo  {journal}
  {Phys. Atom. Nucl.}\ }\textbf {\bibinfo {volume} {67}},\ \bibinfo {pages}
  {1559} (\bibinfo {year} {2004})},\ \Eprint
  {http://arxiv.org/abs/hep-ph/0211432} {arXiv:hep-ph/0211432} \BibitemShut
  {NoStop}%
\bibitem [{\citenamefont {Abada}\ \emph {et~al.}(2019)\citenamefont {Abada}
  \emph {et~al.}}]{FCC:2018byv}%
  \BibitemOpen
  \bibfield  {author} {\bibinfo {author} {\bibfnamefont {A.}~\bibnamefont
  {Abada}} \emph {et~al.} (\bibinfo {collaboration} {FCC}),\ }\href {\doibase
  10.1140/epjc/s10052-019-6904-3} {\bibfield  {journal} {\bibinfo  {journal}
  {Eur. Phys. J. C}\ }\textbf {\bibinfo {volume} {79}},\ \bibinfo {pages} {474}
  (\bibinfo {year} {2019})}\BibitemShut {NoStop}%
\bibitem [{\citenamefont {Fleischer}\ and\ \citenamefont
  {Wyler}(2000)}]{PhysRevD.62.057503}%
  \BibitemOpen
  \bibfield  {author} {\bibinfo {author} {\bibfnamefont {R.}~\bibnamefont
  {Fleischer}}\ and\ \bibinfo {author} {\bibfnamefont {D.}~\bibnamefont
  {Wyler}},\ }\href {\doibase 10.1103/PhysRevD.62.057503} {\bibfield  {journal}
  {\bibinfo  {journal} {Phys. Rev. D}\ }\textbf {\bibinfo {volume} {62}},\
  \bibinfo {pages} {057503} (\bibinfo {year} {2000})}\BibitemShut {NoStop}%
\bibitem [{\citenamefont {Alonso-\'Alvarez}\ \emph
  {et~al.}(2021{\natexlab{b}})\citenamefont {Alonso-\'Alvarez}, \citenamefont
  {Elor}, \citenamefont {Escudero}, \citenamefont {Fornal}, \citenamefont
  {Grinstein},\ and\ \citenamefont {Camalich}}]{Alonso-Alvarez:2021oaj}%
  \BibitemOpen
  \bibfield  {author} {\bibinfo {author} {\bibfnamefont {G.}~\bibnamefont
  {Alonso-\'Alvarez}}, \bibinfo {author} {\bibfnamefont {G.}~\bibnamefont
  {Elor}}, \bibinfo {author} {\bibfnamefont {M.}~\bibnamefont {Escudero}},
  \bibinfo {author} {\bibfnamefont {B.}~\bibnamefont {Fornal}}, \bibinfo
  {author} {\bibfnamefont {B.}~\bibnamefont {Grinstein}}, \ and\ \bibinfo
  {author} {\bibfnamefont {J.~M.}\ \bibnamefont {Camalich}},\ }\href@noop {} {\
   (\bibinfo {year} {2021}{\natexlab{b}})},\ \Eprint
  {http://arxiv.org/abs/2111.12712} {arXiv:2111.12712 [hep-ph]} \BibitemShut
  {NoStop}%
\bibitem [{\citenamefont {Dror}\ \emph
  {et~al.}(2020{\natexlab{a}})\citenamefont {Dror}, \citenamefont {Elor},\ and\
  \citenamefont {McGehee}}]{Dror:2019onn}%
  \BibitemOpen
  \bibfield  {author} {\bibinfo {author} {\bibfnamefont {J.~A.}\ \bibnamefont
  {Dror}}, \bibinfo {author} {\bibfnamefont {G.}~\bibnamefont {Elor}}, \ and\
  \bibinfo {author} {\bibfnamefont {R.}~\bibnamefont {McGehee}},\ }\href
  {\doibase 10.1103/PhysRevLett.124.181301} {\bibfield  {journal} {\bibinfo
  {journal} {Phys. Rev. Lett.}\ }\textbf {\bibinfo {volume} {124}},\ \bibinfo
  {pages} {18} (\bibinfo {year} {2020}{\natexlab{a}})},\ \Eprint
  {http://arxiv.org/abs/1905.12635} {arXiv:1905.12635 [hep-ph]} \BibitemShut
  {NoStop}%
\bibitem [{\citenamefont {Dror}\ \emph
  {et~al.}(2020{\natexlab{b}})\citenamefont {Dror}, \citenamefont {Elor},\ and\
  \citenamefont {McGehee}}]{Dror:2019dib}%
  \BibitemOpen
  \bibfield  {author} {\bibinfo {author} {\bibfnamefont {J.~A.}\ \bibnamefont
  {Dror}}, \bibinfo {author} {\bibfnamefont {G.}~\bibnamefont {Elor}}, \ and\
  \bibinfo {author} {\bibfnamefont {R.}~\bibnamefont {McGehee}},\ }\href
  {\doibase 10.1007/JHEP02(2020)134} {\bibfield  {journal} {\bibinfo  {journal}
  {JHEP}\ }\textbf {\bibinfo {volume} {02}},\ \bibinfo {pages} {134} (\bibinfo
  {year} {2020}{\natexlab{b}})},\ \Eprint {http://arxiv.org/abs/1908.10861}
  {arXiv:1908.10861 [hep-ph]} \BibitemShut {NoStop}%
\bibitem [{\citenamefont {Beneke}\ and\ \citenamefont
  {Jager}(2006)}]{Beneke:2005vv}%
  \BibitemOpen
  \bibfield  {author} {\bibinfo {author} {\bibfnamefont {M.}~\bibnamefont
  {Beneke}}\ and\ \bibinfo {author} {\bibfnamefont {S.}~\bibnamefont {Jager}},\
  }\href {\doibase 10.1016/j.nuclphysb.2006.06.010} {\bibfield  {journal}
  {\bibinfo  {journal} {Nucl. Phys. B}\ }\textbf {\bibinfo {volume} {751}},\
  \bibinfo {pages} {160} (\bibinfo {year} {2006})},\ \Eprint
  {http://arxiv.org/abs/hep-ph/0512351} {arXiv:hep-ph/0512351} \BibitemShut
  {NoStop}%
\bibitem [{\citenamefont {Shrock}(1981)}]{PhysRevD.24.1232}%
  \BibitemOpen
  \bibfield  {author} {\bibinfo {author} {\bibfnamefont {R.~E.}\ \bibnamefont
  {Shrock}},\ }\href {\doibase 10.1103/PhysRevD.24.1232} {\bibfield  {journal}
  {\bibinfo  {journal} {Phys. Rev. D}\ }\textbf {\bibinfo {volume} {24}},\
  \bibinfo {pages} {1232} (\bibinfo {year} {1981})}\BibitemShut {NoStop}%
\bibitem [{\citenamefont {Bryman}\ and\ \citenamefont
  {Shrock}(2019)}]{Bryman:2019bjg}%
  \BibitemOpen
  \bibfield  {author} {\bibinfo {author} {\bibfnamefont {D.~A.}\ \bibnamefont
  {Bryman}}\ and\ \bibinfo {author} {\bibfnamefont {R.}~\bibnamefont
  {Shrock}},\ }\href {\doibase 10.1103/PhysRevD.100.073011} {\bibfield
  {journal} {\bibinfo  {journal} {Phys. Rev. D}\ }\textbf {\bibinfo {volume}
  {100}},\ \bibinfo {pages} {073011} (\bibinfo {year} {2019})},\ \Eprint
  {http://arxiv.org/abs/1909.11198} {arXiv:1909.11198 [hep-ph]} \BibitemShut
  {NoStop}%
\bibitem [{\citenamefont {Aguilar-Arevalo}\ \emph {et~al.}(2018)\citenamefont
  {Aguilar-Arevalo} \emph {et~al.}}]{Aguilar-Arevalo:2017vlf}%
  \BibitemOpen
  \bibfield  {author} {\bibinfo {author} {\bibfnamefont {A.}~\bibnamefont
  {Aguilar-Arevalo}} \emph {et~al.} (\bibinfo {collaboration} {PIENU}),\ }\href
  {\doibase 10.1103/PhysRevD.97.072012} {\bibfield  {journal} {\bibinfo
  {journal} {Phys. Rev. D}\ }\textbf {\bibinfo {volume} {97}},\ \bibinfo
  {pages} {072012} (\bibinfo {year} {2018})},\ \Eprint
  {http://arxiv.org/abs/1712.03275} {arXiv:1712.03275 [hep-ex]} \BibitemShut
  {NoStop}%
\bibitem [{\citenamefont {Aguilar-Arevalo}\ \emph {et~al.}(2019)\citenamefont
  {Aguilar-Arevalo} \emph {et~al.}}]{Aguilar-Arevalo:2019owf}%
  \BibitemOpen
  \bibfield  {author} {\bibinfo {author} {\bibfnamefont {A.}~\bibnamefont
  {Aguilar-Arevalo}} \emph {et~al.} (\bibinfo {collaboration} {PIENU}),\ }\href
  {\doibase 10.1016/j.physletb.2019.134980} {\bibfield  {journal} {\bibinfo
  {journal} {Phys. Lett.}\ }\textbf {\bibinfo {volume} {B798}},\ \bibinfo
  {pages} {134980} (\bibinfo {year} {2019})},\ \Eprint
  {http://arxiv.org/abs/1904.03269} {arXiv:1904.03269 [hep-ex]} \BibitemShut
  {NoStop}%
%%CITATION = ARXIV:1904.03269;%%
\bibitem [{\citenamefont {Bryman}()}]{Doug}%
  \BibitemOpen
  \bibfield  {author} {\bibinfo {author} {\bibfnamefont {D.}~\bibnamefont
  {Bryman}},\ }\href@noop {} {}\bibinfo {howpublished} {personal
  communication}\BibitemShut {NoStop}%
\bibitem [{\citenamefont {Cortina~Gil}\ \emph {et~al.}(2018)\citenamefont
  {Cortina~Gil} \emph {et~al.}}]{NA62:2017qcd}%
  \BibitemOpen
  \bibfield  {author} {\bibinfo {author} {\bibfnamefont {E.}~\bibnamefont
  {Cortina~Gil}} \emph {et~al.} (\bibinfo {collaboration} {NA62}),\ }\href
  {\doibase 10.1016/j.physletb.2018.01.031} {\bibfield  {journal} {\bibinfo
  {journal} {Phys. Lett. B}\ }\textbf {\bibinfo {volume} {778}},\ \bibinfo
  {pages} {137} (\bibinfo {year} {2018})},\ \Eprint
  {http://arxiv.org/abs/1712.00297} {arXiv:1712.00297 [hep-ex]} \BibitemShut
  {NoStop}%
\bibitem [{\citenamefont {Hayano}\ \emph {et~al.}(1982)\citenamefont {Hayano}
  \emph {et~al.}}]{Hayano:1982wu}%
  \BibitemOpen
  \bibfield  {author} {\bibinfo {author} {\bibfnamefont {R.~S.}\ \bibnamefont
  {Hayano}} \emph {et~al.},\ }\href {\doibase 10.1103/PhysRevLett.49.1305}
  {\bibfield  {journal} {\bibinfo  {journal} {Phys. Rev. Lett.}\ }\textbf
  {\bibinfo {volume} {49}},\ \bibinfo {pages} {1305} (\bibinfo {year}
  {1982})}\BibitemShut {NoStop}%
\bibitem [{\citenamefont {Artamonov}\ \emph {et~al.}(2015)\citenamefont
  {Artamonov} \emph {et~al.}}]{E949:2014gsn}%
  \BibitemOpen
  \bibfield  {author} {\bibinfo {author} {\bibfnamefont {A.~V.}\ \bibnamefont
  {Artamonov}} \emph {et~al.} (\bibinfo {collaboration} {E949}),\ }\href
  {\doibase 10.1103/PhysRevD.91.052001} {\bibfield  {journal} {\bibinfo
  {journal} {Phys. Rev. D}\ }\textbf {\bibinfo {volume} {91}},\ \bibinfo
  {pages} {052001} (\bibinfo {year} {2015})},\ \bibinfo {note} {[Erratum:
  Phys.Rev.D 91, 059903 (2015)]},\ \Eprint {http://arxiv.org/abs/1411.3963}
  {arXiv:1411.3963 [hep-ex]} \BibitemShut {NoStop}%
\bibitem [{\citenamefont {Cortina~Gil}\ \emph {et~al.}(2021)\citenamefont
  {Cortina~Gil} \emph {et~al.}}]{NA62:2021bji}%
  \BibitemOpen
  \bibfield  {author} {\bibinfo {author} {\bibfnamefont {E.}~\bibnamefont
  {Cortina~Gil}} \emph {et~al.} (\bibinfo {collaboration} {NA62}),\ }\href
  {\doibase 10.1016/j.physletb.2021.136259} {\bibfield  {journal} {\bibinfo
  {journal} {Phys. Lett. B}\ }\textbf {\bibinfo {volume} {816}},\ \bibinfo
  {pages} {136259} (\bibinfo {year} {2021})},\ \Eprint
  {http://arxiv.org/abs/2101.12304} {arXiv:2101.12304 [hep-ex]} \BibitemShut
  {NoStop}%
\bibitem [{\citenamefont {Park}\ \emph {et~al.}(2016)\citenamefont {Park} \emph
  {et~al.}}]{Belle:2016nvh}%
  \BibitemOpen
  \bibfield  {author} {\bibinfo {author} {\bibfnamefont {C.~S.}\ \bibnamefont
  {Park}} \emph {et~al.} (\bibinfo {collaboration} {Belle}),\ }\href {\doibase
  10.1103/PhysRevD.94.012003} {\bibfield  {journal} {\bibinfo  {journal} {Phys.
  Rev. D}\ }\textbf {\bibinfo {volume} {94}},\ \bibinfo {pages} {012003}
  (\bibinfo {year} {2016})},\ \Eprint {http://arxiv.org/abs/1605.04430}
  {arXiv:1605.04430 [hep-ex]} \BibitemShut {NoStop}%
\bibitem [{\citenamefont {Alexander}\ \emph {et~al.}(2009)\citenamefont
  {Alexander} \emph {et~al.}}]{CLEO:2009lvj}%
  \BibitemOpen
  \bibfield  {author} {\bibinfo {author} {\bibfnamefont {J.~P.}\ \bibnamefont
  {Alexander}} \emph {et~al.} (\bibinfo {collaboration} {CLEO}),\ }\href
  {\doibase 10.1103/PhysRevD.79.052001} {\bibfield  {journal} {\bibinfo
  {journal} {Phys. Rev. D}\ }\textbf {\bibinfo {volume} {79}},\ \bibinfo
  {pages} {052001} (\bibinfo {year} {2009})},\ \Eprint
  {http://arxiv.org/abs/0901.1216} {arXiv:0901.1216 [hep-ex]} \BibitemShut
  {NoStop}%
\bibitem [{\citenamefont {del Amo~Sanchez}\ \emph
  {et~al.}(2010{\natexlab{a}})\citenamefont {del Amo~Sanchez} \emph
  {et~al.}}]{BaBar:2010ixw}%
  \BibitemOpen
  \bibfield  {author} {\bibinfo {author} {\bibfnamefont {P.}~\bibnamefont {del
  Amo~Sanchez}} \emph {et~al.} (\bibinfo {collaboration} {BaBar}),\ }\href
  {\doibase 10.1103/PhysRevD.82.091103} {\bibfield  {journal} {\bibinfo
  {journal} {Phys. Rev. D}\ }\textbf {\bibinfo {volume} {82}},\ \bibinfo
  {pages} {091103} (\bibinfo {year} {2010}{\natexlab{a}})},\ \bibinfo {note}
  {[Erratum: Phys.Rev.D 91, 019901 (2015)]},\ \Eprint
  {http://arxiv.org/abs/1008.4080} {arXiv:1008.4080 [hep-ex]} \BibitemShut
  {NoStop}%
\bibitem [{\citenamefont {Zupanc}\ \emph {et~al.}(2013)\citenamefont {Zupanc}
  \emph {et~al.}}]{Belle:2013isi}%
  \BibitemOpen
  \bibfield  {author} {\bibinfo {author} {\bibfnamefont {A.}~\bibnamefont
  {Zupanc}} \emph {et~al.} (\bibinfo {collaboration} {Belle}),\ }\href
  {\doibase 10.1007/JHEP09(2013)139} {\bibfield  {journal} {\bibinfo  {journal}
  {JHEP}\ }\textbf {\bibinfo {volume} {09}},\ \bibinfo {pages} {139} (\bibinfo
  {year} {2013})},\ \Eprint {http://arxiv.org/abs/1307.6240} {arXiv:1307.6240
  [hep-ex]} \BibitemShut {NoStop}%
\bibitem [{\citenamefont {Kato}\ \emph
  {et~al.}(2018{\natexlab{a}})\citenamefont {Kato} \emph
  {et~al.}}]{Belle:2017psv}%
  \BibitemOpen
  \bibfield  {author} {\bibinfo {author} {\bibfnamefont {Y.}~\bibnamefont
  {Kato}} \emph {et~al.} (\bibinfo {collaboration} {Belle}),\ }\href {\doibase
  10.1103/PhysRevD.97.012005} {\bibfield  {journal} {\bibinfo  {journal} {Phys.
  Rev. D}\ }\textbf {\bibinfo {volume} {97}},\ \bibinfo {pages} {012005}
  (\bibinfo {year} {2018}{\natexlab{a}})},\ \Eprint
  {http://arxiv.org/abs/1709.06108} {arXiv:1709.06108 [hep-ex]} \BibitemShut
  {NoStop}%
\bibitem [{\citenamefont {Aubert}\ \emph
  {et~al.}(2007{\natexlab{a}})\citenamefont {Aubert} \emph
  {et~al.}}]{BaBar:2006rof}%
  \BibitemOpen
  \bibfield  {author} {\bibinfo {author} {\bibfnamefont {B.}~\bibnamefont
  {Aubert}} \emph {et~al.} (\bibinfo {collaboration} {BaBar}),\ }\href
  {\doibase 10.1103/PhysRevD.75.031101} {\bibfield  {journal} {\bibinfo
  {journal} {Phys. Rev. D}\ }\textbf {\bibinfo {volume} {75}},\ \bibinfo
  {pages} {031101} (\bibinfo {year} {2007}{\natexlab{a}})},\ \Eprint
  {http://arxiv.org/abs/hep-ex/0610027} {arXiv:hep-ex/0610027} \BibitemShut
  {NoStop}%
\bibitem [{\citenamefont {Abe}\ \emph {et~al.}(2006)\citenamefont {Abe} \emph
  {et~al.}}]{Belle:2006cuz}%
  \BibitemOpen
  \bibfield  {author} {\bibinfo {author} {\bibfnamefont {K.}~\bibnamefont
  {Abe}} \emph {et~al.} (\bibinfo {collaboration} {Belle}),\ }\href {\doibase
  10.1103/PhysRevD.73.051106} {\bibfield  {journal} {\bibinfo  {journal} {Phys.
  Rev. D}\ }\textbf {\bibinfo {volume} {73}},\ \bibinfo {pages} {051106}
  (\bibinfo {year} {2006})},\ \Eprint {http://arxiv.org/abs/hep-ex/0601032}
  {arXiv:hep-ex/0601032} \BibitemShut {NoStop}%
\bibitem [{\citenamefont {Lees}\ \emph {et~al.}(2017)\citenamefont {Lees} \emph
  {et~al.}}]{BaBar:2015pwa}%
  \BibitemOpen
  \bibfield  {author} {\bibinfo {author} {\bibfnamefont {J.~P.}\ \bibnamefont
  {Lees}} \emph {et~al.} (\bibinfo {collaboration} {BaBar}),\ }\href {\doibase
  10.1103/PhysRevD.96.072001} {\bibfield  {journal} {\bibinfo  {journal} {Phys.
  Rev. D}\ }\textbf {\bibinfo {volume} {96}},\ \bibinfo {pages} {072001}
  (\bibinfo {year} {2017})},\ \Eprint {http://arxiv.org/abs/1501.00705}
  {arXiv:1501.00705 [hep-ex]} \BibitemShut {NoStop}%
\bibitem [{\citenamefont {Eckhart}\ \emph {et~al.}(2002)\citenamefont {Eckhart}
  \emph {et~al.}}]{CLEO:2002jwu}%
  \BibitemOpen
  \bibfield  {author} {\bibinfo {author} {\bibfnamefont {E.}~\bibnamefont
  {Eckhart}} \emph {et~al.} (\bibinfo {collaboration} {CLEO}),\ }\href
  {\doibase 10.1103/PhysRevLett.89.251801} {\bibfield  {journal} {\bibinfo
  {journal} {Phys. Rev. Lett.}\ }\textbf {\bibinfo {volume} {89}},\ \bibinfo
  {pages} {251801} (\bibinfo {year} {2002})},\ \Eprint
  {http://arxiv.org/abs/hep-ex/0206024} {arXiv:hep-ex/0206024} \BibitemShut
  {NoStop}%
\bibitem [{\citenamefont {Aubert}\ \emph
  {et~al.}(2009{\natexlab{a}})\citenamefont {Aubert} \emph
  {et~al.}}]{BaBar:2009vfr}%
  \BibitemOpen
  \bibfield  {author} {\bibinfo {author} {\bibfnamefont {B.}~\bibnamefont
  {Aubert}} \emph {et~al.} (\bibinfo {collaboration} {BaBar}),\ }\href
  {\doibase 10.1103/PhysRevD.79.072006} {\bibfield  {journal} {\bibinfo
  {journal} {Phys. Rev. D}\ }\textbf {\bibinfo {volume} {79}},\ \bibinfo
  {pages} {072006} (\bibinfo {year} {2009}{\natexlab{a}})},\ \Eprint
  {http://arxiv.org/abs/0902.2051} {arXiv:0902.2051 [hep-ex]} \BibitemShut
  {NoStop}%
\bibitem [{\citenamefont {Aubert}\ \emph
  {et~al.}(2008{\natexlab{a}})\citenamefont {Aubert} \emph
  {et~al.}}]{BaBar:2008lpx}%
  \BibitemOpen
  \bibfield  {author} {\bibinfo {author} {\bibfnamefont {B.}~\bibnamefont
  {Aubert}} \emph {et~al.} (\bibinfo {collaboration} {BaBar}),\ }\href
  {\doibase 10.1103/PhysRevD.78.012004} {\bibfield  {journal} {\bibinfo
  {journal} {Phys. Rev. D}\ }\textbf {\bibinfo {volume} {78}},\ \bibinfo
  {pages} {012004} (\bibinfo {year} {2008}{\natexlab{a}})},\ \Eprint
  {http://arxiv.org/abs/0803.4451} {arXiv:0803.4451 [hep-ex]} \BibitemShut
  {NoStop}%
\bibitem [{\citenamefont {Garmash}\ \emph {et~al.}(2006)\citenamefont {Garmash}
  \emph {et~al.}}]{Belle:2005rpz}%
  \BibitemOpen
  \bibfield  {author} {\bibinfo {author} {\bibfnamefont {A.}~\bibnamefont
  {Garmash}} \emph {et~al.} (\bibinfo {collaboration} {Belle}),\ }\href
  {\doibase 10.1103/PhysRevLett.96.251803} {\bibfield  {journal} {\bibinfo
  {journal} {Phys. Rev. Lett.}\ }\textbf {\bibinfo {volume} {96}},\ \bibinfo
  {pages} {251803} (\bibinfo {year} {2006})},\ \Eprint
  {http://arxiv.org/abs/hep-ex/0512066} {arXiv:hep-ex/0512066} \BibitemShut
  {NoStop}%
\bibitem [{\citenamefont {Kumar}\ \emph {et~al.}(2006)\citenamefont {Kumar}
  \emph {et~al.}}]{Belle:2006eqz}%
  \BibitemOpen
  \bibfield  {author} {\bibinfo {author} {\bibfnamefont {R.}~\bibnamefont
  {Kumar}} \emph {et~al.} (\bibinfo {collaboration} {Belle}),\ }\href {\doibase
  10.1103/PhysRevD.74.051103} {\bibfield  {journal} {\bibinfo  {journal} {Phys.
  Rev. D}\ }\textbf {\bibinfo {volume} {74}},\ \bibinfo {pages} {051103}
  (\bibinfo {year} {2006})},\ \Eprint {http://arxiv.org/abs/hep-ex/0607008}
  {arXiv:hep-ex/0607008} \BibitemShut {NoStop}%
\bibitem [{\citenamefont {Aaij}\ \emph {et~al.}(2019)\citenamefont {Aaij} \emph
  {et~al.}}]{LHCb:2019xmb}%
  \BibitemOpen
  \bibfield  {author} {\bibinfo {author} {\bibfnamefont {R.}~\bibnamefont
  {Aaij}} \emph {et~al.} (\bibinfo {collaboration} {LHCb}),\ }\href {\doibase
  10.1103/PhysRevLett.123.231802} {\bibfield  {journal} {\bibinfo  {journal}
  {Phys. Rev. Lett.}\ }\textbf {\bibinfo {volume} {123}},\ \bibinfo {pages}
  {231802} (\bibinfo {year} {2019})},\ \Eprint
  {http://arxiv.org/abs/1905.09244} {arXiv:1905.09244 [hep-ex]} \BibitemShut
  {NoStop}%
\bibitem [{\citenamefont {Aubert}\ \emph
  {et~al.}(2005{\natexlab{a}})\citenamefont {Aubert} \emph
  {et~al.}}]{BaBar:2005jqu}%
  \BibitemOpen
  \bibfield  {author} {\bibinfo {author} {\bibfnamefont {B.}~\bibnamefont
  {Aubert}} \emph {et~al.} (\bibinfo {collaboration} {BaBar}),\ }\href
  {\doibase 10.1103/PhysRevD.72.052002} {\bibfield  {journal} {\bibinfo
  {journal} {Phys. Rev. D}\ }\textbf {\bibinfo {volume} {72}},\ \bibinfo
  {pages} {052002} (\bibinfo {year} {2005}{\natexlab{a}})},\ \Eprint
  {http://arxiv.org/abs/hep-ex/0507025} {arXiv:hep-ex/0507025} \BibitemShut
  {NoStop}%
\bibitem [{\citenamefont {Bhardwaj}\ \emph {et~al.}(2008)\citenamefont
  {Bhardwaj} \emph {et~al.}}]{Belle:2008ztp}%
  \BibitemOpen
  \bibfield  {author} {\bibinfo {author} {\bibfnamefont {V.}~\bibnamefont
  {Bhardwaj}} \emph {et~al.} (\bibinfo {collaboration} {Belle}),\ }\href
  {\doibase 10.1103/PhysRevD.78.051104} {\bibfield  {journal} {\bibinfo
  {journal} {Phys. Rev. D}\ }\textbf {\bibinfo {volume} {78}},\ \bibinfo
  {pages} {051104} (\bibinfo {year} {2008})},\ \Eprint
  {http://arxiv.org/abs/0807.2170} {arXiv:0807.2170 [hep-ex]} \BibitemShut
  {NoStop}%
\bibitem [{\citenamefont {Aaij}\ \emph {et~al.}(2020)\citenamefont {Aaij} \emph
  {et~al.}}]{LHCb:2020xcz}%
  \BibitemOpen
  \bibfield  {author} {\bibinfo {author} {\bibfnamefont {R.}~\bibnamefont
  {Aaij}} \emph {et~al.} (\bibinfo {collaboration} {LHCb}),\ }\href {\doibase
  10.1103/PhysRevD.102.112010} {\bibfield  {journal} {\bibinfo  {journal}
  {Phys. Rev. D}\ }\textbf {\bibinfo {volume} {102}},\ \bibinfo {pages}
  {112010} (\bibinfo {year} {2020})},\ \Eprint
  {http://arxiv.org/abs/2010.11802} {arXiv:2010.11802 [hep-ex]} \BibitemShut
  {NoStop}%
\bibitem [{\citenamefont {Aubert}\ \emph
  {et~al.}(2006{\natexlab{a}})\citenamefont {Aubert} \emph
  {et~al.}}]{BaBar:2006zod}%
  \BibitemOpen
  \bibfield  {author} {\bibinfo {author} {\bibfnamefont {B.}~\bibnamefont
  {Aubert}} \emph {et~al.} (\bibinfo {collaboration} {BaBar}),\ }\href
  {\doibase 10.1103/PhysRevD.74.111102} {\bibfield  {journal} {\bibinfo
  {journal} {Phys. Rev. D}\ }\textbf {\bibinfo {volume} {74}},\ \bibinfo
  {pages} {111102} (\bibinfo {year} {2006}{\natexlab{a}})},\ \Eprint
  {http://arxiv.org/abs/hep-ex/0609033} {arXiv:hep-ex/0609033} \BibitemShut
  {NoStop}%
\bibitem [{\citenamefont {Aaij}\ \emph
  {et~al.}(2018{\natexlab{a}})\citenamefont {Aaij} \emph
  {et~al.}}]{LHCb:2017wbt}%
  \BibitemOpen
  \bibfield  {author} {\bibinfo {author} {\bibfnamefont {R.}~\bibnamefont
  {Aaij}} \emph {et~al.} (\bibinfo {collaboration} {LHCb}),\ }\href {\doibase
  10.1016/j.physletb.2017.11.070} {\bibfield  {journal} {\bibinfo  {journal}
  {Phys. Lett. B}\ }\textbf {\bibinfo {volume} {777}},\ \bibinfo {pages} {16}
  (\bibinfo {year} {2018}{\natexlab{a}})},\ \Eprint
  {http://arxiv.org/abs/1708.06370} {arXiv:1708.06370 [hep-ex]} \BibitemShut
  {NoStop}%
\bibitem [{\citenamefont {Aaij}\ \emph
  {et~al.}(2016{\natexlab{a}})\citenamefont {Aaij} \emph
  {et~al.}}]{LHCb:2016gpc}%
  \BibitemOpen
  \bibfield  {author} {\bibinfo {author} {\bibfnamefont {R.}~\bibnamefont
  {Aaij}} \emph {et~al.} (\bibinfo {collaboration} {LHCb}),\ }\href {\doibase
  10.1016/j.physletb.2016.06.022} {\bibfield  {journal} {\bibinfo  {journal}
  {Phys. Lett. B}\ }\textbf {\bibinfo {volume} {760}},\ \bibinfo {pages} {117}
  (\bibinfo {year} {2016}{\natexlab{a}})},\ \Eprint
  {http://arxiv.org/abs/1603.08993} {arXiv:1603.08993 [hep-ex]} \BibitemShut
  {NoStop}%
\bibitem [{\citenamefont {Aaij}\ \emph
  {et~al.}(2013{\natexlab{c}})\citenamefont {Aaij} \emph {et~al.}}]{1308.1277}%
  \BibitemOpen
  \bibfield  {author} {\bibinfo {author} {\bibfnamefont {R.}~\bibnamefont
  {Aaij}} \emph {et~al.} (\bibinfo {collaboration} {LHCb}),\ }\href {\doibase
  10.1016/j.physletb.2013.09.046} {\bibfield  {journal} {\bibinfo  {journal}
  {Phys. Lett. B}\ }\textbf {\bibinfo {volume} {726}},\ \bibinfo {pages} {646}
  (\bibinfo {year} {2013}{\natexlab{c}})},\ \Eprint
  {http://arxiv.org/abs/1308.1277} {arXiv:1308.1277 [hep-ex]} \BibitemShut
  {NoStop}%
\bibitem [{\citenamefont {Aubert}\ \emph
  {et~al.}(2008{\natexlab{b}})\citenamefont {Aubert} \emph
  {et~al.}}]{0807.2408}%
  \BibitemOpen
  \bibfield  {author} {\bibinfo {author} {\bibfnamefont {B.}~\bibnamefont
  {Aubert}} \emph {et~al.} (\bibinfo {collaboration} {BaBar}),\ }\href
  {\doibase 10.1103/PhysRevD.78.092002} {\bibfield  {journal} {\bibinfo
  {journal} {Phys. Rev. D}\ }\textbf {\bibinfo {volume} {78}},\ \bibinfo
  {pages} {092002} (\bibinfo {year} {2008}{\natexlab{b}})},\ \Eprint
  {http://arxiv.org/abs/0807.2408} {arXiv:0807.2408 [hep-ex]} \BibitemShut
  {NoStop}%
\bibitem [{\citenamefont {del Amo~Sanchez}\ \emph
  {et~al.}(2010{\natexlab{b}})\citenamefont {del Amo~Sanchez} \emph
  {et~al.}}]{1007.0504}%
  \BibitemOpen
  \bibfield  {author} {\bibinfo {author} {\bibfnamefont {P.}~\bibnamefont {del
  Amo~Sanchez}} \emph {et~al.} (\bibinfo {collaboration} {BaBar}),\ }\href
  {\doibase 10.1103/PhysRevD.82.072004} {\bibfield  {journal} {\bibinfo
  {journal} {Phys. Rev. D}\ }\textbf {\bibinfo {volume} {82}},\ \bibinfo
  {pages} {072004} (\bibinfo {year} {2010}{\natexlab{b}})},\ \Eprint
  {http://arxiv.org/abs/1007.0504} {arXiv:1007.0504 [hep-ex]} \BibitemShut
  {NoStop}%
\bibitem [{\citenamefont {Aaij}\ \emph
  {et~al.}(2018{\natexlab{b}})\citenamefont {Aaij} \emph
  {et~al.}}]{1708.06370}%
  \BibitemOpen
  \bibfield  {author} {\bibinfo {author} {\bibfnamefont {R.}~\bibnamefont
  {Aaij}} \emph {et~al.} (\bibinfo {collaboration} {LHCb}),\ }\href {\doibase
  10.1016/j.physletb.2017.11.070} {\bibfield  {journal} {\bibinfo  {journal}
  {Phys. Lett. B}\ }\textbf {\bibinfo {volume} {777}},\ \bibinfo {pages} {16}
  (\bibinfo {year} {2018}{\natexlab{b}})},\ \Eprint
  {http://arxiv.org/abs/1708.06370} {arXiv:1708.06370 [hep-ex]} \BibitemShut
  {NoStop}%
\bibitem [{\citenamefont {Aaij}\ \emph
  {et~al.}(2016{\natexlab{b}})\citenamefont {Aaij} \emph
  {et~al.}}]{1603.08993}%
  \BibitemOpen
  \bibfield  {author} {\bibinfo {author} {\bibfnamefont {R.}~\bibnamefont
  {Aaij}} \emph {et~al.} (\bibinfo {collaboration} {LHCb}),\ }\href {\doibase
  10.1016/j.physletb.2016.06.022} {\bibfield  {journal} {\bibinfo  {journal}
  {Phys. Lett. B}\ }\textbf {\bibinfo {volume} {760}},\ \bibinfo {pages} {117}
  (\bibinfo {year} {2016}{\natexlab{b}})},\ \Eprint
  {http://arxiv.org/abs/1603.08993} {arXiv:1603.08993 [hep-ex]} \BibitemShut
  {NoStop}%
\bibitem [{\citenamefont {Lees}\ \emph {et~al.}(2020)\citenamefont {Lees} \emph
  {et~al.}}]{1911.11740}%
  \BibitemOpen
  \bibfield  {author} {\bibinfo {author} {\bibfnamefont {J.~P.}\ \bibnamefont
  {Lees}} \emph {et~al.} (\bibinfo {collaboration} {BaBar}),\ }\href {\doibase
  10.1103/PhysRevLett.124.152001} {\bibfield  {journal} {\bibinfo  {journal}
  {Phys. Rev. Lett.}\ }\textbf {\bibinfo {volume} {124}},\ \bibinfo {pages}
  {152001} (\bibinfo {year} {2020})},\ \Eprint
  {http://arxiv.org/abs/1911.11740} {arXiv:1911.11740 [hep-ex]} \BibitemShut
  {NoStop}%
\bibitem [{\citenamefont {Bhardwaj}\ \emph {et~al.}(2011)\citenamefont
  {Bhardwaj} \emph {et~al.}}]{1105.0177}%
  \BibitemOpen
  \bibfield  {author} {\bibinfo {author} {\bibfnamefont {V.}~\bibnamefont
  {Bhardwaj}} \emph {et~al.} (\bibinfo {collaboration} {Belle}),\ }\href
  {\doibase 10.1103/PhysRevLett.107.091803} {\bibfield  {journal} {\bibinfo
  {journal} {Phys. Rev. Lett.}\ }\textbf {\bibinfo {volume} {107}},\ \bibinfo
  {pages} {091803} (\bibinfo {year} {2011})},\ \Eprint
  {http://arxiv.org/abs/1105.0177} {arXiv:1105.0177 [hep-ex]} \BibitemShut
  {NoStop}%
\bibitem [{\citenamefont {Aubert}\ \emph
  {et~al.}(2007{\natexlab{b}})\citenamefont {Aubert} \emph
  {et~al.}}]{0707.4561}%
  \BibitemOpen
  \bibfield  {author} {\bibinfo {author} {\bibfnamefont {B.}~\bibnamefont
  {Aubert}} \emph {et~al.} (\bibinfo {collaboration} {BaBar}),\ }\href
  {\doibase 10.1103/PhysRevLett.99.241803} {\bibfield  {journal} {\bibinfo
  {journal} {Phys. Rev. Lett.}\ }\textbf {\bibinfo {volume} {99}},\ \bibinfo
  {pages} {241803} (\bibinfo {year} {2007}{\natexlab{b}})},\ \Eprint
  {http://arxiv.org/abs/0707.4561} {arXiv:0707.4561 [hep-ex]} \BibitemShut
  {NoStop}%
\bibitem [{\citenamefont {Aaij}\ \emph
  {et~al.}(2012{\natexlab{c}})\citenamefont {Aaij} \emph
  {et~al.}}]{LHCb:2012xkq}%
  \BibitemOpen
  \bibfield  {author} {\bibinfo {author} {\bibfnamefont {R.}~\bibnamefont
  {Aaij}} \emph {et~al.} (\bibinfo {collaboration} {LHCb}),\ }\href {\doibase
  10.1016/j.physletb.2012.04.060} {\bibfield  {journal} {\bibinfo  {journal}
  {Phys. Lett. B}\ }\textbf {\bibinfo {volume} {712}},\ \bibinfo {pages} {203}
  (\bibinfo {year} {2012}{\natexlab{c}})},\ \bibinfo {note} {[Erratum:
  Phys.Lett.B 713, 351 (2012)]},\ \Eprint {http://arxiv.org/abs/1203.3662}
  {arXiv:1203.3662 [hep-ex]} \BibitemShut {NoStop}%
\bibitem [{\citenamefont {Aubert}\ \emph
  {et~al.}(2009{\natexlab{b}})\citenamefont {Aubert} \emph
  {et~al.}}]{BaBar:2008cqt}%
  \BibitemOpen
  \bibfield  {author} {\bibinfo {author} {\bibfnamefont {B.}~\bibnamefont
  {Aubert}} \emph {et~al.} (\bibinfo {collaboration} {BaBar}),\ }\href
  {\doibase 10.1103/PhysRevD.79.011102} {\bibfield  {journal} {\bibinfo
  {journal} {Phys. Rev. D}\ }\textbf {\bibinfo {volume} {79}},\ \bibinfo
  {pages} {011102} (\bibinfo {year} {2009}{\natexlab{b}})},\ \Eprint
  {http://arxiv.org/abs/0805.1317} {arXiv:0805.1317 [hep-ex]} \BibitemShut
  {NoStop}%
\bibitem [{\citenamefont {Nishida}\ \emph {et~al.}(2005)\citenamefont {Nishida}
  \emph {et~al.}}]{Belle:2004oww}%
  \BibitemOpen
  \bibfield  {author} {\bibinfo {author} {\bibfnamefont {S.}~\bibnamefont
  {Nishida}} \emph {et~al.} (\bibinfo {collaboration} {Belle}),\ }\href
  {\doibase 10.1016/j.physletb.2005.01.097} {\bibfield  {journal} {\bibinfo
  {journal} {Phys. Lett. B}\ }\textbf {\bibinfo {volume} {610}},\ \bibinfo
  {pages} {23} (\bibinfo {year} {2005})},\ \Eprint
  {http://arxiv.org/abs/hep-ex/0411065} {arXiv:hep-ex/0411065} \BibitemShut
  {NoStop}%
\bibitem [{\citenamefont {Duh}\ \emph {et~al.}(2013)\citenamefont {Duh},
  \citenamefont {Wu}, \citenamefont {Chang}, \citenamefont {Mohanty},
  \citenamefont {Unno}, \citenamefont {Adachi}, \citenamefont {Aihara},
  \citenamefont {Asner}, \citenamefont {Aulchenko}, \citenamefont {Aushev},\
  and\ \citenamefont {et~al.}}]{1210.1348}%
  \BibitemOpen
  \bibfield  {author} {\bibinfo {author} {\bibfnamefont {Y.-T.}\ \bibnamefont
  {Duh}}, \bibinfo {author} {\bibfnamefont {T.-Y.}\ \bibnamefont {Wu}},
  \bibinfo {author} {\bibfnamefont {P.}~\bibnamefont {Chang}}, \bibinfo
  {author} {\bibfnamefont {G.~B.}\ \bibnamefont {Mohanty}}, \bibinfo {author}
  {\bibfnamefont {Y.}~\bibnamefont {Unno}}, \bibinfo {author} {\bibfnamefont
  {I.}~\bibnamefont {Adachi}}, \bibinfo {author} {\bibfnamefont
  {H.}~\bibnamefont {Aihara}}, \bibinfo {author} {\bibfnamefont {D.~M.}\
  \bibnamefont {Asner}}, \bibinfo {author} {\bibfnamefont {V.}~\bibnamefont
  {Aulchenko}}, \bibinfo {author} {\bibfnamefont {T.}~\bibnamefont {Aushev}}, \
  and\ \bibinfo {author} {\bibnamefont {et~al.}},\ }\href {\doibase
  10.1103/physrevd.87.031103} {\bibfield  {journal} {\bibinfo  {journal}
  {Physical Review D}\ }\textbf {\bibinfo {volume} {87}} (\bibinfo {year}
  {2013}),\ 10.1103/physrevd.87.031103}\BibitemShut {NoStop}%
\bibitem [{\citenamefont {Lees}\ \emph
  {et~al.}(2012{\natexlab{a}})\citenamefont {Lees} \emph
  {et~al.}}]{BaBar:2012iuj}%
  \BibitemOpen
  \bibfield  {author} {\bibinfo {author} {\bibfnamefont {J.~P.}\ \bibnamefont
  {Lees}} \emph {et~al.} (\bibinfo {collaboration} {BaBar}),\ }\href {\doibase
  10.1103/PhysRevD.85.112010} {\bibfield  {journal} {\bibinfo  {journal} {Phys.
  Rev. D}\ }\textbf {\bibinfo {volume} {85}},\ \bibinfo {pages} {112010}
  (\bibinfo {year} {2012}{\natexlab{a}})},\ \Eprint
  {http://arxiv.org/abs/1201.5897} {arXiv:1201.5897 [hep-ex]} \BibitemShut
  {NoStop}%
\bibitem [{\citenamefont {Aubert}\ \emph
  {et~al.}(2008{\natexlab{c}})\citenamefont {Aubert} \emph
  {et~al.}}]{0803.4451}%
  \BibitemOpen
  \bibfield  {author} {\bibinfo {author} {\bibfnamefont {B.}~\bibnamefont
  {Aubert}} \emph {et~al.} (\bibinfo {collaboration} {BaBar}),\ }\href
  {\doibase 10.1103/PhysRevD.78.012004} {\bibfield  {journal} {\bibinfo
  {journal} {Phys. Rev. D}\ }\textbf {\bibinfo {volume} {78}},\ \bibinfo
  {pages} {012004} (\bibinfo {year} {2008}{\natexlab{c}})},\ \Eprint
  {http://arxiv.org/abs/0803.4451} {arXiv:0803.4451 [hep-ex]} \BibitemShut
  {NoStop}%
\bibitem [{\citenamefont {Lees}\ \emph
  {et~al.}(2012{\natexlab{b}})\citenamefont {Lees} \emph {et~al.}}]{1201.5897}%
  \BibitemOpen
  \bibfield  {author} {\bibinfo {author} {\bibfnamefont {J.~P.}\ \bibnamefont
  {Lees}} \emph {et~al.} (\bibinfo {collaboration} {BaBar}),\ }\href {\doibase
  10.1103/PhysRevD.85.112010} {\bibfield  {journal} {\bibinfo  {journal} {Phys.
  Rev. D}\ }\textbf {\bibinfo {volume} {85}},\ \bibinfo {pages} {112010}
  (\bibinfo {year} {2012}{\natexlab{b}})},\ \Eprint
  {http://arxiv.org/abs/1201.5897} {arXiv:1201.5897 [hep-ex]} \BibitemShut
  {NoStop}%
\bibitem [{\citenamefont {Garmash}\ \emph {et~al.}(2005)\citenamefont {Garmash}
  \emph {et~al.}}]{Belle:2004drb}%
  \BibitemOpen
  \bibfield  {author} {\bibinfo {author} {\bibfnamefont {A.}~\bibnamefont
  {Garmash}} \emph {et~al.} (\bibinfo {collaboration} {Belle}),\ }\href
  {\doibase 10.1103/PhysRevD.71.092003} {\bibfield  {journal} {\bibinfo
  {journal} {Phys. Rev. D}\ }\textbf {\bibinfo {volume} {71}},\ \bibinfo
  {pages} {092003} (\bibinfo {year} {2005})},\ \Eprint
  {http://arxiv.org/abs/hep-ex/0412066} {arXiv:hep-ex/0412066} \BibitemShut
  {NoStop}%
\bibitem [{\citenamefont {Jessop}\ \emph {et~al.}(1997)\citenamefont {Jessop}
  \emph {et~al.}}]{CLEO:1997ilq}%
  \BibitemOpen
  \bibfield  {author} {\bibinfo {author} {\bibfnamefont {C.~P.}\ \bibnamefont
  {Jessop}} \emph {et~al.} (\bibinfo {collaboration} {CLEO}),\ }\href {\doibase
  10.1103/PhysRevLett.79.4533} {\bibfield  {journal} {\bibinfo  {journal}
  {Phys. Rev. Lett.}\ }\textbf {\bibinfo {volume} {79}},\ \bibinfo {pages}
  {4533} (\bibinfo {year} {1997})},\ \Eprint
  {http://arxiv.org/abs/hep-ex/9702013} {arXiv:hep-ex/9702013} \BibitemShut
  {NoStop}%
\bibitem [{\citenamefont {Aubert}\ \emph
  {et~al.}(2005{\natexlab{b}})\citenamefont {Aubert} \emph
  {et~al.}}]{BaBar:2005sdl}%
  \BibitemOpen
  \bibfield  {author} {\bibinfo {author} {\bibfnamefont {B.}~\bibnamefont
  {Aubert}} \emph {et~al.} (\bibinfo {collaboration} {BaBar}),\ }\href
  {\doibase 10.1103/PhysRevD.72.051101} {\bibfield  {journal} {\bibinfo
  {journal} {Phys. Rev. D}\ }\textbf {\bibinfo {volume} {72}},\ \bibinfo
  {pages} {051101} (\bibinfo {year} {2005}{\natexlab{b}})},\ \Eprint
  {http://arxiv.org/abs/hep-ex/0507012} {arXiv:hep-ex/0507012} \BibitemShut
  {NoStop}%
\bibitem [{\citenamefont {Choudhury}\ \emph {et~al.}(2021)\citenamefont
  {Choudhury} \emph {et~al.}}]{1908.01848}%
  \BibitemOpen
  \bibfield  {author} {\bibinfo {author} {\bibfnamefont {S.}~\bibnamefont
  {Choudhury}} \emph {et~al.} (\bibinfo {collaboration} {BELLE}),\ }\href
  {\doibase 10.1007/JHEP03(2021)105} {\bibfield  {journal} {\bibinfo  {journal}
  {JHEP}\ }\textbf {\bibinfo {volume} {03}},\ \bibinfo {pages} {105} (\bibinfo
  {year} {2021})},\ \Eprint {http://arxiv.org/abs/1908.01848} {arXiv:1908.01848
  [hep-ex]} \BibitemShut {NoStop}%
\bibitem [{\citenamefont {Chilikin}\ \emph {et~al.}(2019)\citenamefont
  {Chilikin} \emph {et~al.}}]{1903.06414}%
  \BibitemOpen
  \bibfield  {author} {\bibinfo {author} {\bibfnamefont {K.}~\bibnamefont
  {Chilikin}} \emph {et~al.} (\bibinfo {collaboration} {Belle}),\ }\href
  {\doibase 10.1103/PhysRevD.100.012001} {\bibfield  {journal} {\bibinfo
  {journal} {Phys. Rev. D}\ }\textbf {\bibinfo {volume} {100}},\ \bibinfo
  {pages} {012001} (\bibinfo {year} {2019})},\ \Eprint
  {http://arxiv.org/abs/1903.06414} {arXiv:1903.06414 [hep-ex]} \BibitemShut
  {NoStop}%
\bibitem [{\citenamefont {Kato}\ \emph
  {et~al.}(2018{\natexlab{b}})\citenamefont {Kato} \emph
  {et~al.}}]{1709.06108}%
  \BibitemOpen
  \bibfield  {author} {\bibinfo {author} {\bibfnamefont {Y.}~\bibnamefont
  {Kato}} \emph {et~al.} (\bibinfo {collaboration} {Belle}),\ }\href {\doibase
  10.1103/PhysRevD.97.012005} {\bibfield  {journal} {\bibinfo  {journal} {Phys.
  Rev. D}\ }\textbf {\bibinfo {volume} {97}},\ \bibinfo {pages} {012005}
  (\bibinfo {year} {2018}{\natexlab{b}})},\ \Eprint
  {http://arxiv.org/abs/1709.06108} {arXiv:1709.06108 [hep-ex]} \BibitemShut
  {NoStop}%
\bibitem [{\citenamefont {Aubert}\ \emph
  {et~al.}(2006{\natexlab{b}})\citenamefont {Aubert} \emph
  {et~al.}}]{BaBar:2005pcw}%
  \BibitemOpen
  \bibfield  {author} {\bibinfo {author} {\bibfnamefont {B.}~\bibnamefont
  {Aubert}} \emph {et~al.} (\bibinfo {collaboration} {BaBar}),\ }\href
  {\doibase 10.1103/PhysRevLett.96.052002} {\bibfield  {journal} {\bibinfo
  {journal} {Phys. Rev. Lett.}\ }\textbf {\bibinfo {volume} {96}},\ \bibinfo
  {pages} {052002} (\bibinfo {year} {2006}{\natexlab{b}})},\ \Eprint
  {http://arxiv.org/abs/hep-ex/0510070} {arXiv:hep-ex/0510070} \BibitemShut
  {NoStop}%
\bibitem [{\citenamefont {Aubert}\ \emph
  {et~al.}(2005{\natexlab{c}})\citenamefont {Aubert} \emph
  {et~al.}}]{BaBar:2004htr}%
  \BibitemOpen
  \bibfield  {author} {\bibinfo {author} {\bibfnamefont {B.}~\bibnamefont
  {Aubert}} \emph {et~al.} (\bibinfo {collaboration} {BaBar}),\ }\href
  {\doibase 10.1103/PhysRevLett.94.141801} {\bibfield  {journal} {\bibinfo
  {journal} {Phys. Rev. Lett.}\ }\textbf {\bibinfo {volume} {94}},\ \bibinfo
  {pages} {141801} (\bibinfo {year} {2005}{\natexlab{c}})},\ \Eprint
  {http://arxiv.org/abs/hep-ex/0412062} {arXiv:hep-ex/0412062} \BibitemShut
  {NoStop}%
\bibitem [{\citenamefont {Aubert}\ \emph
  {et~al.}(2006{\natexlab{c}})\citenamefont {Aubert} \emph
  {et~al.}}]{BaBar:2006qhm}%
  \BibitemOpen
  \bibfield  {author} {\bibinfo {author} {\bibfnamefont {B.}~\bibnamefont
  {Aubert}} \emph {et~al.} (\bibinfo {collaboration} {BaBar}),\ }\href
  {\doibase 10.1103/PhysRevD.74.051104} {\bibfield  {journal} {\bibinfo
  {journal} {Phys. Rev. D}\ }\textbf {\bibinfo {volume} {74}},\ \bibinfo
  {pages} {051104} (\bibinfo {year} {2006}{\natexlab{c}})},\ \Eprint
  {http://arxiv.org/abs/hep-ex/0607113} {arXiv:hep-ex/0607113} \BibitemShut
  {NoStop}%
\bibitem [{\citenamefont {Aubert}\ \emph
  {et~al.}(2008{\natexlab{d}})\citenamefont {Aubert} \emph
  {et~al.}}]{BaBar:2008qcq}%
  \BibitemOpen
  \bibfield  {author} {\bibinfo {author} {\bibfnamefont {B.}~\bibnamefont
  {Aubert}} \emph {et~al.} (\bibinfo {collaboration} {BaBar}),\ }\href
  {\doibase 10.1103/PhysRevD.78.092002} {\bibfield  {journal} {\bibinfo
  {journal} {Phys. Rev. D}\ }\textbf {\bibinfo {volume} {78}},\ \bibinfo
  {pages} {092002} (\bibinfo {year} {2008}{\natexlab{d}})},\ \Eprint
  {http://arxiv.org/abs/0807.2408} {arXiv:0807.2408 [hep-ex]} \BibitemShut
  {NoStop}%
\bibitem [{\citenamefont {Aaij}\ \emph
  {et~al.}(2018{\natexlab{c}})\citenamefont {Aaij} \emph
  {et~al.}}]{LHCb:2018uli}%
  \BibitemOpen
  \bibfield  {author} {\bibinfo {author} {\bibfnamefont {R.}~\bibnamefont
  {Aaij}} \emph {et~al.} (\bibinfo {collaboration} {LHCb}),\ }\href {\doibase
  10.1007/JHEP05(2018)160} {\bibfield  {journal} {\bibinfo  {journal} {JHEP}\
  }\textbf {\bibinfo {volume} {05}},\ \bibinfo {pages} {160} (\bibinfo {year}
  {2018}{\natexlab{c}})},\ \Eprint {http://arxiv.org/abs/1803.10990}
  {arXiv:1803.10990 [hep-ex]} \BibitemShut {NoStop}%
\bibitem [{\citenamefont {Aaij}\ \emph
  {et~al.}(2018{\natexlab{d}})\citenamefont {Aaij} \emph
  {et~al.}}]{LHCb:2017vtv}%
  \BibitemOpen
  \bibfield  {author} {\bibinfo {author} {\bibfnamefont {R.}~\bibnamefont
  {Aaij}} \emph {et~al.} (\bibinfo {collaboration} {LHCb}),\ }\href {\doibase
  10.1007/JHEP01(2018)131} {\bibfield  {journal} {\bibinfo  {journal} {JHEP}\
  }\textbf {\bibinfo {volume} {01}},\ \bibinfo {pages} {131} (\bibinfo {year}
  {2018}{\natexlab{d}})},\ \Eprint {http://arxiv.org/abs/1711.05637}
  {arXiv:1711.05637 [hep-ex]} \BibitemShut {NoStop}%
\bibitem [{\citenamefont {Aubert}\ \emph
  {et~al.}(2006{\natexlab{d}})\citenamefont {Aubert} \emph
  {et~al.}}]{BaBar:2006uih}%
  \BibitemOpen
  \bibfield  {author} {\bibinfo {author} {\bibfnamefont {B.}~\bibnamefont
  {Aubert}} \emph {et~al.} (\bibinfo {collaboration} {BaBar}),\ }\href
  {\doibase 10.1103/PhysRevD.73.112004} {\bibfield  {journal} {\bibinfo
  {journal} {Phys. Rev. D}\ }\textbf {\bibinfo {volume} {73}},\ \bibinfo
  {pages} {112004} (\bibinfo {year} {2006}{\natexlab{d}})},\ \Eprint
  {http://arxiv.org/abs/hep-ex/0604037} {arXiv:hep-ex/0604037} \BibitemShut
  {NoStop}%
\bibitem [{\citenamefont {Adachi}\ \emph {et~al.}(2008)\citenamefont {Adachi}
  \emph {et~al.}}]{Belle:2008doh}%
  \BibitemOpen
  \bibfield  {author} {\bibinfo {author} {\bibfnamefont {I.}~\bibnamefont
  {Adachi}} \emph {et~al.} (\bibinfo {collaboration} {Belle}),\ }\href
  {\doibase 10.1103/PhysRevD.77.091101} {\bibfield  {journal} {\bibinfo
  {journal} {Phys. Rev. D}\ }\textbf {\bibinfo {volume} {77}},\ \bibinfo
  {pages} {091101} (\bibinfo {year} {2008})},\ \Eprint
  {http://arxiv.org/abs/0802.2988} {arXiv:0802.2988 [hep-ex]} \BibitemShut
  {NoStop}%
\bibitem [{\citenamefont {Aubert}\ \emph
  {et~al.}(2009{\natexlab{c}})\citenamefont {Aubert} \emph
  {et~al.}}]{BaBar:2008flx}%
  \BibitemOpen
  \bibfield  {author} {\bibinfo {author} {\bibfnamefont {B.}~\bibnamefont
  {Aubert}} \emph {et~al.} (\bibinfo {collaboration} {BaBar}),\ }\href
  {\doibase 10.1103/PhysRevLett.102.132001} {\bibfield  {journal} {\bibinfo
  {journal} {Phys. Rev. Lett.}\ }\textbf {\bibinfo {volume} {102}},\ \bibinfo
  {pages} {132001} (\bibinfo {year} {2009}{\natexlab{c}})},\ \Eprint
  {http://arxiv.org/abs/0809.0042} {arXiv:0809.0042 [hep-ex]} \BibitemShut
  {NoStop}%
\bibitem [{\citenamefont {Soni}\ \emph {et~al.}(2006)\citenamefont {Soni} \emph
  {et~al.}}]{Belle:2005eoz}%
  \BibitemOpen
  \bibfield  {author} {\bibinfo {author} {\bibfnamefont {N.}~\bibnamefont
  {Soni}} \emph {et~al.} (\bibinfo {collaboration} {Belle}),\ }\href {\doibase
  10.1016/j.physletb.2006.01.013} {\bibfield  {journal} {\bibinfo  {journal}
  {Phys. Lett. B}\ }\textbf {\bibinfo {volume} {634}},\ \bibinfo {pages} {155}
  (\bibinfo {year} {2006})},\ \Eprint {http://arxiv.org/abs/hep-ex/0508032}
  {arXiv:hep-ex/0508032} \BibitemShut {NoStop}%
\bibitem [{\citenamefont {Richichi}\ \emph {et~al.}(2001)\citenamefont
  {Richichi} \emph {et~al.}}]{CLEO:2000ped}%
  \BibitemOpen
  \bibfield  {author} {\bibinfo {author} {\bibfnamefont {S.~J.}\ \bibnamefont
  {Richichi}} \emph {et~al.} (\bibinfo {collaboration} {CLEO}),\ }\href
  {\doibase 10.1103/PhysRevD.63.031103} {\bibfield  {journal} {\bibinfo
  {journal} {Phys. Rev. D}\ }\textbf {\bibinfo {volume} {63}},\ \bibinfo
  {pages} {031103} (\bibinfo {year} {2001})},\ \Eprint
  {http://arxiv.org/abs/hep-ex/0010036} {arXiv:hep-ex/0010036} \BibitemShut
  {NoStop}%
\bibitem [{\citenamefont {Aaij}\ \emph
  {et~al.}(2017{\natexlab{b}})\citenamefont {Aaij} \emph
  {et~al.}}]{LHCb:2017egy}%
  \BibitemOpen
  \bibfield  {author} {\bibinfo {author} {\bibfnamefont {R.}~\bibnamefont
  {Aaij}} \emph {et~al.} (\bibinfo {collaboration} {LHCb}),\ }\href {\doibase
  10.1007/JHEP11(2017)156} {\bibfield  {journal} {\bibinfo  {journal} {JHEP}\
  }\textbf {\bibinfo {volume} {11}},\ \bibinfo {pages} {156} (\bibinfo {year}
  {2017}{\natexlab{b}})},\ \bibinfo {note} {[Erratum: JHEP 05, 067 (2018)]},\
  \Eprint {http://arxiv.org/abs/1709.05855} {arXiv:1709.05855 [hep-ex]}
  \BibitemShut {NoStop}%
\bibitem [{\citenamefont {Aubert}\ \emph
  {et~al.}(2009{\natexlab{d}})\citenamefont {Aubert} \emph
  {et~al.}}]{BaBar:2009dzx}%
  \BibitemOpen
  \bibfield  {author} {\bibinfo {author} {\bibfnamefont {B.}~\bibnamefont
  {Aubert}} \emph {et~al.} (\bibinfo {collaboration} {BaBar}),\ }\href
  {\doibase 10.1103/PhysRevD.80.092001} {\bibfield  {journal} {\bibinfo
  {journal} {Phys. Rev. D}\ }\textbf {\bibinfo {volume} {80}},\ \bibinfo
  {pages} {092001} (\bibinfo {year} {2009}{\natexlab{d}})},\ \Eprint
  {http://arxiv.org/abs/0909.3981} {arXiv:0909.3981 [hep-ex]} \BibitemShut
  {NoStop}%
\bibitem [{\citenamefont {del Amo~Sanchez}\ \emph
  {et~al.}(2010{\natexlab{c}})\citenamefont {del Amo~Sanchez} \emph
  {et~al.}}]{BaBar:2010cpi}%
  \BibitemOpen
  \bibfield  {author} {\bibinfo {author} {\bibfnamefont {P.}~\bibnamefont {del
  Amo~Sanchez}} \emph {et~al.} (\bibinfo {collaboration} {BaBar}),\ }\href
  {\doibase 10.1103/PhysRevD.82.011502} {\bibfield  {journal} {\bibinfo
  {journal} {Phys. Rev. D}\ }\textbf {\bibinfo {volume} {82}},\ \bibinfo
  {pages} {011502} (\bibinfo {year} {2010}{\natexlab{c}})},\ \Eprint
  {http://arxiv.org/abs/1004.0240} {arXiv:1004.0240 [hep-ex]} \BibitemShut
  {NoStop}%
\bibitem [{\citenamefont {Aubert}\ \emph
  {et~al.}(2009{\natexlab{e}})\citenamefont {Aubert} \emph
  {et~al.}}]{BaBar:2009mcf}%
  \BibitemOpen
  \bibfield  {author} {\bibinfo {author} {\bibfnamefont {B.}~\bibnamefont
  {Aubert}} \emph {et~al.} (\bibinfo {collaboration} {BaBar}),\ }\href
  {\doibase 10.1103/PhysRevD.79.052005} {\bibfield  {journal} {\bibinfo
  {journal} {Phys. Rev. D}\ }\textbf {\bibinfo {volume} {79}},\ \bibinfo
  {pages} {052005} (\bibinfo {year} {2009}{\natexlab{e}})},\ \Eprint
  {http://arxiv.org/abs/0901.3703} {arXiv:0901.3703 [hep-ex]} \BibitemShut
  {NoStop}%
\bibitem [{\citenamefont {Lees}\ \emph {et~al.}(2011)\citenamefont {Lees} \emph
  {et~al.}}]{BaBar:2011cmh}%
  \BibitemOpen
  \bibfield  {author} {\bibinfo {author} {\bibfnamefont {J.~P.}\ \bibnamefont
  {Lees}} \emph {et~al.} (\bibinfo {collaboration} {BaBar}),\ }\href {\doibase
  10.1103/PhysRevD.84.092007} {\bibfield  {journal} {\bibinfo  {journal} {Phys.
  Rev. D}\ }\textbf {\bibinfo {volume} {84}},\ \bibinfo {pages} {092007}
  (\bibinfo {year} {2011})},\ \Eprint {http://arxiv.org/abs/1109.0143}
  {arXiv:1109.0143 [hep-ex]} \BibitemShut {NoStop}%
\bibitem [{\citenamefont {Aubert}\ \emph
  {et~al.}(2006{\natexlab{e}})\citenamefont {Aubert} \emph
  {et~al.}}]{BaBar:2006kjz}%
  \BibitemOpen
  \bibfield  {author} {\bibinfo {author} {\bibfnamefont {B.}~\bibnamefont
  {Aubert}} \emph {et~al.} (\bibinfo {collaboration} {BaBar}),\ }\href
  {\doibase 10.1103/PhysRevD.73.111104} {\bibfield  {journal} {\bibinfo
  {journal} {Phys. Rev. D}\ }\textbf {\bibinfo {volume} {73}},\ \bibinfo
  {pages} {111104} (\bibinfo {year} {2006}{\natexlab{e}})},\ \Eprint
  {http://arxiv.org/abs/hep-ex/0604017} {arXiv:hep-ex/0604017} \BibitemShut
  {NoStop}%
\bibitem [{\citenamefont {Mahapatra}\ \emph {et~al.}(2002)\citenamefont
  {Mahapatra} \emph {et~al.}}]{CLEO:2001wbc}%
  \BibitemOpen
  \bibfield  {author} {\bibinfo {author} {\bibfnamefont {R.}~\bibnamefont
  {Mahapatra}} \emph {et~al.} (\bibinfo {collaboration} {CLEO}),\ }\href
  {\doibase 10.1103/PhysRevLett.88.101803} {\bibfield  {journal} {\bibinfo
  {journal} {Phys. Rev. Lett.}\ }\textbf {\bibinfo {volume} {88}},\ \bibinfo
  {pages} {101803} (\bibinfo {year} {2002})},\ \Eprint
  {http://arxiv.org/abs/hep-ex/0112033} {arXiv:hep-ex/0112033} \BibitemShut
  {NoStop}%
\bibitem [{\citenamefont {Aubert}\ \emph
  {et~al.}(2006{\natexlab{f}})\citenamefont {Aubert} \emph
  {et~al.}}]{BaBar:2006ltn}%
  \BibitemOpen
  \bibfield  {author} {\bibinfo {author} {\bibfnamefont {B.}~\bibnamefont
  {Aubert}} \emph {et~al.} (\bibinfo {collaboration} {BaBar}),\ }\href
  {\doibase 10.1103/PhysRevLett.97.201802} {\bibfield  {journal} {\bibinfo
  {journal} {Phys. Rev. Lett.}\ }\textbf {\bibinfo {volume} {97}},\ \bibinfo
  {pages} {201802} (\bibinfo {year} {2006}{\natexlab{f}})},\ \Eprint
  {http://arxiv.org/abs/hep-ex/0608005} {arXiv:hep-ex/0608005} \BibitemShut
  {NoStop}%
\bibitem [{\citenamefont {Abe}\ \emph {et~al.}(2002)\citenamefont {Abe} \emph
  {et~al.}}]{Belle:2002otd}%
  \BibitemOpen
  \bibfield  {author} {\bibinfo {author} {\bibfnamefont {K.}~\bibnamefont
  {Abe}} \emph {et~al.} (\bibinfo {collaboration} {Belle}),\ }\href {\doibase
  10.1016/S0370-2693(02)01969-X} {\bibfield  {journal} {\bibinfo  {journal}
  {Phys. Lett. B}\ }\textbf {\bibinfo {volume} {538}},\ \bibinfo {pages} {11}
  (\bibinfo {year} {2002})},\ \Eprint {http://arxiv.org/abs/hep-ex/0205021}
  {arXiv:hep-ex/0205021} \BibitemShut {NoStop}%
\bibitem [{\citenamefont {Aubert}\ \emph
  {et~al.}(2007{\natexlab{c}})\citenamefont {Aubert} \emph
  {et~al.}}]{BaBar:2007esv}%
  \BibitemOpen
  \bibfield  {author} {\bibinfo {author} {\bibfnamefont {B.}~\bibnamefont
  {Aubert}} \emph {et~al.} (\bibinfo {collaboration} {BaBar}),\ }\href
  {\doibase 10.1103/PhysRevD.76.092004} {\bibfield  {journal} {\bibinfo
  {journal} {Phys. Rev. D}\ }\textbf {\bibinfo {volume} {76}},\ \bibinfo
  {pages} {092004} (\bibinfo {year} {2007}{\natexlab{c}})},\ \Eprint
  {http://arxiv.org/abs/0707.1648} {arXiv:0707.1648 [hep-ex]} \BibitemShut
  {NoStop}%
\end{thebibliography}%

\end{document}